\documentclass[english]{JHEP3}
\usepackage[T1]{fontenc}
\usepackage[latin1]{inputenc}
\usepackage{float}
\usepackage{textcomp}
\usepackage{amsmath}
\usepackage{graphicx}
\usepackage{amssymb}

\makeatletter

\newcommand{\lyxdot}{.}

\usepackage{float}

\makeatletter

\makeatletter

\newcommand{\HWPP}{\textsf{Herwig++}}

\usepackage{cite}

\usepackage[mathscr]{euscript}

\title{A Positive-Weight Next-to-Leading Order Monte Carlo Simulation for Higgs Boson Production}
\author{Keith Hamilton \\ Centre for Particle Physics and Phenomenology (CP3), \\ Universit\'{e} Catholique de Louvain, \\ Chemin du Cyclotron 2, 1348 Louvain-la-Neuve, Belgium \\ Email: \email{keith.hamilton@uclouvain.be}}
\author{Peter Richardson\\
   Institute of Particle Physics Phenomenology, Department of Physics \\ University of Durham,  Durham, DH1 3LE, UK \\Email: \email{peter.richardson@durham.ac.uk}}
\author{Jon Tully \\ Institute of Particle Physics Phenomenology, Department of Physics \\ University of Durham,  Durham, DH1 3LE, UK \\ Email: \email{j.m.tully@durham.ac.uk}} 
\preprint{ HERWIG/09/01 \\ MCnet/09/08 \\ IPPP/09/20 \\ CPT/09/40 \\ CP3-09-09 }

\abstract{
In this article we describe simulations of Higgs boson production via the gluon fusion and Higgs-strahlung processes, using  the positive weight next-to-leading-order\,(NLO) matching scheme, {\sf POWHEG}, in the {\sf Herwig++ 2.3} event generator. This formalism consistently incorporates the full NLO corrections to these processes within the parton shower simulation, without the production of negative weight events. 
These simulations include a full implementation of the truncated shower required
to correctly model soft emissions in an angular-ordered parton shower. We present a thorough validation of these simulations, comparing them to other methods and calculations. The results obtained for the gluon fusion process corroborate, and provide detailed explanations for, findings reported by Alioli \emph{et al} using an independent {\sf POWHEG} simulation, neglecting truncated shower effects, released at the same time as our code.}

\maketitle

\def\ds#1{\setbox0=\hbox{$#1$}#1\hskip-\wd0\hbox to\wd0{\hss\sl/\/\hss}}
\def\slashb#1{\setbox0=\hbox{$#1$}#1\hskip-\wd0\dimen0=5pt\advance
       \dimen0 by-\ht0\advance\dimen0 by\dp0\lower0.5\dimen0\hbox
         to\wd0{\hss\sl/\/\hss}}

\newcommand{\splusminus}{{\mathchoice%
{\vplusminus\displaystyle}%
{\vplusminus\scriptstyle}%
{\vplusminus\scriptscriptstyle}%
{\vplusminus\scriptscriptstyle}%
}}
\newcommand{\sminusplus}{{\mathchoice%
{\vminusplus\displaystyle}%
{\vminusplus\scriptstyle}%
{\vminusplus\scriptscriptstyle}%
{\vminusplus\scriptscriptstyle}%
}}
\newdimen\hbigcirc
\newdimen\wbigcirc
\newcommand{\vplusminus}[1]{%
\settoheight{\hbigcirc}{$#1\bigcirc$}%
\settowidth{\wbigcirc}{$#1\bigcirc$}%
\makebox[\wbigcirc]{%
\makebox[0pt]{\rule[0.4\hbigcirc]{0.5\wbigcirc}{0.05\hbigcirc}}%
\makebox[0pt]{\rule[0.1\hbigcirc]{0.5\wbigcirc}{0.05\hbigcirc}}%
\makebox[0pt]{\rule[0.1\hbigcirc]{0.05\wbigcirc}{0.6\hbigcirc}}%
\makebox[0pt]{$#1\bigcirc$}}%
}
\newcommand{\vminusplus}[1]{%
\settoheight{\hbigcirc}{$#1\bigcirc$}%
\settowidth{\wbigcirc}{$#1\bigcirc$}%
\makebox[\wbigcirc]{%
\makebox[0pt]{\rule[0.2\hbigcirc]{0.5\wbigcirc}{0.05\hbigcirc}}%
\makebox[0pt]{\rule[0.5\hbigcirc]{0.5\wbigcirc}{0.05\hbigcirc}}%
\makebox[0pt]{\rule[-0.1\hbigcirc]{0.05\wbigcirc}{0.6\hbigcirc}}%
\makebox[0pt]{$#1\bigcirc$}}%
}
\makeatother

\makeatother

\makeatother

\makeatother

\makeatother

\makeatother

\usepackage{babel}

\begin{document}

\section{Introduction}

The primary objective of the ongoing and imminent physics programmes
at the Tevatron and LHC, is to elucidate the nature of electroweak
symmetry breaking. The great majority of the effort in this direction
is devoted to the hunt for the Higgs boson, the origin of this symmetry
breaking in the Standard Model (SM) \cite{Englert:1964et,Higgs:1964pj,Guralnik:1964eu,Kibble:1967sv}.

Of all the ways in which the SM Higgs boson can be produced, the gluon fusion 
process \cite{Georgi:1977gs}, in which it couples to colliding gluons via a 
top quark loop, has the largest cross section for Higgs boson masses less than 
$\sim$1 TeV. This process is of great importance for the detection of the Higgs 
boson at the Tevatron and, more so, at the LHC, particularly in the low mass 
region, favoured by the latest fits of the Standard Model to electroweak 
precision data \cite{EWWG:2008}, where the decay of the Higgs boson into two 
photons is known to give a clean experimental signal. Although observing a 
narrow resonance in the diphoton invariant mass spectrum should be possible 
using only the experimental data \cite{Djouadi:2005gi}, determining the quantum 
numbers and couplings of the resonance \emph{i.e.} determining that it really 
is a fundamental scalar and, moreover, the SM Higgs boson, will involve a 
comprehensive analysis of a number of channels, using accurate, flexible, Monte 
Carlo simulations to predict distributions for signals and backgrounds.

A key part of that identification procedure will be the measurement 
the couplings of the Higgs boson to the weak gauge bosons and the top quark. 
Although the gluon fusion process directly probes the latter, associated QCD 
radiation renders it a significant background to the vector boson fusion process 
(VBF), in which the Higgs boson originates from $HWW$ and $HZZ$ vertices. 
Precise simulations of the gluon fusion process are then also required to model 
the extent to which these events contaminate the VBF 
signal~\cite{Andersen:2008gc,Campbell:2006xx,DelDuca:2006hk}.

Another direct probe of Higgs boson interactions with electroweak gauge bosons
is the \emph{Higgs-strahlung} process~\cite{Glashow:1978ab,Eichten:1984eu}. At 
leading order this consists of a quark anti-quark annihilation producing a 
virtual vector boson $\left(V=W/Z\right)$, which becomes on-shell by radiating 
a Higgs boson. This channel is particularly important for Higgs boson searches at the 
Tevatron~\cite{Abazov:2008eb}. The cross section for Higgs-strahlung processes 
at the LHC is approximately one order of magnitude larger than at the Tevatron, 
however, the backgrounds scale up by a much greater factor. In spite of these 
difficulties the Higgs-strahlung processes can still be observed, notably in 
the $\gamma\gamma$ and $W^{+}W^{-}$ decay channels, where significant background 
rejection can be achieved~(see for example Ref.~\cite{Ball:2007zza} for recent 
experimental studies). Interest in the $b\bar{b}$ plus leptons decay channel has 
also recently been revived through the observation that a highly collimated 
$b\bar{b}$ jet, from a boosted Higgs boson, could provide a clean signature 
\cite{Butterworth:2008iy,Butterworth:2008sd}. These processes were recommended 
for inclusion in studies to determine Higgs boson couplings in 
Ref.~\cite{Zeppenfeld:2000td}. As with the gluon fusion process, accurate Monte 
Carlo simulations will be central to such studies.

Lately combined Tevatron analyses of the gluon fusion and Higgs-strahlung 
channels have begun to show sensitivity to Higgs boson 
production~\cite{Phenomena:2009pt}. Presently these studies exclude, at the 
95\% confidence level, the existence of a SM Higgs boson with a mass in the 
range 160 - 170 GeV. This exclusion limit largely follows from analysis of the 
gluon fusion channel in which the Higgs decays into a $W^{+}W^{-}$ pair. Monte 
Carlo simulations of Higgs-strahlung and gluon fusion processes were essential in obtaining these measurements.

In recent years research in Monte Carlo simulations has seen major
progress, most significantly in the extension of existing parton shower
simulations to consistently include exact next-to-leading order corrections~\cite{Frixione:2002ik,Frixione:2003ei,Frixione:2005vw,Frixione:2006gn,Frixione:2007zp,Frixione:2008yi,LatundeDada:2007jg,Nason:2004rx,Nason:2006hfa,Frixione:2007nu,Frixione:2007vw,Frixione:2007nw,LatundeDada:2006gx,Hamilton:2008pd,Alioli:2008gx,LatundeDada:2008bv}
and, separately, in the consistent combination of parton shower simulations
and high multiplicity tree-level matrix element generators~\cite{Catani:2001cc,Krauss:2002up,Schalicke:2005nv,Lonnblad:2001iq,Mangano:2001xp,Mrenna:2003if}.
The first successful NLO matching scheme was \textsf{MC@NLO}~\cite{Frixione:2002ik,Frixione:2003ei,Frixione:2005vw,Frixione:2006gn,Frixione:2007zp,Frixione:2008yi}
which has been realised using the \textsf{HERWIG} event generator
for many processes. This method has two draw backs; first, it involves
subtracting the parton shower approximation from the NLO calculation,
which can lead to unphysical negative weight events, second, the implementation
of the method is fundamentally dependent on the details of the parton
shower algorithm. In 2004 a new formalism known as \textsf{POWHEG}
(POsitive Weight Hardest Emission Generator \cite{Nason:2004rx})
was derived, achieving the same aims as \textsf{MC@NLO} but with the
added benefits of only generating physical, positive weight, events
as they occur in nature, and of being independent of the details of
the parton shower algorithm. This method has been already successfully
applied to a number of phenomenologically important processes \cite{Nason:2006hfa,LatundeDada:2006gx,Frixione:2007nw,LatundeDada:2007jg,Alioli:2008gx,Hamilton:2008pd,Alioli:2008tz,LatundeDada:2008bv,Papaefstathiou:2009sr}.

In this paper we describe the application of the \textsf{POWHEG} method
to Monte Carlo simulations for Higgs boson production via gluon fusion
and Higgs-strahlung processes, within the \HWPP\ \cite{Bahr:2008tx,Bahr:2008pv}
event generator. We include a complete description of truncated shower
effects.%
\footnote{Currently, only the\textsf{ Herwig++} \textsf{POWHEG} simulation of
the Drell-Yan process includes a complete treatment of truncated shower
effects \cite{Hamilton:2008pd}.%
} Our primary aim is to present the ingredients used in the simulations
and to validate them, where appropriate, against existing calculations.

The structure of the paper is as follows. In Sect.~\ref{sec:The-POWHEG-method}
we briefly review the main features of the \textsf{POWHEG} method.
In Sect.~\ref{sec:Next-to-leading-order-cross} we collect the essential
formulae relating to the NLO cross sections, for implementation in
the program. In Section~\ref{sec:Implementation} we give details
of the event generation process for the hard configurations. This
is accompanied by a description of how the hard configurations are
subsequently reproduced by the angular-ordered parton shower, including
truncated showering effects, in respect of the colour coherence phenomenon
manifest in soft wide angle gluon emissions. In Sect.~\ref{sec:Results}
we present the results of our implementation, comparing it to the
MCFM and \textsf{MC@NLO} generators, before summarizing our findings
in Sect.~\ref{sec:Conclusion}.

\section{The \textsf{POWHEG} formalism\label{sec:The-POWHEG-method}}

The central formula in the \textsf{POWHEG} approach is derived by
manipulating and modifying, through the inclusion of formally higher
order terms, the NLO differential cross section, such that it has
the same form as that given by the parton shower~\cite{Nason:2004rx}.
For a given \emph{N}-body process this is \begin{equation}
\mathrm{d}\sigma=\overline{B}\left(\Phi_{B}\right)\,\mathrm{d}\Phi_{B}\,\left[\Delta_{\hat{R}}\left(0\right)+\frac{\widehat{R}\left(\Phi_{B},\Phi_{R}\right)}{B\left(\Phi_{B}\right)}\,\Delta_{\hat{R}}\left(k_{T}\left(\Phi_{B},\Phi_{R}\right)\right)\,\mathrm{d}\Phi_{R}\right],\label{eq:powheg_1}\end{equation}
 where $\overline{B}\left(\Phi_{B}\right)$ is defined as \begin{equation}
\overline{B}\left(\Phi_{B}\right)=B\left(\Phi_{B}\right)+V\left(\Phi_{B}\right)+\int\,\mathrm{d}\Phi_{R}\,\left[\widehat{R}\left(\Phi_{B},\Phi_{R}\right)-\sum_{i}\, C_{i}\left(\Phi_{B},\Phi_{R}\right)\right]\,,\label{eq:powheg_2}\end{equation}
 with $B\left(\Phi_{B}\right)$ the leading-order contribution, dependent
on the \emph{Born variables}, $\Phi_{B}$, which specify a point in
the \emph{N-}body phase space. $V\left(\Phi_{B}\right)$ is a finite
contribution arising from the combination of unresolvable, real emission
and virtual loop corrections. The remaining terms are due to the \emph{N}+1-body
real emission processes, hence they have an additional dependence
on the \emph{radiative variables}, $\Phi_{R}$, which parametrize
the phase space associated with the extra parton. The real emission
term, $\widehat{R}\left(\Phi_{B},\Phi_{R}\right)$, is given by the
product of the parton flux factors with the relevant squared real
emission matrix element, summed over each channel contributing to
the NLO cross section. $C_{i}\left(\Phi_{B},\Phi_{R}\right)$ denotes
a combination of \emph{real counterterms}/\emph{counter-event weights},
regulating the singularities in $\widehat{R}\left(\Phi_{B},\Phi_{R}\right)$.
Finally, the \textsf{POWHEG} Sudakov form factor, $\Delta_{\hat{R}}$,
is defined as´\begin{equation}
\Delta_{\hat{R}}\left(p_{T}\right)=\exp\left[-\int\mathrm{d}\Phi_{R}\,\frac{\widehat{R}\left(\Phi_{B},\Phi_{R}\right)}{B\left(\Phi_{B}\right)}\,\theta\left(k_{T}\left(\Phi_{B},\Phi_{R}\right)-p_{T}\right)\right],\label{eq:powheg_3}\end{equation}
 where $k_{T}\left(\Phi_{B},\Phi_{R}\right)$ tends to the transverse
momentum of the emitted parton in the soft and collinear limits.

To $\mathcal{O}\left(\alpha_{S}\right)$ Eq.\,\ref{eq:powheg_1}
is just the usual NLO differential cross section. The analogous parton
shower expression is given by replacing $\overline{B}\left(\Phi_{B}\right)\rightarrow B$$\left(\Phi_{B}\right)$
and the real emission corrections $\widehat{R}\left(\Phi_{B},\Phi_{R}\right)$
by their collinear approximation, \emph{i.e.} replacing the argument
of the Sudakov form factor by a sum of Altarelli-Parisi kernels. Typically
parton shower simulations generate an \emph{N}-body configuration
according to $B\left(\Phi_{B}\right)$ and then shower it using such
a Sudakov form factor. In the \textsf{POWHEG} formalism the initial
\emph{N}-body configuration is instead generated according to $\overline{B}\left(\Phi_{B}\right)$,
and retained with probability $\Delta_{\hat{R}}\left(0\right)$ as
a \emph{non-radiative} event, or, showered to give the hardest emission
with $p_{T}=k_{T}\left(\Phi_{B},\Phi_{R}\right)$, with probability
$\Delta_{\hat{R}}\left(p_{T}\right)$ (Eq.\,\ref{eq:powheg_1}).
Further, lower $p_{T}$, emissions represent terms of next-to-next-to-leading-order~(NNLO)
and beyond, hence we can return to the usual parton shower formulation
to simulate these. Provided the perturbative approach is valid, \emph{i.e.}
provided the next-to-leading order terms are smaller than the leading
order ones, it is clear from Eq.\,\ref{eq:powheg_2} that $\overline{B}\left(\Phi_{B}\right)$
is positive, therefore no negative weights arise.

Furthermore, it is well known that when a bunch of collimated QCD
charges emit a gluon at wide angle, the intensity of the radiation
is proportional to the \emph{coherent} \emph{sum} of emissions from
the constituents, \emph{viz.} the jet parent. This effect is manifest
in the perturbative series as large soft logarithms. The other major
triumph of the \textsf{POWHEG}~\cite{Nason:2004rx} approach, besides
avoiding negative weights, is in rigorously decomposing the angular-ordered
parton shower into a \emph{truncated shower}, describing soft wide
angle radiation, the \emph{hardest emission}, as described above,
and a further \emph{vetoed shower} comprised of lower $p_{T}$, increasingly
collinear emissions.

We implement the \textsf{POWHEG} formalism in fullness according to
the following procedure:
\begin{itemize}
\item generate \emph{N}- and \emph{N}+1-body configurations according to
Eq.\,\ref{eq:powheg_1}; 
\item directly hadronize any \emph{N}-body, non-radiative, events; 
\item map the radiative variables parametrizing the emission into the evolution
scale, momentum fraction and azimuthal angle $\left(\tilde{q}_{h},\, z_{h},\,\phi_{h}\right)$,
from which the parton shower would reconstruct identical momenta; 
\item take the initial \emph{N}-body configuration, $\Phi_{B}$, generated
from $\overline{B}\left(\Phi_{B}\right)$, and evolve the emitting
leg from the default initial scale down to $\tilde{q}_{h}$ using
the truncated shower; 
\item insert a branching with parameters~\mbox{$\left(\tilde{q}_{h},\, z_{h},\,\phi_{h}\right)$}
into the shower when the evolution scale reaches $\tilde{q}_{h}$; 
\item generate $p_{T}$ vetoed showers from all external legs. 
\end{itemize}

\section{Next-to-Leading Order Cross Sections\label{sec:Next-to-leading-order-cross}}

The NLO ingredients needed to implement the \textsf{POWHEG} method
for Higgs-strahlung processes can be performed in the same way as
Ref.\,\cite{Hamilton:2008pd} due to the careful organization of
the Drell-Yan differential cross section in our earlier work. There
we explicitly factorized the real emission corrections to the leading-order
$q+\bar{q}\rightarrow l+\bar{l}$ process at the level of the phase
space \emph{and} the real emission matrix elements. This meant that
the radiative variables could be generated completely independently
from the details of the decay of the off-shell vector boson. We refer
the reader to Ref.\,\cite{Hamilton:2008pd} for details of those
matrix elements.

Since the diagrams involved in the NLO corrections to Drell-Yan and
Higgs-strahlung processes are identical up to replacing the final-state
lepton pair with a vector boson and a Higgs boson, the factorized
NLO differential cross section is exactly the same as that in Ref.\,\cite{Hamilton:2008pd}
for the Drell-Yan process; one simply replaces the $q+\bar{q}\rightarrow l+\bar{l}$
leading-order matrix element with that for $q+\bar{q}\rightarrow V+H$
throughout. No information regarding spin correlations is lost in
this procedure and the full NLO distribution is generated without
approximation. This method is originally due to Kleiss \cite{Kleiss:1986re,Seymour:1994we}
and was extended for use in our \textsf{POWHEG} simulation, with a
view to making other processes easier to implement. Due to the complexity
of the 4-body final state (we include the decays of $V$ and $H$)
and the subtleties of the Monte Carlo algorithm, we have carried out
detailed comparisons against the parton level NLO simulation MCFM
\cite{Campbell:2000bg}. Results of these comparisons are given later
in Sect.\,\ref{sec:Results}.

The NLO cross section for Higgs boson production via gluon fusion
was first calculated in Refs.\,\cite{Dawson:1990zj,Djouadi:1991tka},
in the infinite top-quark mass limit. Later this calculation was improved
to exactly include the effects of the finite top mass in \cite{Spira:1995rr},
where it was found that the infinite top-quark mass limit provided
an excellent approximation to the full result; below the $t\bar{t}$
threshold the exact and approximate calculations agree at the level
of 10\%~\ \cite{Djouadi:2005gi}. In the last decade the NNLO corrections 
were also fully computed by four independent groups, in the infinite top-quark 
mass limit \cite{Catani:2001ic,Catani:2001cr,Harlander:2002wh,Anastasiou:2002yz,Ravindran:2003um}. 
More recently a parton level Monte Carlo program accurate to NNLO has become 
available \cite{Catani:2007vq,Grazzini:2008tf}.
A key point arising from this theoretical activity is that the dominant
perturbative corrections to the total cross section originate from
virtual and soft-gluon corrections, which explains the surprising
accuracy of the infinite top-quark mass approximation for $m_{H}\le2m_{t}$
\cite{Kramer:1996iq}.

In our simulation we use the infinite top-quark mass limit. Although
the NLO formulae in \cite{Dawson:1990zj,Djouadi:1991tka} are helpful,
they were only used to compute the most inclusive of measurements,
namely the total cross section. Further work is needed to obtain a
form suitable for a fully differential Monte Carlo simulation. In
this section we collect the ingredients that arise in the NLO calculation
for $g+g\rightarrow H$, necessary for the implementation of the \textsf{POWHEG}
method.

\subsection{Kinematics and phase space\label{sub:Kinematics-and-phase}}

Here we restrict ourselves to considering leading-order processes
of the type,\linebreak\mbox{$\bar{p}_{\oplus}+\bar{p}_{\ominus}\rightarrow\bar{p}_{1}+...+\bar{p}_{N}$},
in which all the particles in the \emph{N}-body final state are either
massive or colourless. For such processes the NLO corrections may
contain soft singularities and initial-state collinear singularities
only. The incoming hadron momenta are labeled $P_{\splusminus}$,
for hadrons incident from the $\pm z$ directions, respectively. It
therefore follows that the momenta of the colliding, massless partons,
with momentum fractions $\bar{x}_{\oplus}$ and $\bar{x}_{\ominus}$,
are given by $\bar{p}_{\splusminus}=\bar{x}_{\splusminus}P_{\splusminus}$.
The momentum of the $i$th final-state parton produced in the leading-order
process is denoted $\bar{p}_{i}$.

Using $\bar{p}$ to represent the sum of all $\bar{p}_{i}$, and $\bar{\mathrm{y}}$
to signify the rapidity of $\bar{p}$, the phase space for the leading-order
process can be written simply as \begin{equation}
\mathrm{d}\Phi_{B}\mbox{ }=\mbox{ }\mathrm{d}\bar{x}_{\oplus}\,\mathrm{d}\bar{x}_{\ominus}\,\mathrm{d}\hat{\Phi}_{B}\mbox{ }=\mbox{ }\frac{1}{s}\,\mathrm{d}\bar{p}^{2}\,\mathrm{d}\bar{\mathrm{y}}\,\mathrm{d}\hat{\Phi}_{B},\label{eq:nlo_1_2}\end{equation}
 where $\mathrm{d}\hat{\Phi}_{B}$ is the Lorentz invariant phase
space for the partonic $2\rightarrow N$ process and $s$ the hadronic
centre-of-mass energy. We refer to the variables $\Phi_{B}=\left\{ \bar{p}^{2},\,\bar{\mathrm{y}},\,\hat{\Phi}_{B}\right\} $
as the \emph{Born variables}, where $\hat{\Phi}_{B}$ parametrizes
the \emph{N}-body phase space in the partonic centre-of-mass frame;
for the simple case of a decaying scalar, like the Higgs boson, the
matrix element does not depend on $\hat{\Phi}_{B}$, which could therefore
be trivially integrated out at this point.

The NLO real emission corrections to the leading-order process consist
of~$2\rightarrow N+1$ processes, $p_{\oplus}+p_{\ominus}\rightarrow p_{1}+...+p_{N}+k$,
where we denote the momenta of the same \emph{N} particles produced
in the leading-order process $p_{i}$ and that of the extra colour
charged parton by $k$. For these processes we introduce the Mandelstam
variables $\hat{s}$, $\hat{t}$, $\hat{u}$ and the related \emph{radiative
variables} $\Phi_{R}=\left\{ x,\, y\right\} $, which parametrize
the extra emission: \begin{subequations} \begin{eqnarray}
\hat{s} & = & \left(p_{\oplus}+p_{\ominus}\right)^{2}=\frac{p^{2}}{x}\,,\label{eq:nlo_1_4_a}\\
\hat{t} & = & \left(p_{\oplus}-k\right)^{2}\ \;=-\frac{1}{2}\,\frac{p^{2}}{x}\,\left(1-x\right)\left(1-y\right)\,,\label{eq:nlo_1_4_b}\\
\hat{u} & = & \left(p_{\ominus}-k\right)^{2}\ \;=-\frac{1}{2}\,\frac{p^{2}}{x}\,\left(1-x\right)\left(1+y\right)\,,\label{eq:nlo_1_4_c}\end{eqnarray}
 \label{eq:nlo_1_4}\end{subequations} where $p=\sum\, p_{i}$. We
do not explicitly include an azimuthal angle for the gluon about the
$z$ axis, instead it is used to define the $+y$ axis relative to
which the azimuthal angle of the other final-state particles is measured;
ultimately all generated events are randomly rotated about the $z$
axis in the hadronic centre-of-mass frame.

To perform a simultaneous Monte Carlo sampling of the \emph{N}- and
\emph{N}+1-body phase spaces one has to specify the integration variables.
We choose two of these to be the mass and rapidity of the system of
colourless particles, therefore $\bar{p}^{2}\equiv p^{2}$ and $\bar{\mathrm{y}}\equiv\mathrm{y}$,
where $\mathrm{y}$ is the rapidity of $p$.%
\footnote{Henceforth we will always refer to these variables as $p^{2}$ and
$\mathrm{y}$.%
} An immediate consequence of this partial mapping of the \emph{N}
and \emph{N}+1 body phase spaces is that the momentum fractions, $x_{\splusminus}$,
of the incident partons in the $2\rightarrow N+1$ processes are related
to those of the $2\rightarrow N$ process by \begin{align}
x_{\oplus} & =\frac{\bar{x}_{\oplus}}{\sqrt{x}}\sqrt{\frac{2-\left(1-x\right)\left(1-y\right)}{2-\left(1-x\right)\left(1+y\right)}}\,, & x_{\ominus} & =\frac{\bar{x}_{\ominus}}{\sqrt{x}}\sqrt{\frac{2-\left(1-x\right)\left(1+y\right)}{2-\left(1-x\right)\left(1-y\right)}},\label{eq:nlo_1_5}\end{align}
 where, by definition, $p_{\splusminus}=x_{\splusminus}P_{\splusminus}$.

Since we restrict ourselves to processes for which the NLO corrections
contain at most soft and initial-state collinear singularities, the
product of $\hat{t}\,\hat{u}$ with the squared real emission matrix
elements will be finite throughout the radiative phase space. Working
in conventional dimensional regularization, in $n=4-2\epsilon$ space-time
dimensions, we are then able to perform an expansion in $\epsilon$
of the \emph{N}+1-body phase-space measure, $\mathrm{d}\Phi_{N+1}$,
using similar manipulations to those in Refs.\,\cite{Mele:1990bq,Frixione:1992pj,Frixione:1993yp,Mangano:1991jk},
giving \begin{equation}
\mathrm{d}\Phi_{N+1}=\mathrm{d}\Phi_{B}\,\mathrm{d}\Phi_{R}\,\frac{p^{2}}{\left(4\pi\right)^{2}x^{2}}\,\left(\frac{1}{p^{2}}\right)^{\epsilon}c_{\Gamma}\,\mathcal{J}\left(x,y\right),\label{eq:nlo_1_14}\end{equation}
 where here the Born variables $\hat{\Phi}_{B}$ specify a configuration
in the rest frame of $p$ rather than $\bar{p}$. The function $\mathcal{J}\left(x,y\right)$
is given by \[
\mathcal{J}\left(x,y\right)=\left[\mathcal{S}\,\delta\left(1-x\right)+\mathcal{C}\left(x\right)\,\left(2\delta\left(1+y\right)+2\delta\left(1-y\right)\right)+\mathcal{H}\left(x,y\right)\right]\,\frac{\hat{t}\,\hat{u}}{\hat{s}^{2}}\,,\]
 where \begin{subequations} \begin{eqnarray}
\mathcal{S} & = & \frac{1}{\epsilon^{2}}-\frac{\pi^{2}}{6}-\frac{4}{\epsilon}\ln\eta+8\ln^{2}\eta\,,\label{eq:nlo_1_12_a}\\
\mathcal{C}\left(x\right) & = & -\frac{1}{\epsilon}\frac{1}{\left(1-x\right)_{\rho}}-\frac{1}{\left(1-x\right)_{\rho}}\ln x+2\left(\frac{\ln\left(1-x\right)}{1-x}\right)_{\rho}\,,\label{eq:nlo_1_12_b}\\
\mathcal{H}\left(x,y\right) & = & \frac{2}{\left(1-x\right)_{\rho}}\left[\left(\frac{1}{1+y}\right)_{+}+\left(\frac{1}{1-y}\right)_{+}\right]\,,\label{eq:nlo_1_12_c}\end{eqnarray}
 \label{eq:nlo_1_12}\end{subequations} with $\rho=\left(\sum_{i}\sqrt{p_{i}^{2}}\right)^{2}/\hat{s}$
and $\eta=\sqrt{1-\rho}$. The constant $c_{\Gamma}$, which appears
due to the use of dimensional regularization, is given by\begin{eqnarray*}
c_{\Gamma} & = & \left(4\pi\right)^{\epsilon}\frac{\Gamma\left(1-\epsilon\right)^{2}\Gamma\left(1+\epsilon\right)}{\Gamma\left(1-2\epsilon\right)}\,,\end{eqnarray*}
 and the $\rho$-distributions are defined by the relation\[
\int_{\rho}^{1}\mathrm{d}x\, h\left(x\right)\left(\frac{\ln^{n}\left(1-x\right)}{1-x}\right)_{\rho}=\int_{\rho}^{1}\mathrm{d}x\,\left(h\left(x\right)-h\left(1\right)\right)\frac{\ln^{n}\left(1-x\right)}{1-x}\,,\]
for any sufficiently regular test function $h\left(x\right)$. The
radiative phase space can be parametrized in terms of the radiative
variables as \begin{equation}
\begin{array}{lcl}
\mathrm{d}\Phi_{R} & = & \frac{1}{2}\mathrm{d}y\,\mathrm{d}x\,.\end{array}\label{eq:nlo_1_11}\end{equation}

The precise definition of $\hat{\Phi}_{B}$ in the context of the
radiative event is, in general, given by a series of boosts and rotations,
denoted by $\mathbb{B}$, which \emph{embed} the \emph{N}-particle,
leading-order final state in the \emph{N}+1-particle radiative events.
To assemble a radiative event configuration we first construct the
\emph{N} final-state momenta of the leading-order configuration according
to the definition of $\hat{\Phi}_{B}$ and, separately, the momenta
of the incident partons and the radiated parton $k$ in the \emph{hadronic}
centre-of-mass frame:\begin{align}
p_{\oplus} & =\frac{1}{2}\sqrt{s}\left(x_{\oplus},0,0,+x_{\oplus}\right)\,, & p & =\left(E_{T}\cosh\mathrm{y},\,0,\,-p_{T},\, E_{T}\sinh\mathrm{y}\right)\,,\label{eq:nlo_1_15}\\
p_{\ominus} & =\frac{1}{2}\sqrt{s}\left(x_{\ominus},0,0,-x_{\ominus}\right)\,, & k & =\left(p_{T}\cosh\mathrm{y}_{k},\,0,\,\mbox{ }\mbox{ }p_{T},\, p_{T}\sinh\mathrm{y}_{k}\right)\,,\label{eq:nlo_1_16}\end{align}
 where $E_{T}=\sqrt{p_{T}^{2}+p^{2}},$ and \begin{align}
p_{T}^{2} & =\frac{p^{2}}{x}\,\frac{1}{4}\left(1-y^{2}\right)\left(1-x\right)^{2}\,, & \mathrm{y}_{k} & =\mathrm{y}+\frac{1}{2}\ln\left(\frac{1+y}{1-y}\right)+\frac{1}{2}\ln\left(\frac{2-\left(1-x\right)\left(1-y\right)}{2-\left(1-x\right)\left(1+y\right)}\right).\label{eq:nlo_1_17}\end{align}
 To embed the \emph{N} final-state particles of the leading-order
process in the radiative event we first transform them with a longitudinal
boost, $\mathbb{B}_{\mathrm{y}}$, taking us to their rest frame.
Should we wish to adhere to the conventions in Refs.\,\cite{Mele:1990bq,Frixione:1992pj,Frixione:1993yp,Mangano:1991jk},
we then apply a rotation to them, $\mathbb{R}$, defined such that,
ultimately, $\mathbb{B}$ will preserve the direction of $p_{\oplus}$,
\emph{i.e.} with $p_{\oplus}$ defining the $+z$ axis in the $p$
rest frame, and also with the transverse momentum of $k$ defining
the $y$ axis. However, for the simple case at hand, no such rotation
is necessary since the Higgs boson decays isotropically in its rest
frame.%
\footnote{As previously noted, the Higgs-strahlung process was simulated in
a special way using the Kleiss trick as described in \cite{Hamilton:2008pd}.%
} A further transverse boost, $\mathbb{B}_{T}$, parallel with the
$y$ axis, is then carried out, where $\mathbb{B}_{T}^{-1}$ is a
transformation from the lab to the frame in which $p$ has no transverse
component. Finally, the inverse boost $\mathbb{B}_{\mathrm{y}}^{-1}$
is applied to the \emph{N} particles, which clearly returns them to
a frame where their total rapidity is $\mathrm{y}=\bar{\mathrm{y}}$.
Altogether the embedding boost $\mathbb{B}$, to be applied to the
leading-order final-state momenta, is given as\begin{equation}
\mathbb{B}=\mathbb{B}_{\mathrm{y}}^{-1}\,\mathbb{B}_{T}\,\mathbb{R}\,\mathbb{B}_{\mathrm{y}}\,,\label{eq:nlo_1_18}\end{equation}
 which, combined with $p_{\oplus}$, $p_{\ominus}$ and $k$ in Eqs.\,\ref{eq:nlo_1_15},
\ref{eq:nlo_1_16}, completely specifies the radiative kinematics.

\subsection{Differential cross section for $a+b\rightarrow n$\label{sub:Differential-cross-section}}

In this section we restrict ourselves to discussing the general form
of the NLO differential cross section for processes of the type $a+b\rightarrow i_{1}+...+i_{N}$,
where $n=\sum_{j}\, i_{j}$ are colourless particles, which we collectively
refer to as \emph{neutrals}. In Sect.\,\ref{sub:ggH_nlo_xsec} we
give the squared matrix elements for $g+g\rightarrow H$ which were
ultimately inserted in these formulae. We reiterate that analogous
Higgs-strahlung ingredients are \emph{identical} to those in Ref.\,\cite{Hamilton:2008pd}
\emph{modulo} the substitution of the $q+\bar{q}\rightarrow l+\bar{l}$
leading-order matrix element with that of a given Higgs-strahlung
process.

We define the combined incident flux of partons of type $a$ from
hadron $A$, and partons of type $b$ from hadron $B$, with respective
momentum fractions $x_{\oplus}$ and $x_{\ominus}$, at a scale $\mu^{2}$,
as \begin{equation}
\mathcal{L}_{ab}\left(x_{\oplus},x_{\ominus}\right)=f_{a}^{A}\left(x_{\oplus},\mu^{2}\right)f_{b}^{B}\left(x_{\ominus},\mu^{2}\right)\,,\label{eq:nlo_3_0}\end{equation}
 where $f_{i}^{I}\left(x_{i},\mu^{2}\right)$ are the relevant parton
distribution functions (PDFs). To obtain the differential cross section
for a real emission process $a+b\rightarrow n+c$, we simply multiply
the relevant squared matrix element by the phase-space measure in
(Eq.\,\ref{eq:nlo_1_14}), $\mathcal{L}_{ab}$ and the flux factor
$1/2\hat{s}$\,:\begin{eqnarray}
\mathrm{d}\sigma_{ab}^{N+1} & = & \frac{1}{2\hat{s}}\,\mathcal{M}_{ab}^{N+1}\left(p_{\oplus},\, p_{\ominus}\right)\,\mathcal{L}_{ab}\left(x_{\oplus},x_{\ominus}\right)\,\mathrm{d}\Phi_{N+1}.\label{eq:nlo_2_1}\end{eqnarray}

In general each real emission process makes three contributions to
the NLO differential cross section, one for each term in the phase-space
measure Eq.\,\ref{eq:nlo_1_12}: a soft contribution $\mathrm{d}\sigma_{ab}^{S_{0}}$
proportional to $\delta\left(1-x\right)$, collinear contributions
$\mathrm{d}\sigma_{ab}^{C_{0}\pm}$ proportional to $\delta\left(1\pm y\right)$,
and a finite, \emph{hard}, contribution $\mathrm{d}\sigma_{ab}^{H}$
proportional to $\mathcal{H}\left(x,y\right)$. The subscript $0$
in $\mathrm{d}\sigma_{ab}^{S_{0}}$ and $\mathrm{d}\sigma_{ab}^{C_{0}\splusminus}$
reflects the fact that they are bare, divergent, quantities.

For leading-order processes of the type $a+b\rightarrow n$ the two
initial-state partons must be either a pair of gluons or a quark and
an antiquark. The squared matrix elements for the real emission corrections
to this process, in which a gluon is emitted from an initial-state
parton, $a+b\rightarrow n+g$, factorize in the limit that the gluon
is soft $\left(x\rightarrow1\right)$, according to\begin{equation}
\lim_{x\rightarrow1}\mathcal{M}_{ab}^{N+1}\left(p_{\oplus},p_{\ominus}\right)=8\pi\alpha_{S}\mu^{2\epsilon}\,\frac{1}{p^{2}}\,2\, C_{ab}\,\frac{\hat{s}^{2}}{\hat{t}\,\hat{u}}\,\mathcal{M}_{ab}^{N}\left(p_{\oplus},\, p_{\ominus}\right)\,.\label{eq:nlo_2_2}\end{equation}
 The colour factor $C_{ab}$ is equal to $C_{A}$ if $a$ and $b$
are gluons, and $C_{F}$ if $a$ is a quark and $b$ is an antiquark
or \emph{vice} \emph{versa}. The real emission processes $a+b\rightarrow n+g$
are the only ones contributing to the cross section in the limit $x\rightarrow1$,
all other real emission matrix elements are finite in this limit,
hence the product of $\hat{t}\,\hat{u}$ with the squared matrix element
vanishes there. Hence, the soft contribution to the NLO differential
cross section is \begin{equation}
\mathrm{d}\sigma_{ab}^{S_{0}}=\frac{\alpha_{S}c_{\Gamma}}{2\pi}\,\left(\frac{\mu^{2}}{p^{2}}\right)^{\epsilon}\, C_{ab}\,\left(\frac{2}{\epsilon^{2}}-\frac{\pi^{2}}{3}-\frac{8}{\epsilon}\ln\eta+16\ln^{2}\eta\right)\, B\left(\Phi_{B}\right)\,\mathrm{d}\Phi_{B}\,,\label{eq:nlo_2_3}\end{equation}
 where $B\left(\Phi_{B}\right)$ is the differential cross section
for the leading-order process \begin{align}
\mathrm{d}\sigma_{ab}^{B} & =B\left(\Phi_{B}\right)\,\mathrm{d}\Phi_{B}\,, & B\left(\Phi_{B}\right) & =\frac{1}{2p^{2}}\,\mathcal{M}_{ab}^{N}\left(\bar{p}_{\oplus},\,\bar{p}_{\ominus}\right)\mathcal{L}_{ab}\left(\bar{x}_{\oplus},\,\bar{x}_{\ominus}\right)\,.\label{eq:nlo_2_4}\end{align}

For an arbitrary process in which an initial-state parton $a$, with
momentum $p_{\oplus}$, branches to produce a collinear time-like
daughter parton with momentum $k=\left(1-x\right)p_{\oplus}$ and
its space-like sister parton $\widetilde{ac}$, with momentum $xp_{\oplus}$,
the squared matrix element factorizes according to \cite{Catani:1996vz}
\begin{equation}
\lim_{y\rightarrow+1}\mathcal{M}_{ab}^{N+1}\left(p_{\oplus},p_{\ominus}\right)=\frac{1}{p^{2}}\,\frac{\hat{s}^{2}}{\hat{t}\,\hat{u}}\,8\pi\alpha_{S}\mu^{2\epsilon}\,\left(1-x\right)\,\hat{P}_{a\,\widetilde{ac}}\left(x\,;\epsilon\right)\,\mathcal{M}_{ab}^{N}\left(xp_{\oplus},\, p_{\ominus}\right)\,,\label{eq:nlo_2_5}\end{equation}
 where we explicitly show the dependence of the \emph{N} and \emph{N}+1
parton squared matrix elements on the incident parton momenta. Replacing
$xp_{\splusminus}\leftrightarrow p_{\sminusplus}$, $a\rightarrow b$
in the splitting function we obtain the analogous formula for the
case that $c$ is emitted collinear to $b$ $\left(y\rightarrow-1\right)$.
Using this factorization of the matrix element, the collinear contributions
to the real emission cross section, arising from initial-state radiation,
are seen to have the general form, \begin{eqnarray}
\mathrm{d}\sigma_{ab}^{C_{0}\splusminus} & = & \mathrm{d}\sigma_{ab}^{S\, C\splusminus}+\mathrm{d}\sigma_{ab}^{C\splusminus}-\mathrm{d}\sigma_{ab}^{CT\splusminus},\nonumber \\
\mathrm{d}\sigma_{ab}^{S\, C\splusminus} & = & \frac{\alpha_{S}c_{\Gamma}}{2\pi}\,\left(\frac{\mu^{2}}{p^{2}}\right)^{\epsilon}\, C_{i\,\widetilde{ic}}\left(p_{i\,\widetilde{ic}}+4\ln\eta\right)\left(\frac{1}{\epsilon}+\ln\left(\frac{p^{2}}{\mu^{2}}\right)\right)\, B\left(\Phi_{B}\right)\,\mathrm{d}\Phi_{B},\nonumber \\
\mathrm{d}\sigma_{ab}^{C\splusminus} & = & \frac{\alpha_{S}}{2\pi}\,\mathcal{C}_{i\,\widetilde{ic}}^{\splusminus}\left(x\right)\, B^{\splusminus}\left(\Phi_{B}\right)\,\widehat{\mathcal{L}}_{ab}^{\splusminus}\left(x_{\oplus},\, x_{\ominus}\right)\,\frac{1}{x}\,\mathrm{d}\Phi_{B}\,\mathrm{d}x,\label{eq:nlo_2_7}\\
\mathcal{C}_{i\,\widetilde{ic}}^{\splusminus}\left(x\right) & = & \left[\frac{1}{\left(1-x\right)_{\rho}}\ln\left(\frac{p^{2}}{\mu^{2}x}\right)+2\left(\frac{\ln\left(1-x\right)}{1-x}\right)_{\rho}\right]\,\left(1-x\right)\hat{P}_{i\,\widetilde{ic}}\left(x\right)-\hat{P}_{i\,\widetilde{ic}}^{\epsilon}\left(x\right),\nonumber \\
\mathrm{d}\sigma_{ab}^{CT\splusminus} & = & \frac{1}{\bar{\epsilon}}\,\frac{\alpha_{S}}{2\pi}\, P_{i\,\widetilde{ic}}\left(x\right)\, B^{\splusminus}\left(\Phi_{B}\right)\,\widehat{\mathcal{L}}_{ab}^{\splusminus}\left(x_{\oplus},\, x_{\ominus}\right)\,\frac{1}{x}\,\mathrm{d}\Phi_{B}\,\mathrm{d}x,\nonumber \end{eqnarray}
 where $i=a$ in the case that parton $a$ splits to produce parton
$c$, and $i=b$ for the case that parton $b$ branches to produce
$c$. $\hat{P}_{i\,\widetilde{ic}}\left(x;\epsilon\right)$ and $P_{i\,\widetilde{ic}}\left(x\right)$
are the bare and regularized Altarelli-Parisi kernels given in Appendix\,\ref{sec:Splitting-functions};
$C_{i\,\widetilde{ic}}\, p_{i\,\widetilde{ic}}$ is equal to the coefficient
of the $\delta\left(1-x\right)$ term in the latter, for the case
$\rho=0$. The $B^{\splusminus}$ and $\widehat{\mathcal{L}}^{\splusminus}$
functions are related to the leading-order differential cross section
and parton flux: \begin{equation}
\begin{array}{rclcrcl}
B^{\oplus}\left(\Phi_{B}\right) & = & \frac{1}{2p^{2}}\,\mathcal{M}_{ab}^{N}\left(xp_{\oplus},\, p_{\ominus}\right)\mathcal{L}_{ab}\left(\bar{x}_{\oplus},\,\bar{x}_{\ominus}\right)\,, &  & \widehat{\mathcal{L}}_{ab}^{\oplus}\left(x_{\oplus},\, x_{\ominus}\right) & = & \frac{\mathcal{L}_{ab}\left(\frac{\bar{x}_{\oplus}}{x},\,\bar{x}_{\ominus}\right)}{\mathcal{L}_{ab}\left(\bar{x}_{\oplus},\,\bar{x}_{\ominus}\right)}\,,\\
B^{\ominus}\left(\Phi_{B}\right) & = & \frac{1}{2p^{2}}\,\mathcal{M}_{ab}^{N}\left(p_{\oplus},\, xp_{\ominus}\right)\mathcal{L}_{ab}\left(\bar{x}_{\oplus},\,\bar{x}_{\ominus}\right)\,, &  & \widehat{\mathcal{L}}_{ab}^{\ominus}\left(x_{\oplus},\, x_{\ominus}\right) & = & \frac{\mathcal{L}_{ab}\left(\bar{x}_{\oplus},\,\frac{\bar{x}_{\ominus}}{x}\right)}{\mathcal{L}_{ab}\left(\bar{x}_{\oplus},\,\bar{x}_{\ominus}\right)}\,.\end{array}\label{eq:nlo_2_8}\end{equation}

In the $\overline{\mathrm{MS}}$ scheme each collinear singular $\mathrm{d}\sigma_{ab}^{CT\splusminus}$
term in the cross section is exactly compensated for by an additive
collinear counter term identical to it (hence the $CT$ labeling).
This amounts to absorbing the divergence in the PDFs, renormalizing
them, hence we now omit them. The only remaining divergences are soft
and collinear terms $\mathrm{d}\sigma_{ab}^{S\, C\splusminus}$, which
we absorb in the soft contribution to the cross section $\mathrm{d}\sigma_{ab}^{S_{0}}$.

Absorbing the soft and collinear terms $\mathrm{d}\sigma_{ab}^{S\, C\splusminus}$
in the soft contribution to the cross section Eq.\,\ref{eq:nlo_2_3},
redefines it as\begin{eqnarray}
\mathrm{d}\sigma_{ab}^{S_{0}} & = & \mathcal{S}_{0}\, B\left(\Phi_{B}\right)\,\mathrm{d}\Phi_{B}\,,\label{eq:nlo_2_9}\end{eqnarray}
 where\begin{equation}
\mathcal{S}_{0}=\frac{\alpha_{S}c_{\Gamma}}{2\pi}\,\left(\frac{\mu^{2}}{p^{2}}\right)^{\epsilon}\, C_{ab}\left[\frac{2}{\epsilon^{2}}+\frac{2}{\epsilon}\, p_{a\,\widetilde{ag}}+\ln\left(\frac{p^{2}}{\mu^{2}}\right)\left(2p_{a\,\widetilde{ag}}+8\ln\eta\right)+16\ln^{2}\eta-\frac{\pi^{2}}{3}\right].\label{eq:nlo_2_10}\end{equation}

One must consider also the contribution of virtual corrections to
the leading-order process as well as the real corrections above. Again,
for $a+b\rightarrow n$ processes, these have a simple form (see \emph{e.g.}
Ref.\,\cite{Frixione:2007vw})\begin{eqnarray}
\lim_{x\rightarrow1}\mathcal{M}_{ab}^{N\, V}\left(\bar{p}_{\oplus},\,\bar{p}_{\ominus}\right) & = & \mathcal{V}_{0}\,\mathcal{M}_{ab}^{N}\left(\bar{p}_{\oplus},\,\bar{p}_{\ominus}\right)\label{eq:nlo_2_11}\end{eqnarray}
 with\begin{equation}
\mathcal{V}_{0}=\frac{\alpha_{S}c_{\Gamma}}{2\pi}\,\left(\frac{\mu^{2}}{p^{2}}\right)^{\epsilon}\, C_{ab}\,\left[-\frac{2}{\epsilon^{2}}-\frac{2}{\epsilon}\, p_{a\,\widetilde{ag}}-\frac{\pi^{2}}{3}\right]+\widehat{\mathcal{M}}_{ab}^{N\, V_{\mathrm{reg}}}\left(\bar{p}_{\oplus},\,\bar{p}_{\ominus}\right)\,,\label{eq:nlo_2_12}\end{equation}
 where $\widehat{\mathcal{M}}_{ab}^{N\, V_{\mathrm{reg}}}$ is the
remainder of the virtual correction, regular as $\epsilon\rightarrow0$,
divided by the leading-order squared matrix element,\begin{equation}
\widehat{\mathcal{M}}_{ab}^{N\, V_{\mathrm{reg}}}\left(\bar{p}_{\oplus},\,\bar{p}_{\ominus}\right)=\frac{\mathcal{M}_{ab}^{N\, V_{\mathrm{reg}}}\left(\bar{p}_{\oplus},\,\bar{p}_{\ominus}\right)}{\mathcal{M}_{ab}^{N}\left(\bar{p}_{\oplus},\,\bar{p}_{\ominus}\right)}\,.\label{eq:nlo_2_13}\end{equation}
 The poles in $\epsilon$ in $\mathcal{V}_{0}$ will exactly cancel
those in $\mathcal{S}_{0}$ making the differential cross section
finite for $\epsilon\rightarrow0$.

The cross section for the $a+b\rightarrow n$ leading-order process,
combined with the virtual corrections and $a+b\rightarrow n+g$ real
emission corrections, may then be written as \begin{equation}
\mathrm{d}\sigma_{ab}=B\left(\Phi_{B}\right)\,\mathrm{d}\Phi_{B}+V\left(\Phi_{B}\right)\,\mathrm{d}\Phi_{B}+R_{ab}\left(\Phi_{B},\Phi_{R}\right)\,\mathrm{d}\Phi_{B}\,\mathrm{d}\Phi_{R},\label{eq:nlo_2_14}\end{equation}
 where $V\left(\Phi_{B}\right)$ results from the combination of the
soft and virtual corrections to the cross section (Eqs.\,\ref{eq:nlo_2_9},
\ref{eq:nlo_2_10}), \begin{align}
V\left(\Phi_{B}\right) & =\mathcal{V}\, B\left(\Phi_{B}\right)\,, & \mathcal{V} & =\left.\mathcal{S}_{0}+\mathcal{V}_{0}\right|_{\epsilon=0}\,,\label{eq:nlo_2_15}\end{align}
 and $R\left(\Phi_{B},\Phi_{R}\right)$ results from the remaining
collinear and hard real emission corrections, \begin{eqnarray}
R_{ab}\left(\Phi_{B},\Phi_{R}\right) & = & \frac{\alpha_{S}}{2\pi}\,\frac{1}{x}\,\mathcal{R}_{ab}\,\widehat{\mathcal{L}}_{ab}\left(x_{\oplus},x_{\ominus}\right)\, B\left(\Phi_{B}\right)\,,\nonumber \\
\mathcal{R}_{ab} & = & 2\,\mathcal{C}_{a\,\widetilde{ac}}^{\oplus}\left(x\right)\delta\left(1-y\right)+2\,\mathcal{C}_{b\,\widetilde{bc}}^{\ominus}\left(x\right)\delta\left(1+y\right)+\mathcal{H}_{ab}\,,\label{eq:nlo_2_16}\\
\mathcal{H}_{ab} & = & \frac{x\,\hat{t}\,\hat{u}}{\hat{s}}\,\frac{1}{8\pi\alpha_{S}}\,\frac{\mathcal{M}_{ab}^{N+1}\left(p_{\oplus},\, p_{\ominus}\right)}{\mathcal{M}_{ab}^{N}\left(\bar{p}_{\oplus},\,\bar{p}_{\ominus}\right)}\,\mathcal{H}\left(x,y\right)\,.\nonumber \end{eqnarray}
 Note that in writing Eq.\,\ref{eq:nlo_2_14} we have tacitly equated
$\mathcal{M}^{N}\left(xp_{\oplus},p_{\ominus}\right)=\mathcal{M}^{N}\left(p_{\oplus},xp_{\ominus}\right)=\mathcal{M}^{N}\left(\bar{p}_{\oplus},\bar{p}_{\ominus}\right)$
in the collinear term. This is possible due to our defining $p^{2}=\bar{p}^{2}$
and $\mathrm{y}=\bar{\mathrm{y}}$ in Section.\,\ref{sub:Kinematics-and-phase}.

The remaining contributions to the full NLO differential cross section
are due to the production of $n$ via new channels. For $q+\bar{q}\rightarrow n$
processes we add contributions arising from the $q+g\rightarrow n+q$
channel, \begin{eqnarray}
R_{qg}\left(\Phi_{B},\Phi_{R}\right) & = & \frac{\alpha_{S}}{2\pi}\,\frac{1}{x}\,\mathcal{R}_{qg}\,\widehat{\mathcal{L}}_{qg}\left(x_{\oplus},x_{\ominus}\right)\, B\left(\Phi_{B}\right)\,,\label{eq:nlo_2_17}\\
\mathcal{R}_{qg} & = & 2\,\mathcal{C}_{gq}\delta\left(1+y\right)+\mathcal{H}_{qg}\,,\nonumber \\
\mathcal{H}_{qg} & = & \frac{x\,\hat{t}\,\hat{u}}{\hat{s}}\,\frac{1}{8\pi\alpha_{S}}\,\frac{\mathcal{M}_{qg}^{N+1}\left(p_{\oplus},\, p_{\ominus}\right)}{\mathcal{M}_{ab}^{N}\left(\bar{p}_{\oplus},\,\bar{p}_{\ominus}\right)}\,\mathcal{H}\left(x,y\right)\,,\nonumber \end{eqnarray}
 and also a set of terms due to the $\bar{q}+g\rightarrow n+\bar{q}$
channel, identical up to the replacement $q\rightarrow\bar{q}$ on
all terms with the exception of $\mathcal{C}_{gq}$. For the $g+g\rightarrow n$
process we add a contribution from $g+q\rightarrow n+q$, for which
$R_{gq}$ has the same form as Eq.\,\ref{eq:nlo_2_17}, \emph{modulo}
interchanging the subscripts $qg\leftrightarrow gq$, and a further
contribution from the $g+\bar{q}\rightarrow n+\bar{q}$ process, derived
from the former by replacing $q\rightarrow\bar{q}$ on all terms but
$\mathcal{C}_{qg}$.

\subsection{$g+g\rightarrow H$ matrix elements \label{sub:ggH_nlo_xsec}}

The squared, spin and colour averaged, leading-order matrix element
for the $g+g\rightarrow H$ process is given by \begin{equation}
\mathcal{M}_{gg}^{N}\left(\bar{p}_{\oplus},\,\bar{p}_{\ominus}\right)=\mathcal{M}_{gg}^{N}=\frac{\mathcal{N}\, p^{4}}{576\pi\left(1-\epsilon\right)}\,,\label{eq:ggh_Born}\end{equation}
where $\mathcal{N}=\frac{\alpha_{S}^{2}\mu^{2\epsilon}}{\pi v^{2}}$,
with $v$ the vacuum expectation value of the Higgs field and the
scale $\mu$ emerging from the use of conventional dimensional regularization.
The real emission radiative corrections consist of three processes:
$g+g\rightarrow H+g$; $q+g\rightarrow H+q$; and $q+\bar{q}\rightarrow H+g$.
As noted above, the singular, soft and/or collinear, limits of the
matrix elements, associated to the $\epsilon$ poles in $\mathrm{d}\Phi_{N+1}$,
are universal, therefore a full calculation of the matrix elements
is only needed to multiply the $\mathcal{H}\left(x,y\right)$ term
in $\mathrm{d}\Phi_{N+1}$. Since the term proportional to $\mathcal{H}\left(x,y\right)$
contains no poles in $\epsilon$, the full calculation of the squared
matrix elements can be carried out in four dimensions, giving,\begin{align}
\mathcal{M}_{gg}^{N+1} & =8\pi\alpha_{S}\,\mathcal{M}_{gg}^{N}\,\frac{3}{p^{4}}\,\frac{1}{\hat{s}\hat{t}\hat{u}}\,\left[p^{8}+\hat{s}^{4}+\hat{t}^{4}+\hat{u}^{4}\right]\,, & \mathcal{M}_{gq}^{N+1} & =8\pi\alpha_{S}\,\mathcal{M}_{gg}^{N}\,\frac{-4}{3p^{4}}\,\frac{1}{\hat{u}}\,\left[\hat{s}^{2}+\hat{t}^{2}\right]\,,\label{eq:ggh_Real_MEs}\\
\mathcal{M}_{q\bar{q}}^{N+1} & =8\pi\alpha_{S}\,\mathcal{M}_{gg}^{N}\,\frac{32}{9p^{4}}\,\frac{1}{\hat{s}}\,\left[\hat{u}^{2}+\hat{t}^{2}\right]\,, & \mathcal{M}_{g\bar{q}}^{N+1} & =\mathcal{M}_{gq}^{N+1}\,.\nonumber \end{align}
 The squared matrix element $\mathcal{M}_{qg}^{N+1}$ can be obtained
from $\mathcal{M}_{gq}^{N+1}$ by crossing symmetry simply by interchanging
$\hat{u}\leftrightarrow\hat{t}$.

The $\mathcal{O}\left(\alpha_{S}\right)$ virtual corrections to the
$g+g\rightarrow H$ process consist of purely gluonic swordfish and
triangle vertex corrections, as well as a UV counter term. At NLO
these corrections contribute to the cross section through their interference
with the leading-order amplitude, their form is precisely that in
Eq.\,\ref{eq:nlo_2_12} with, \begin{eqnarray}
\mathcal{M}_{gg}^{V\,\mathrm{reg}} & = & \frac{\alpha_{S}}{2\pi}\, C_{A}\left[\frac{11}{3}+\frac{4}{3}\pi^{2}-\frac{4\pi b_{0}}{C_{A}}\,\ln\left(\frac{p^{2}}{\mu_{R}^{2}}\right)\right]\,\mathcal{M}_{gg}^{B}\,,\label{eq:ggh_Virtual}\end{eqnarray}
where $\mu_{R}$ is the renormalization scale. When used with the
general expressions in Sect.\,\ref{sub:Differential-cross-section},
these are all the ingredients required to write down the $\overline{B}\left(\Phi_{B}\right)$
function and the modified Sudakov form factors for this process (Eqs.\,\ref{eq:powheg_2},
\ref{eq:powheg_3}).

\section{Implementation\label{sec:Implementation}}

\subsection{Generation of the leading-order configuration\label{sub:genlo}}

The first phase of the simulation involves generating a leading-order
configuration by sampling the $\overline{B}\left(\Phi_{B}\right)$
function (see Sect.~\ref{sec:The-POWHEG-method}, Eq.\,\ref{eq:powheg_2}),
which is the NLO differential cross section integrated over the radiative
variables,\begin{equation}
\overline{B}\left(\Phi_{B}\right)=B\left(\Phi_{B}\right)\left[1+\mathcal{V}+\sum_{ab}\int\mathrm{d}\Phi_{R}\,\frac{\alpha_{S}}{2\pi}\,\frac{1}{x}\,\mathcal{R}_{ab}\,\widehat{\mathcal{L}}_{ab}\left(x_{\oplus},x_{\ominus}\right)\right]\,.\label{eq:nlo_3_1}\end{equation}
 Here the summation runs over all real emission processes $a+b\rightarrow n+c$
contributing at NLO. Since the leading-order process is factorized
inside the real emission terms $R_{ab}$, the $\overline{B}\left(\Phi_{B}\right)$
distribution can be generated efficiently by just reweighting the
leading-order cross section.

The dependency of the $x$ integration limits on the $y$ radiative
variable, coupled with the definitions of the $\rho$ and plus distributions,
is a significant obstacle to the numerical implementation. We may
obtain fixed integration limits by a simple change of variables $x\rightarrow\tilde{x}$,
where $\tilde{x}$ is defined through the relation \begin{align}
x\left(\tilde{x},y\right) & =\bar{x}\left(y\right)+\bar{\eta}\left(y\right)^{2}\,\tilde{x}\,, & \bar{\eta}\left(y\right) & =\sqrt{1-\bar{x}\left(y\right)}.\label{eq:nlo_3_3}\end{align}
 Care must be taken in implementing these changes of variables in
the NLO differential cross section, particularly with regard to the
plus and $\rho$ distributions. When the transformation is complete
only plus distributions remain since, unlike $x$, $\tilde{x}$ extends
over the range $\left[0,1\right]$ for all $y$. Numerical implementation
of the $\overline{B}\left(\Phi_{B}\right)$ distribution requires
all plus distributions be replaced by regular functions; due to the
change of variables in Eq.\,\ref{eq:nlo_3_3} this is now trivial
since all plus distributions are to be integrated over exactly the
domains specified in their definition $0\le\tilde{x}\le1$.

After the change of variables, the generation of the \emph{N}-body
configurations is technically carried out in the same way as in Ref.\,\cite{Hamilton:2008pd}:
\emph{N}-body configurations are first generated using the corresponding
leading-order \HWPP\ simulation, after which they are reweighted
and retained with a probability proportional to the integrand of Eq.\,\ref{eq:nlo_3_1},
which is sampled using a VEGAS based algorithm \cite{Lonnblad:2006pt}.

\subsection{Generation of the hardest emission\label{sub:genhard}}

Given an \emph{N}-body configuration generated according to $\overline{B}\left(\Phi_{B}\right)$,
we proceed to generate the largest transverse momentum emission according
to the modified Sudakov form factor in Eq.\,\ref{eq:powheg_3}. The
exponent in the modified Sudakov form factor consists of an integral
over a sum of different contributions, one for each channel, $a+b\rightarrow n+c$,
given by \begin{equation}
W_{ab}\left(\Phi_{R},\Phi_{B}\right)=\frac{\widehat{R}_{ab}\left(\Phi_{B},\Phi_{R}\right)}{B\left(\Phi_{B}\right)}=\frac{\alpha_{S}}{2\pi}\,\frac{1}{x}\,\widehat{\mathcal{H}}_{ab}\,\widehat{\mathcal{L}}_{ab}\left(x_{\oplus},x_{\ominus}\right)\,,\label{eq:hardest_1}\end{equation}
 where $\widehat{\mathcal{H}}_{ab}$ is equal to $\mathcal{H}_{ab}$
with the plus and $\rho$ regularization prescriptions omitted.

Instead of generating the hardest emission in terms of $\Phi_{R}=\left\{ x,\, y\right\} $
we find it more convenient to make a change of variables to $\Phi_{R}^{\prime}=\left\{ p_{T},\,\mathrm{y}_{k}\right\} $,
defined in Eq.\,\ref{eq:nlo_1_17}. Making this change of variables
removes the complicated $\theta$-function in Eq.\,\ref{eq:powheg_3},
replacing it by a lower bound on the integration over $p_{T}$: \begin{equation}
\Delta_{\hat{R}}\left(p_{T}\right)=\exp\left(-\int_{p_{T}}^{p_{T\mathrm{max}}}\mathrm{d}\Phi_{R}\mbox{ }\sum_{ab}\, W_{ab}\left(\Phi_{R},\Phi_{B}\right)\right)\,,\label{eq:hardest_5}\end{equation}
 where the upper bound, $p_{T_{\mathrm{max}}}$, is due to the usual
basic phase-space considerations. The distribution of the transformed
radiative variables arising from $\Delta_{\hat{R}}\left(p_{T}\right)$
(Eq.\,\ref{eq:powheg_3}) is sampled using a \emph{veto algorithm}
\cite{Sjostrand:2006za}, in precisely the same way as was done in
Ref.\,\cite{Hamilton:2008pd}. If an emission is generated it is
reconstructed from the $p_{T}$ and $\mathrm{y}_{k}$ radiative variables
according to Eqs.\,\ref{eq:nlo_1_15},\,\ref{eq:nlo_1_16}.

 When generating the
hardest emission we use a factorization scale of the transverse mass of the
Higgs boson or off-shell vector boson, in gluon fusion and
Higgs-strahlung processes, respectively. In both cases we use the 
$p_T$ of the boson as renormalization scale. This is required to correctly treat
the small $p_T$ region where the \textsf{POWHEG} results should agree with the 
default \textsf{Herwig++} parton shower.

\subsection{Truncated and vetoed parton showers\label{sub:Truncated-and-vetoed}}

The \HWPP\
 shower algorithm \cite{Bahr:2008pv,Gieseke:2003rz} works by evolving
downward in a variable related to the angular separation of parton
branching products, $\tilde{q}$, starting at a scale determined by
the colour flow and kinematics of the underlying hard scattering process,
and ending at an infrared cut-off, beneath which further emissions
are defined to be \emph{unresolvable}. Each branching is specified
by an evolution scale $\tilde{q}$, a light-cone momentum fraction,
$z$, and an azimuthal angle, $\phi$. The momenta of all particles
forming a shower can be uniquely constructed given the $\left\{ \tilde{q},\, z,\,\phi\right\} $
parameters of each branching. Since the showers from each parton in
a given process evolve independently, from what are initially on-shell
particles, generated according to a matrix element, some reshuffling
of these momenta, after the generation of the parton showers, is required
to ensure global energy-momentum conservation.

In order to carry out showering of \emph{N}+1-body final states associated
to the generation of the hardest emission, we treat the \emph{N}+1
momenta as having arisen from the showering of an \emph{N}-body configuration
with the \HWPP\ shower. To this end we calculate the branching parameters
$\left\{ \tilde{q}_{h},\, z_{h},\,\phi_{h}\right\} $ for which the
shower would reconstruct the \emph{N}+1-body system from an initial
\emph{N}-body one. Details of this \emph{inverse reshuffling} calculation
were given already in \cite{Hamilton:2008pd}. The \textsf{POWHEG}
emission is subsequently regenerated in the course of a single \HWPP\
shower as follows:
\begin{enumerate}
\item the shower evolves from the default starting scale to $\tilde{q}_{h}$,
with the imposition that any further emissions be flavour conserving
and of lower $p_{T}$ than that of the hardest emission $\left(p_{T_{h}}\right)$,
the \emph{truncated shower}; 
\item the hardest emission is inserted as a set of branching parameters
$\left\{ \tilde{q}_{h},\, z_{h},\,\phi_{h}\right\} $; 
\item the evolution continues down to the cut-off scale, vetoing any emissions
whose transverse momentum exceeds $p_{T_{h}}$, the \emph{vetoed shower}. 
\end{enumerate}
Should the hardest emission occur in an area of phase space that the
shower cannot populate, \emph{i.e.} the wide angle/high $p_{T}$
\emph{dead zone} (Sect.\,\ref{sub:The-dead-zone}), subsequent emissions
will have sufficient resolving power to \emph{see} the widely separated
emitters individually. It follows that no truncated shower is then
required, since this models coherent, large angle emission from more
collimated configurations of partons, and so we proceed directly to
the vetoed shower.

\section{Results\label{sec:Results}}

In this section we present predictions from our \textsf{POWHEG} simulations
of the $g+g\rightarrow H$ and $q+\bar{q}\rightarrow V+H$ processes
within the \HWPP\ event generator. By comparing our results to other
predictions, based on independent calculations and methods, we aim
to validate these realisations of the formalism.

In Sect.\,\ref{sub:Powheg-differential-cross-section-and-Bbar} we
seek to check the calculation and implementation of the \textsf{POWHEG}
NLO differential cross section and $\overline{B}\left(\Phi_{B}\right)$
functions, Eqs.~\ref{eq:powheg_1}-\ref{eq:powheg_2}, we thereby
check the NLO accuracy of the calculation and in particular the generation
of the Born variables (Sect.\,\ref{sub:genlo}). Technically this
is the most delicate part of the simulation, requiring a full calculation
and numerical implementation of the NLO differential cross section.
We compare our results to the NLO parton level Monte Carlo program
MCFM \cite{Campbell:2000bg} to this end.

In Sect.\,\ref{sub:Hardest-emission-generation-results} we move
to focus on distributions sensitive to the generation of the hardest
emission (Sect.\,\ref{sub:genhard}) and the subsequent merging with
the shower algorithm. Here we compare our results to three different
approaches: the bare angular-ordered parton shower, the parton shower
including \emph{matrix element corrections} and also \textsf{MC@NLO}.
While \textsf{MC@NLO} consistently combines NLO calculations with
the \textsf{HERWIG} parton shower, matrix element corrections work
to adjust the distribution of the hardest emission from the \HWPP\
parton shower to be equal to that of the \emph{real part} of the NLO
contribution. Matrix element corrections also serve to populate an
area of the real emission phase space which the shower cannot ordinarily
reach, the so-called \emph{dead-zone}, which we describe fully in
Sect.\,\ref{sub:Hardest-emission-generation-results}. Since the
effects of the NLO contributions on the normalisation of the results
are examined in detail in Sect.\,\ref{sub:Powheg-differential-cross-section-and-Bbar}
and, moreover, the predictions from the shower, with and without matrix
element corrections, have leading-order normalisations, in Sect.\,\ref{sub:Hardest-emission-generation-results}
we concentrate on the shapes of these distributions.

Note that we use the notation 
$\mathcal{O}\left(\alpha_{S}^{n}\right)$ to denote terms of order 
$\alpha_{S}^{n}$ \emph{relative} to the leading order contribution.

\subsection{\textsf{POWHEG} differential cross sections and $\overline{B}\left(\Phi_{B}\right)$\label{sub:Powheg-differential-cross-section-and-Bbar}}

In order to check the calculation of the \textsf{POWHEG} differential
cross section and $\overline{B}\left(\Phi_{B}\right)$ functions,
Eqs.~\ref{eq:powheg_1}-\ref{eq:powheg_2}, total cross sections
and differential distributions were compared between our \textsf{POWHEG}
implementation and the parton level NLO program MCFM \cite{Campbell:2000bg}.
Since MCFM computes the effects of fixed order (NLO) corrections to
the processes in question, these comparisons are performed prior to
any showering of the \textsf{POWHEG} NLO configurations by \HWPP.

In carrying out these comparisons both \HWPP\ and MCFM were run
using the MRST2001 NLO \cite{Martin:2002dr} parton distribution set
via the LHAPDF interface \cite{Whalley:2005nh}. A fixed, constant,
factorization and renormalization scale of 100 GeV was used, in order
to eliminate small variations in the treatment of the running coupling
and PDF evolution as a possible source of discrepancy. Also, for this
part of the validation, we have assumed the Higgs boson mass to be
115 GeV. In all cases the total cross sections from MCFM and our \textsf{POWHEG}
implementation agreed to within 0.5\,\%.

\begin{figure}[H]

\begin{centering}
\includegraphics[width=0.38\textwidth,keepaspectratio,angle=90]{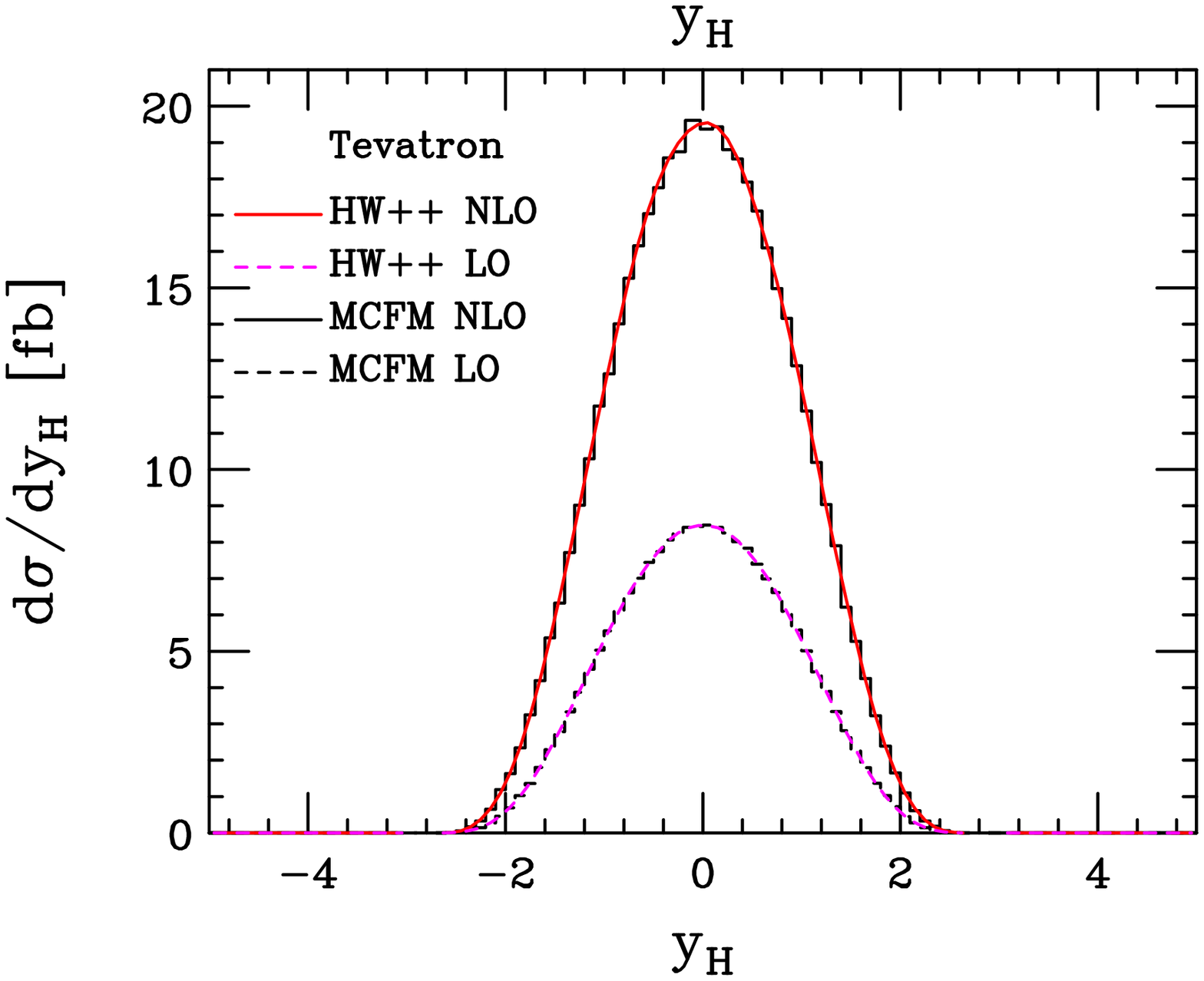}\hfill{}\includegraphics[width=0.38\textwidth,keepaspectratio,angle=90]{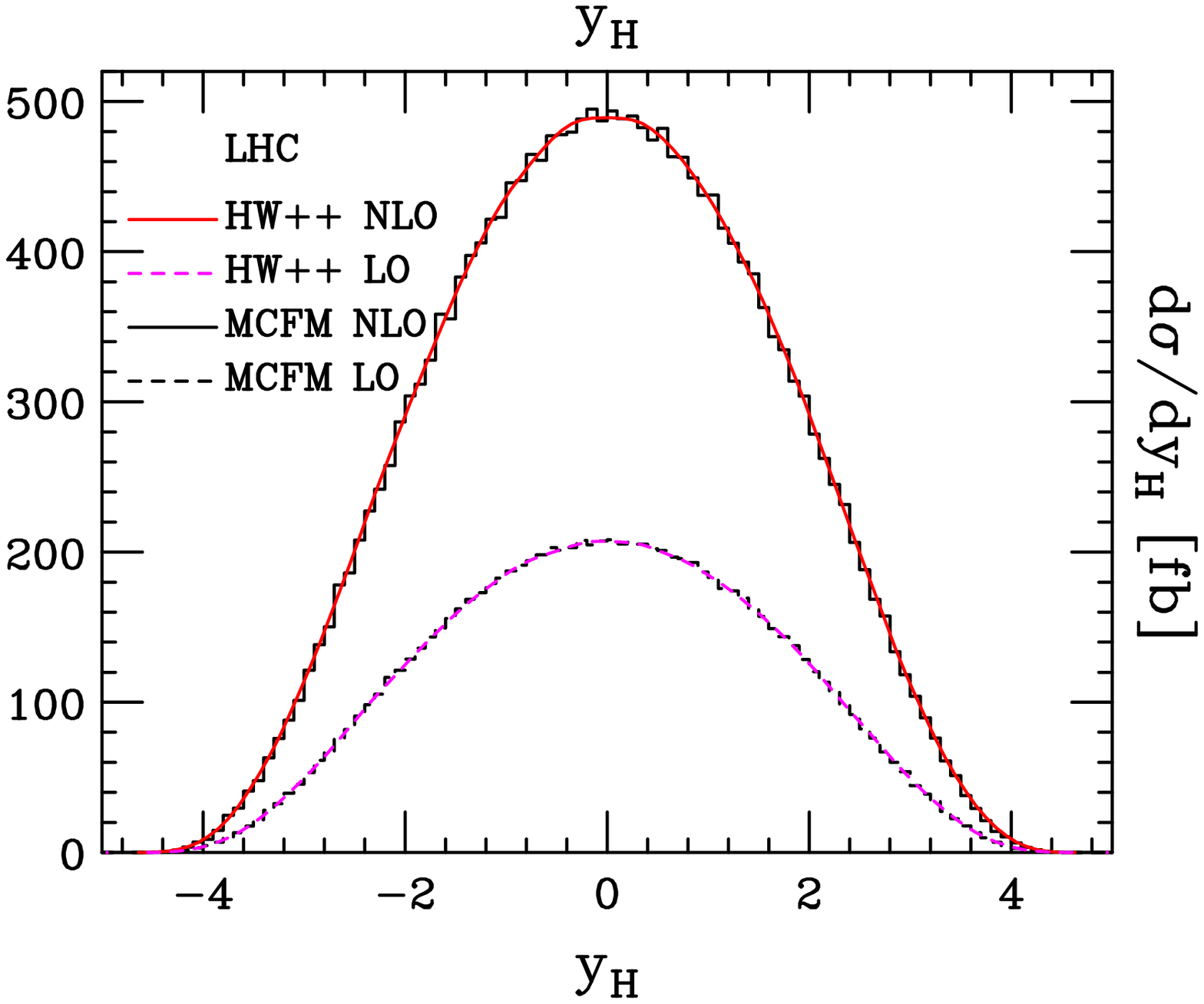}\\
 \vspace{7mm}
 \includegraphics[width=0.38\textwidth,keepaspectratio,angle=90]{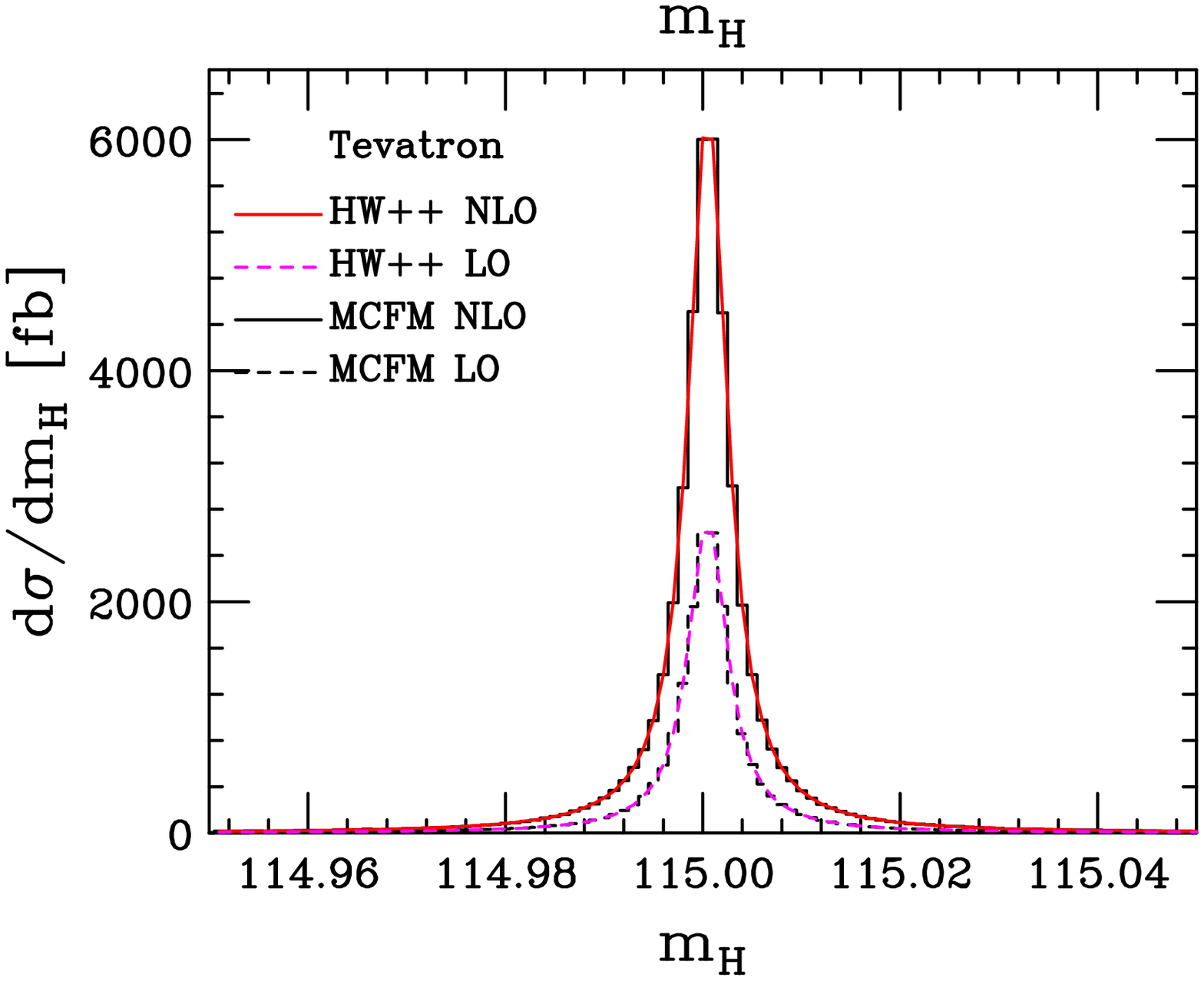}\hfill{}\includegraphics[width=0.38\textwidth,keepaspectratio,angle=90]{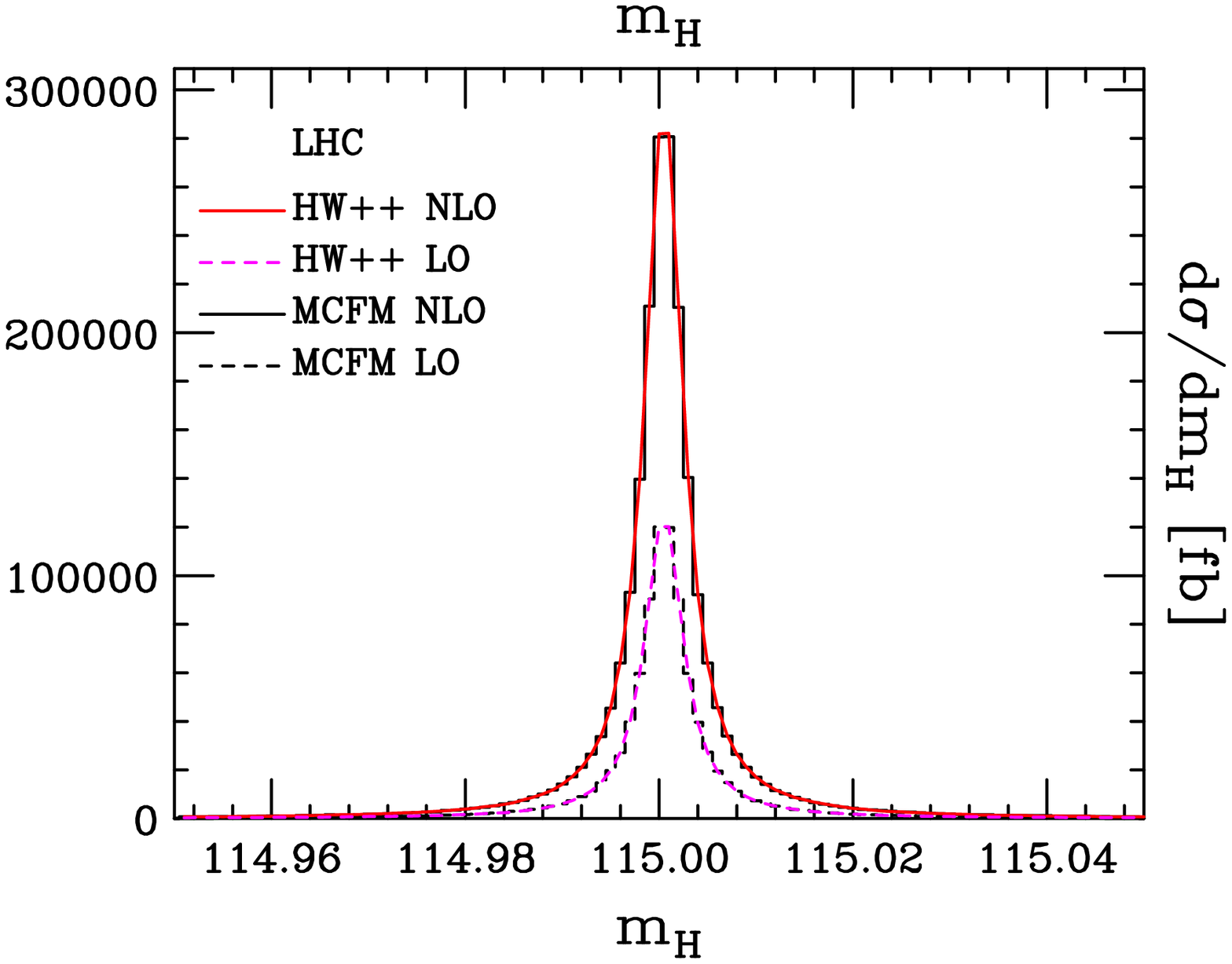}\\
 
\par\end{centering}

\caption{Comparisons of the \textsf{POWHEG} implementation in \HWPP\
and the NLO parton level code MCFM~\cite{Campbell:2000bg}, for the
Higgs boson rapidity ($\mathrm{y}_{\mathrm{H}}$) and mass ($\mathrm{m}_{\mathrm{H}}$)
distributions in the process $g\, g\rightarrow H\rightarrow\tau^{+}\tau^{-}$.
Results on the left are obtained for $p\bar{p}$ collisions at the
Tevatron ($\sqrt{s}=1.96\,\mathrm{TeV}$) while those on the right
are for LHC $pp$ collisions ($\sqrt{s}=14\,\mathrm{TeV}$).}

\label{fig:ggh_born variables} \vspace{5mm}
 %\end{figure}
%\begin{figure}[H]

\begin{centering}
\includegraphics[width=0.38\textwidth,keepaspectratio,angle=90]{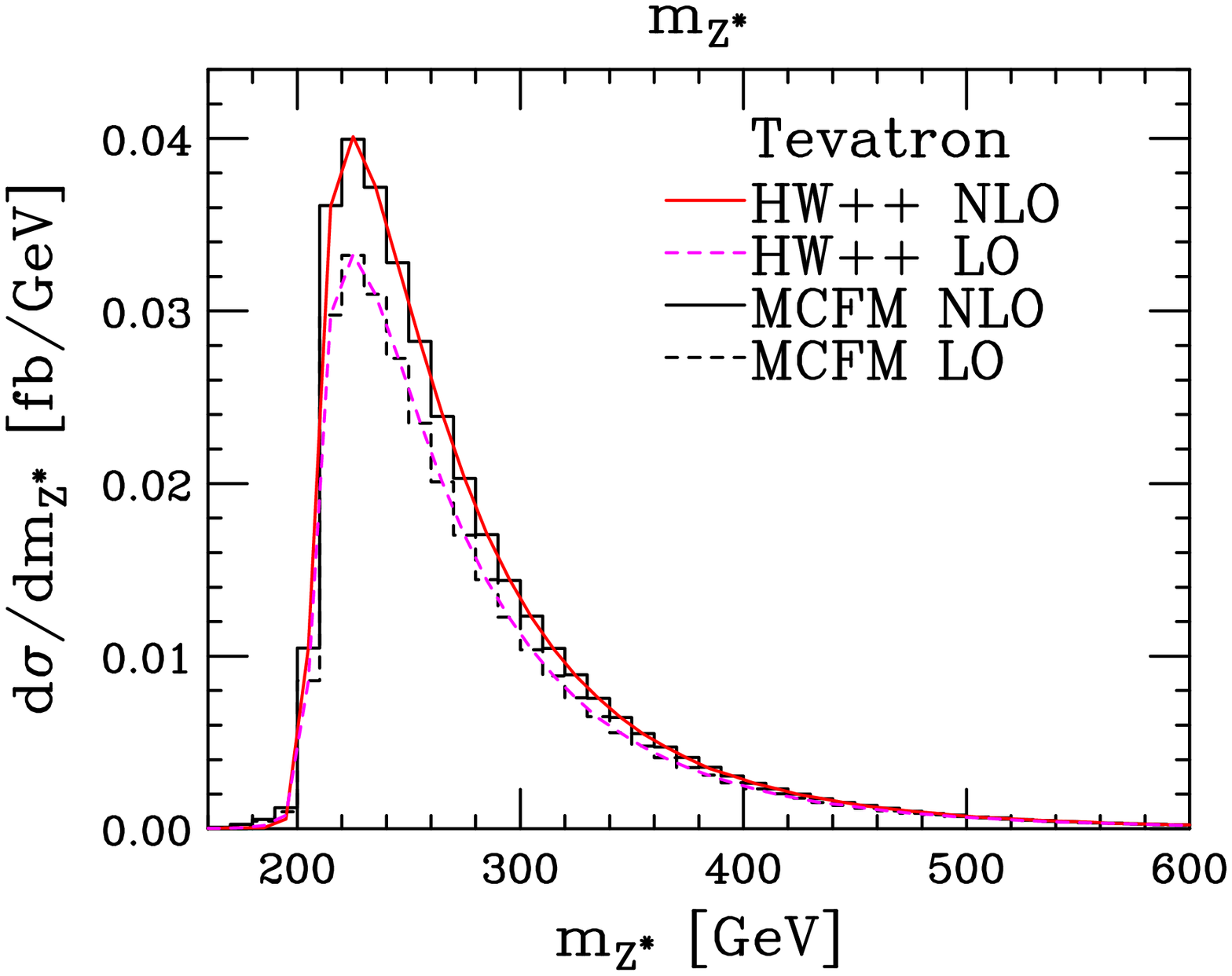}\hfill{}\includegraphics[width=0.38\textwidth,keepaspectratio,angle=90]{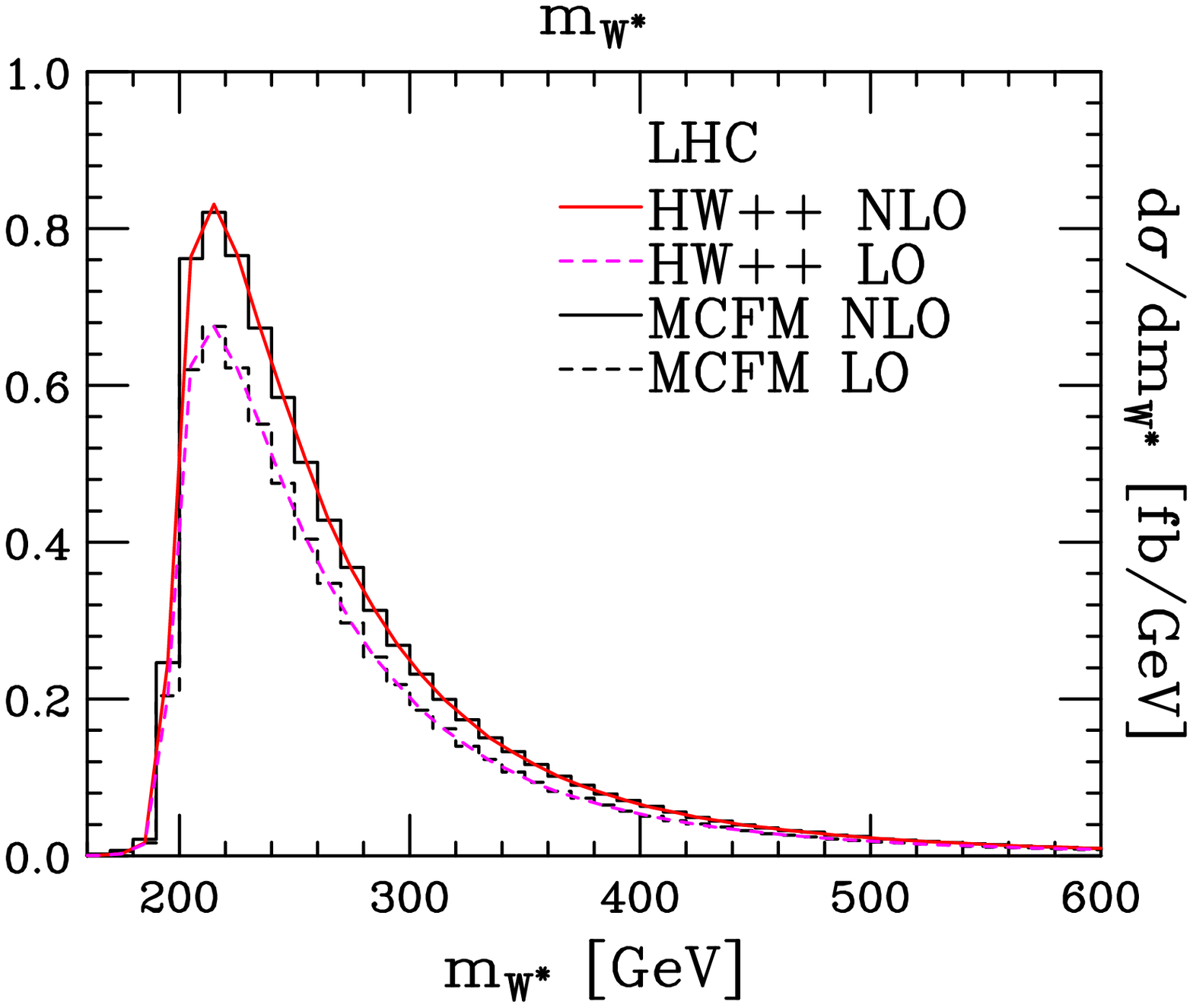}
\par\end{centering}

\caption{Mass of the off-shell virtual vector boson in $q\bar{q}\rightarrow HZ$
and $q\bar{q}\rightarrow HW^{-}$ processes, in the left and right-hand
plots respectively. The $q\bar{q}\rightarrow HZ$ predictions are
for 1960 GeV, Tevatron, proton-antiproton collisions, while the $q\bar{q}\rightarrow HW^{-}$
predictions are for 14 TeV, LHC, proton-proton collisions. This mass
is one of the so-called \emph{Born} \emph{variables} in the NLO differential
cross section (see Sect.\,\ref{sub:Kinematics-and-phase} and Ref.\,\cite{Hamilton:2008pd}).}

\label{fig:higgs-strahlung_mV_star} 
\end{figure}

The gluon fusion process is rather simple in view of the fact that
the scalar Higgs boson decays isotropically in its rest frame. Consequently
the only non-trivial distributions to check are the mass and rapidity
of the Higgs boson. These distributions are plotted for the process%
\footnote{MCFM does not implement the $gg\rightarrow H\rightarrow\gamma\gamma$
process.%
} $gg\rightarrow H\rightarrow\tau^{+}\tau^{-}$ in Fig. \ref{fig:ggh_born variables}
and everywhere exhibit a high level of agreement with MCFM.

The Higgs-strahlung process is more involved than the gluon fusion
process due to the intermediate particle having spin-1 and due to
our re-using the method in our earlier \textsf{POWHEG} work, employing
the Kleiss trick \cite{Kleiss:1986re,Seymour:1994we} to generate
the NLO corrections independently of the details of the decay of the
initially off-shell vector boson. Therefore, to check the correctness
of this method, one must look closely at the distributions of the
final-state particles, particularly, on account of propagating spin
correlation effects, the vector boson and its decay products.

\begin{figure}[H]

\begin{centering}
\includegraphics[width=0.4\textwidth,keepaspectratio,angle=90]{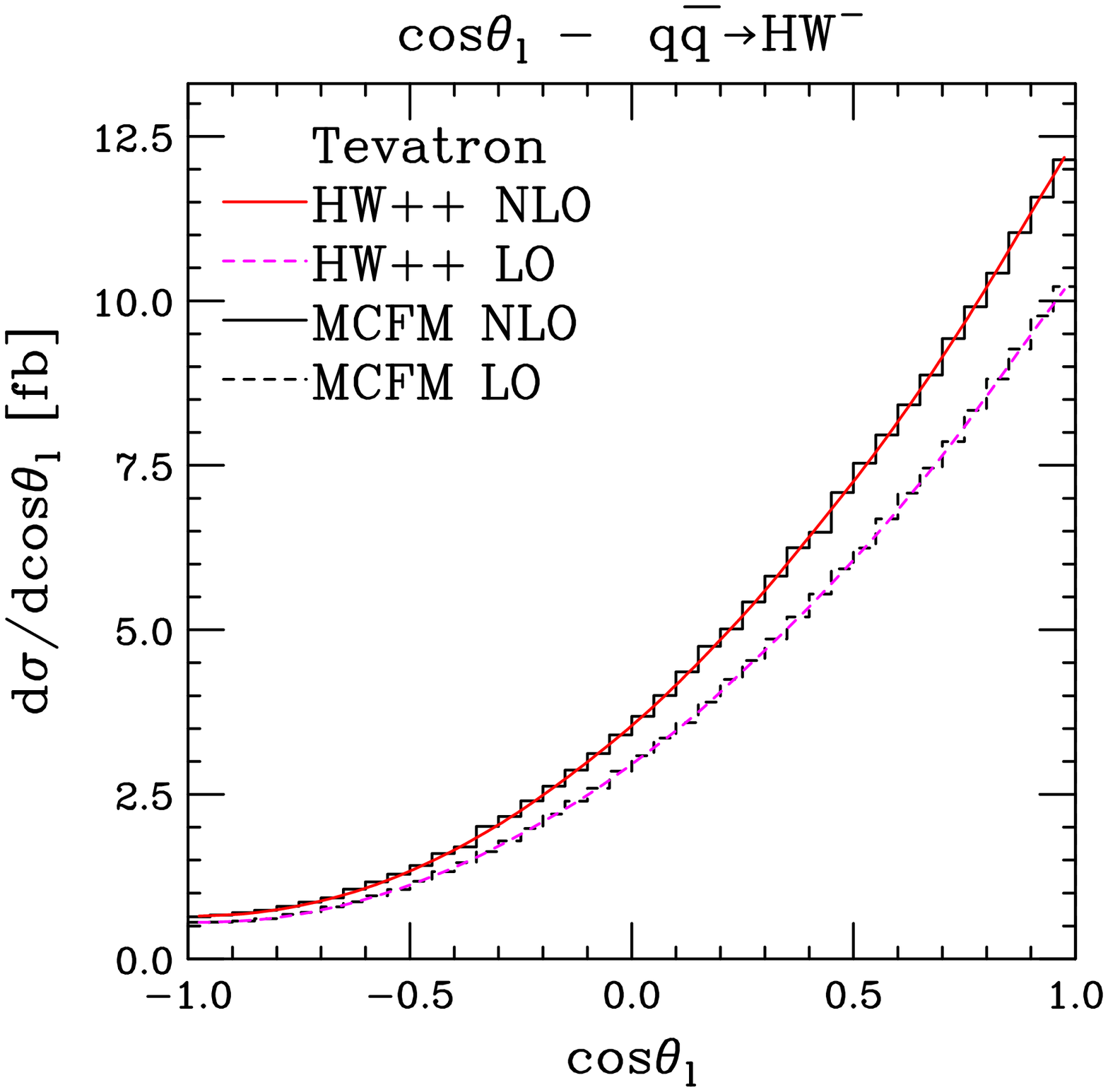}\hfill{}\includegraphics[width=0.4\textwidth,keepaspectratio,angle=90]{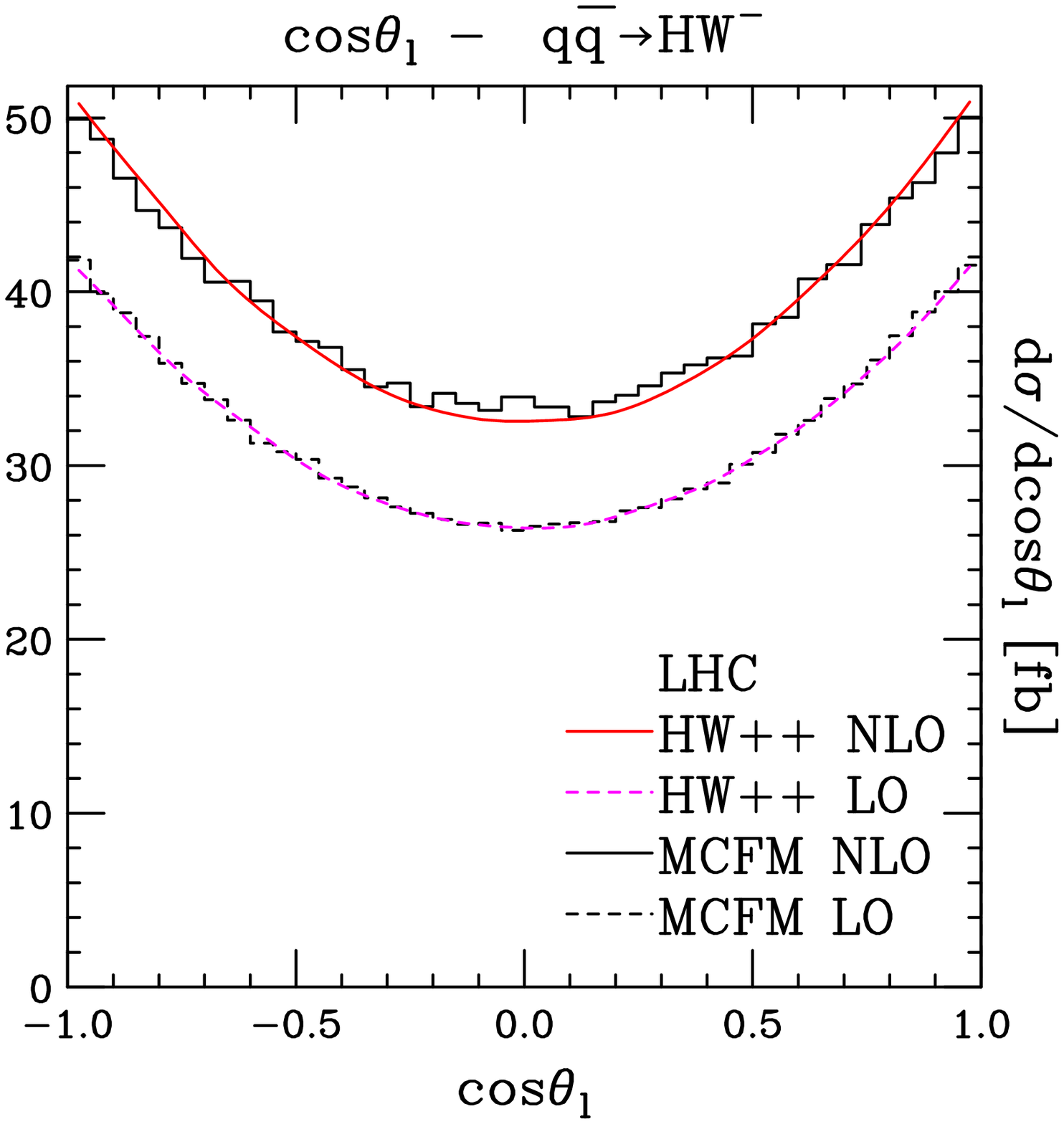}
\par\end{centering}

\vspace{5mm}

\begin{centering}
\includegraphics[width=0.4\textwidth,keepaspectratio,angle=90]{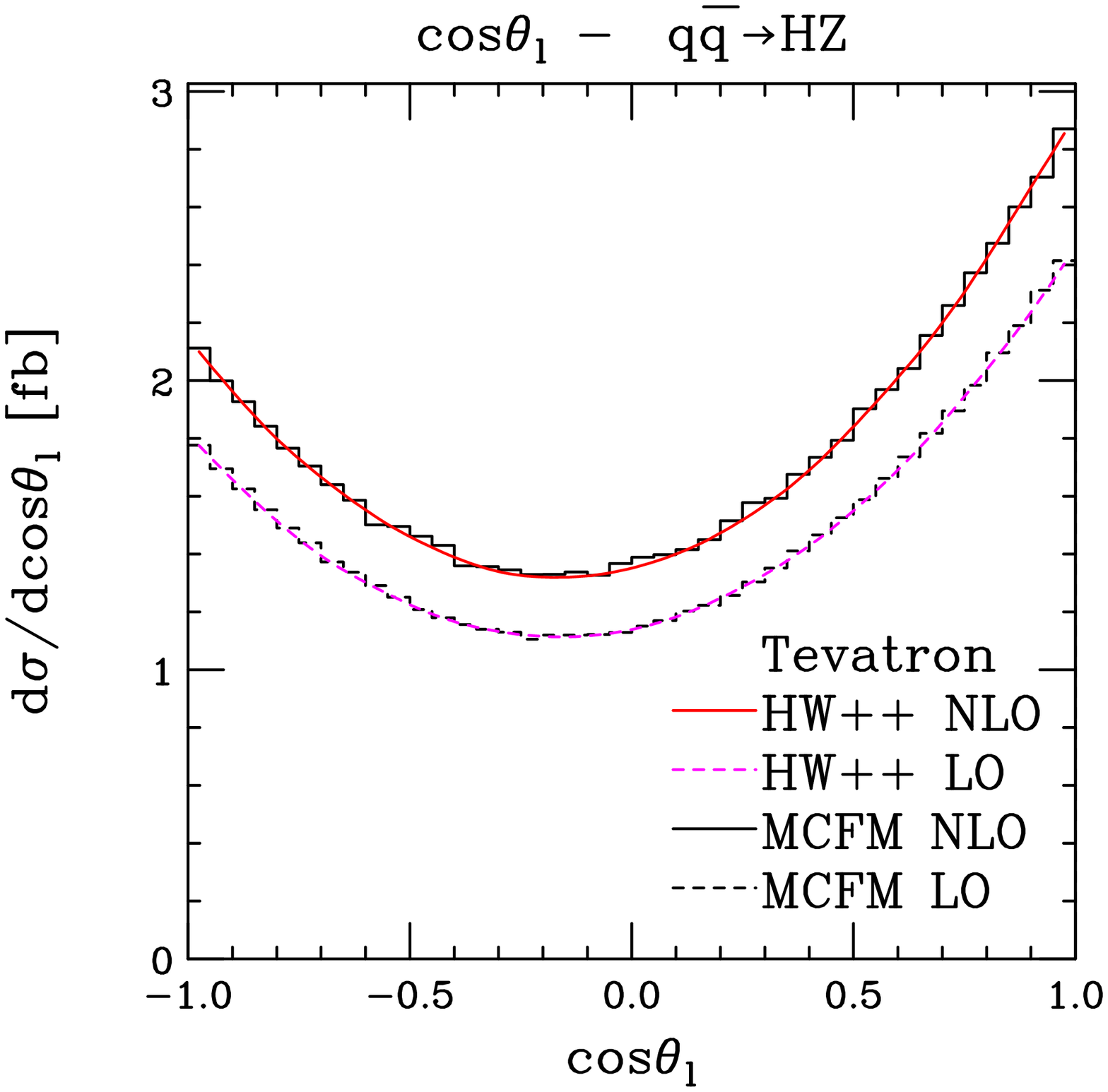}\hfill{}\includegraphics[width=0.4\textwidth,keepaspectratio,angle=90]{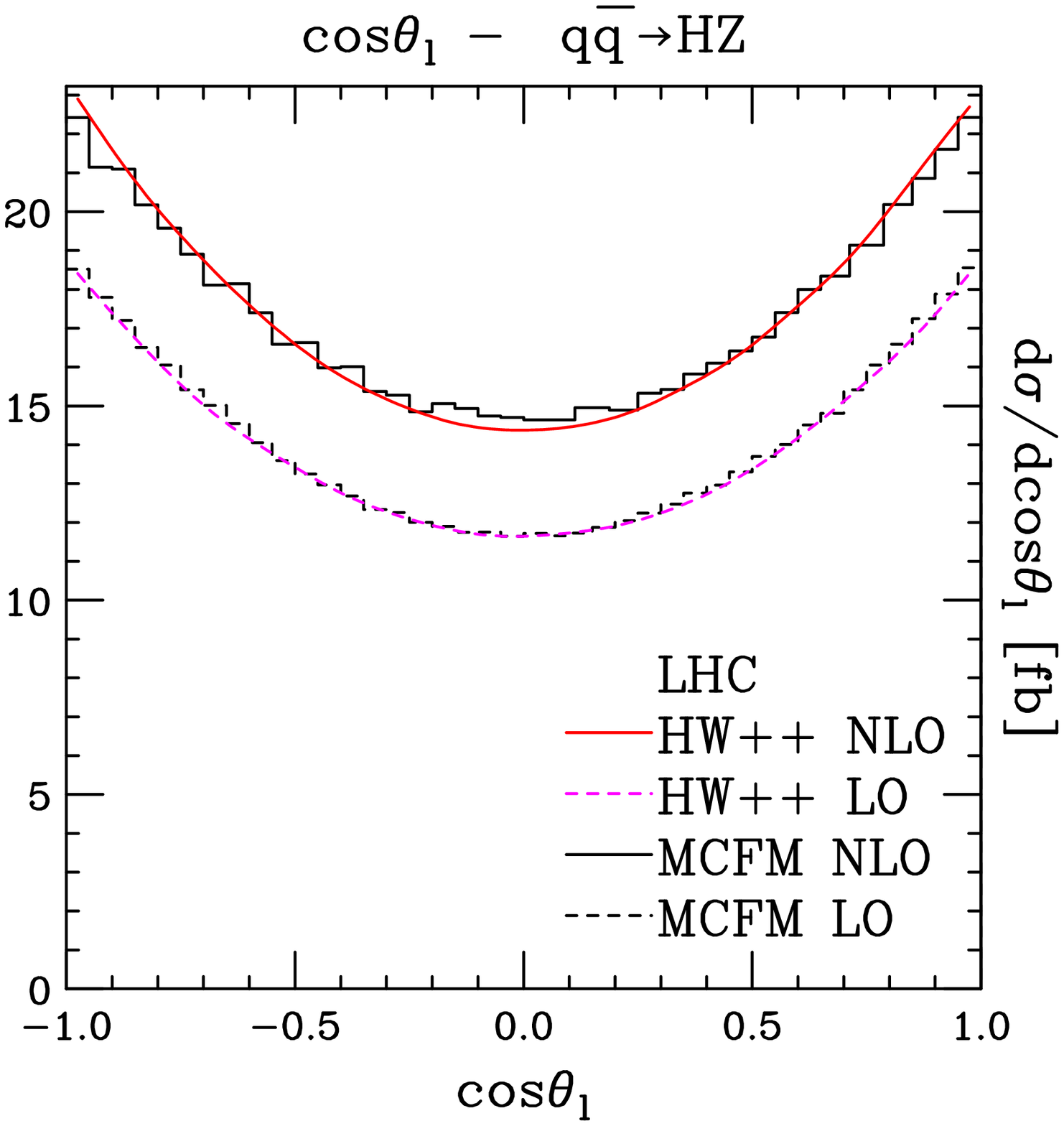}
\par\end{centering}

\caption{The polar angle of the electron produced by the vector boson, in its
rest frame, in Higgs-strahlung processes. The uppermost plots concern
the $q\bar{q}\rightarrow HW^{-}$ process and the lower plots relate
to the $q\bar{q}\rightarrow HZ$ process. For each process we have
displayed the results obtained at the Tevatron on the left and the
LHC on the right. The leading-order (LO) predictions of \HWPP\
and MCFM~\cite{Campbell:2000bg} are shown as dashed lines while
solid lines represent the corresponding NLO predictions. This is a
very important test of the correctness of the Kleiss trick, which
we have used to generate the NLO corrections independently of the
generation of the LO process \cite{Hamilton:2008pd}. }

\label{fig:higgs-strahlung_lepton_angles} 
\end{figure}

\begin{figure}[t]
\begin{centering}
\includegraphics[width=0.45\textwidth,keepaspectratio,angle=90]{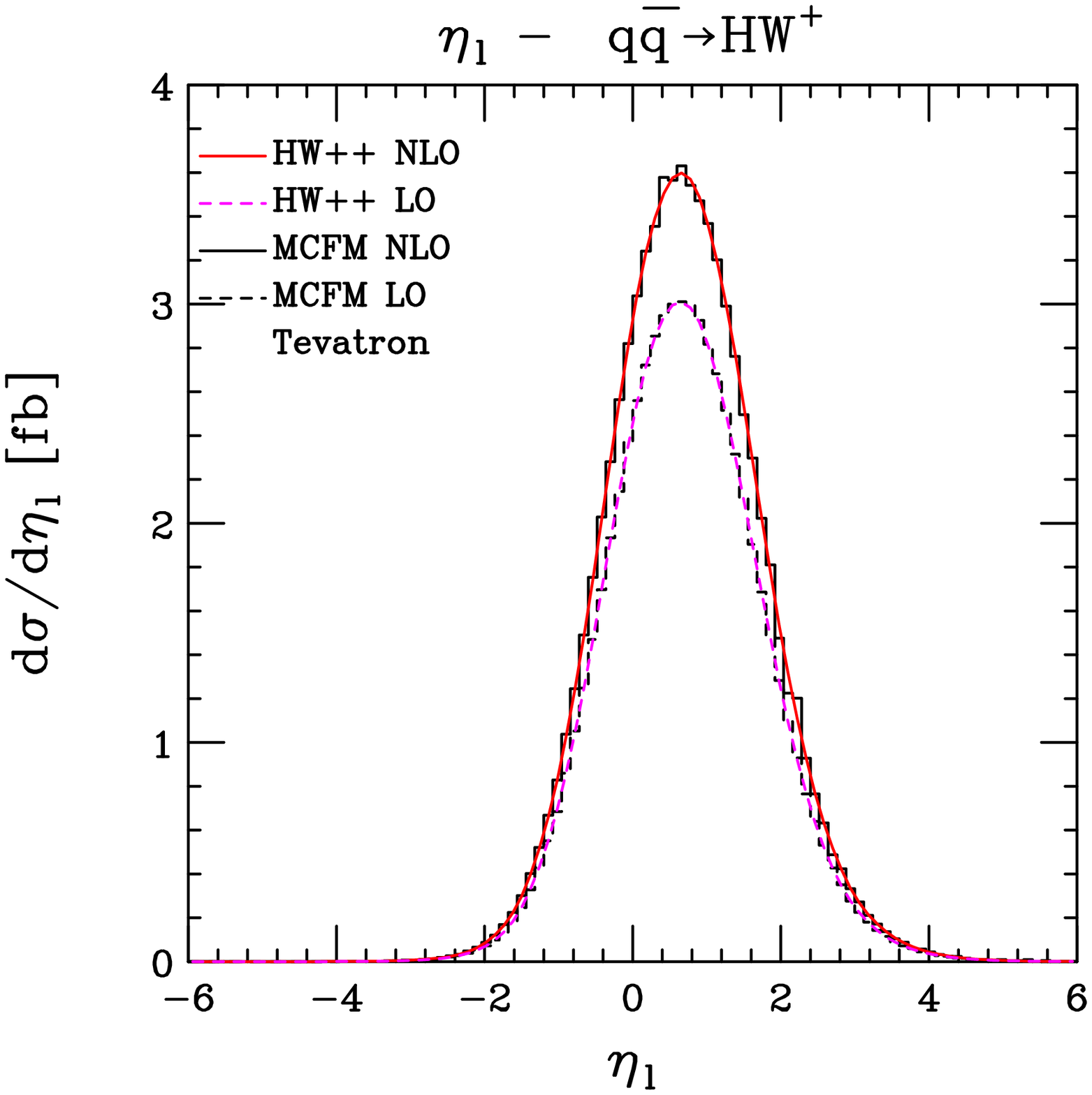}\hfill{}\includegraphics[width=0.45\textwidth,keepaspectratio,angle=90]{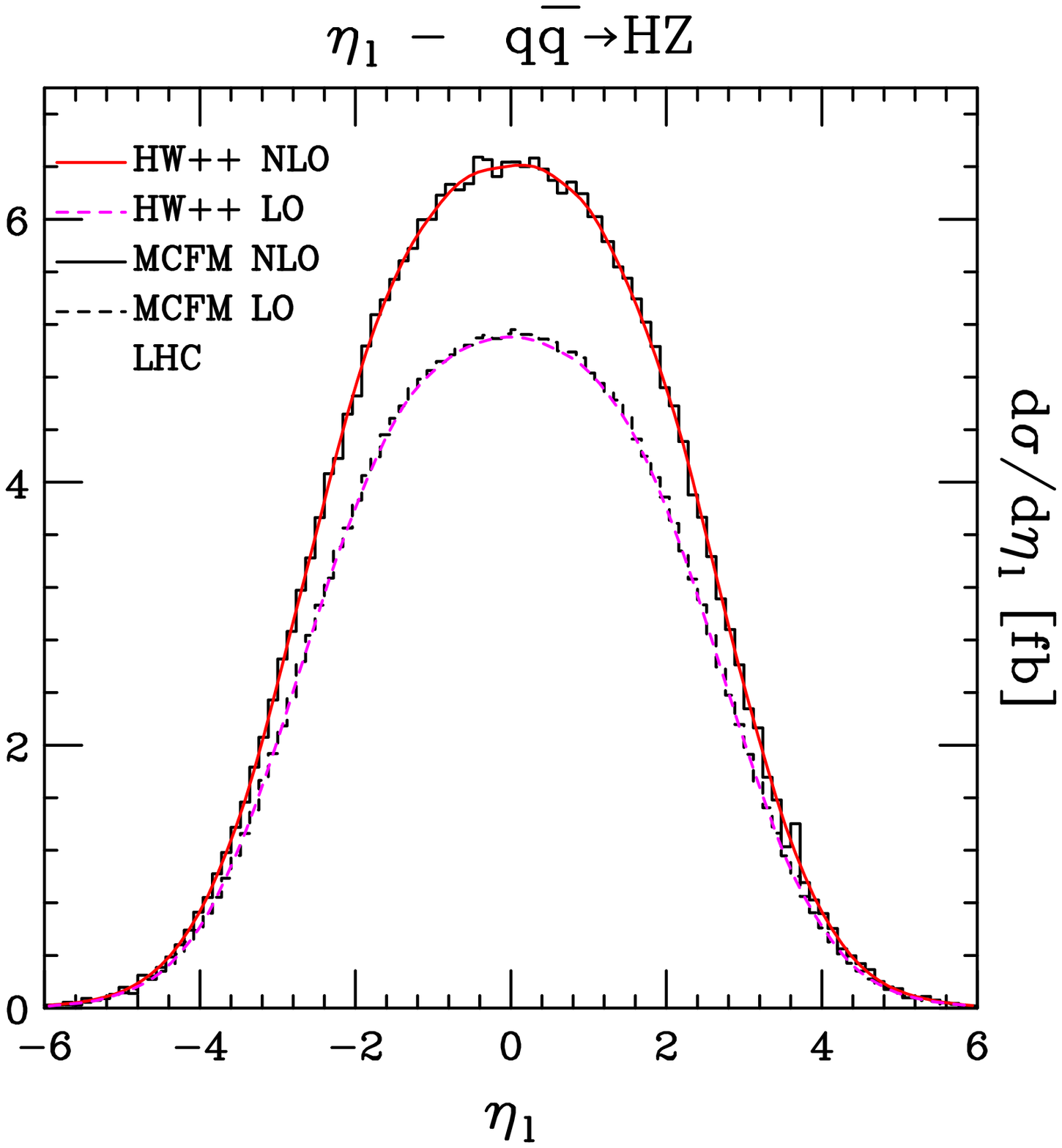}
\par\end{centering}

\caption{The pseudorapidity of the electron produced by the decaying vector
boson, in \mbox{$q\bar{q}\rightarrow HW^{+}$} (left) and $q\bar{q}\rightarrow HZ$
(right) Higgs-strahlung processes, from the \textsf{POWHEG} implementation
and also MCFM. As in Fig.\,\ref{fig:higgs-strahlung_lepton_angles}
the predictions for the Tevatron are given on the left ($p\bar{p},\,\sqrt{s}=1.96\,\mathrm{TeV}$)
and for the LHC on the right ($pp,\,\sqrt{s}=14\,\mathrm{TeV}$).
As well as the distribution of the lepton polar angle, this is a critical
test of the implementation of the Kleiss trick which factorizes the
generation of the LO and NLO calculations \cite{Hamilton:2008pd}.
It shows that all spin correlations have been correctly propagated
through to the vector boson decay products.}

\label{fig:higgs-strahlung_lepton_rapidity} 
\end{figure}

\begin{figure}[t]

\noindent \begin{centering}
\includegraphics[width=0.45\textwidth,keepaspectratio,angle=90]{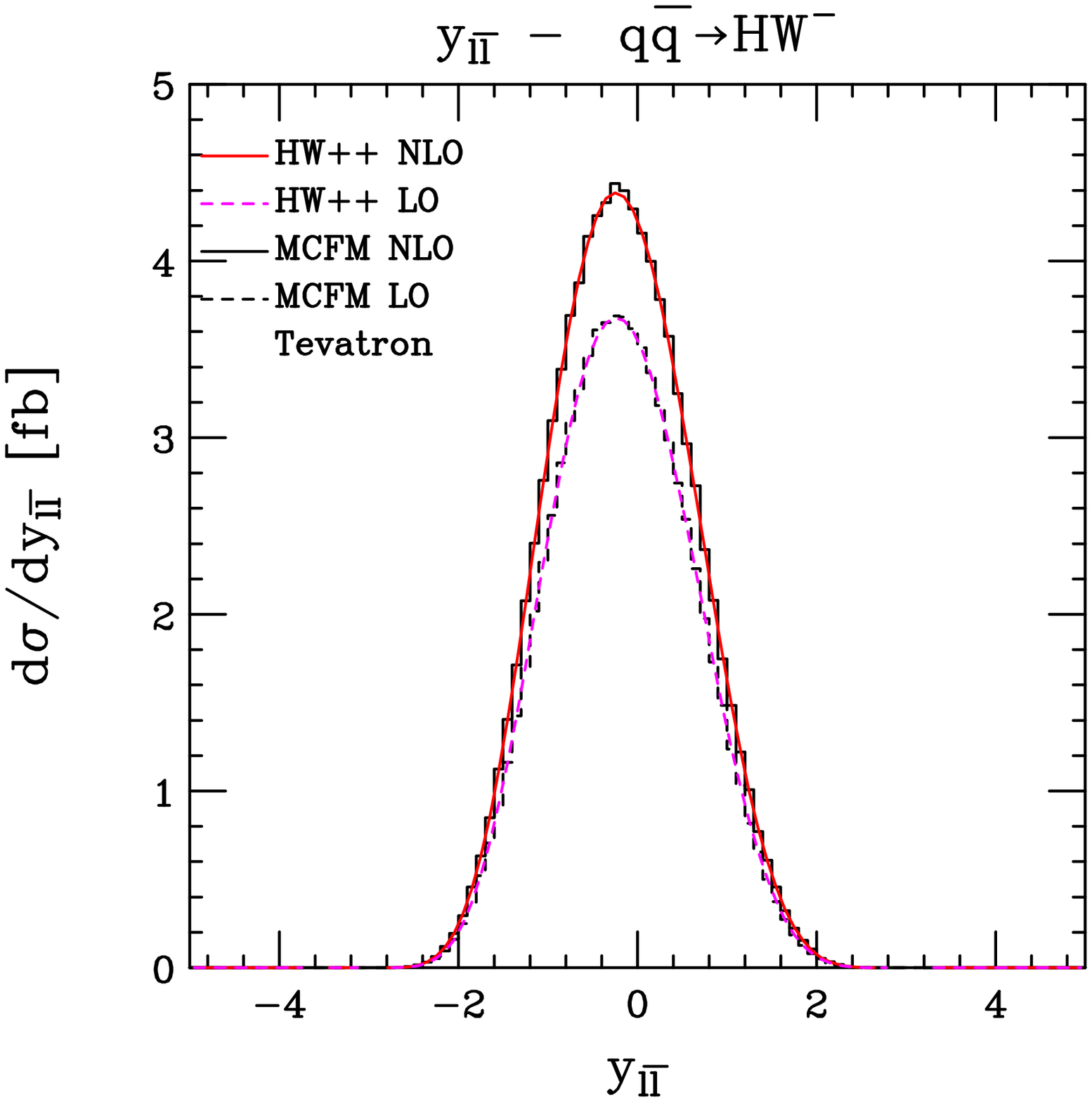}\hfill{}\includegraphics[width=0.45\textwidth,keepaspectratio,angle=90]{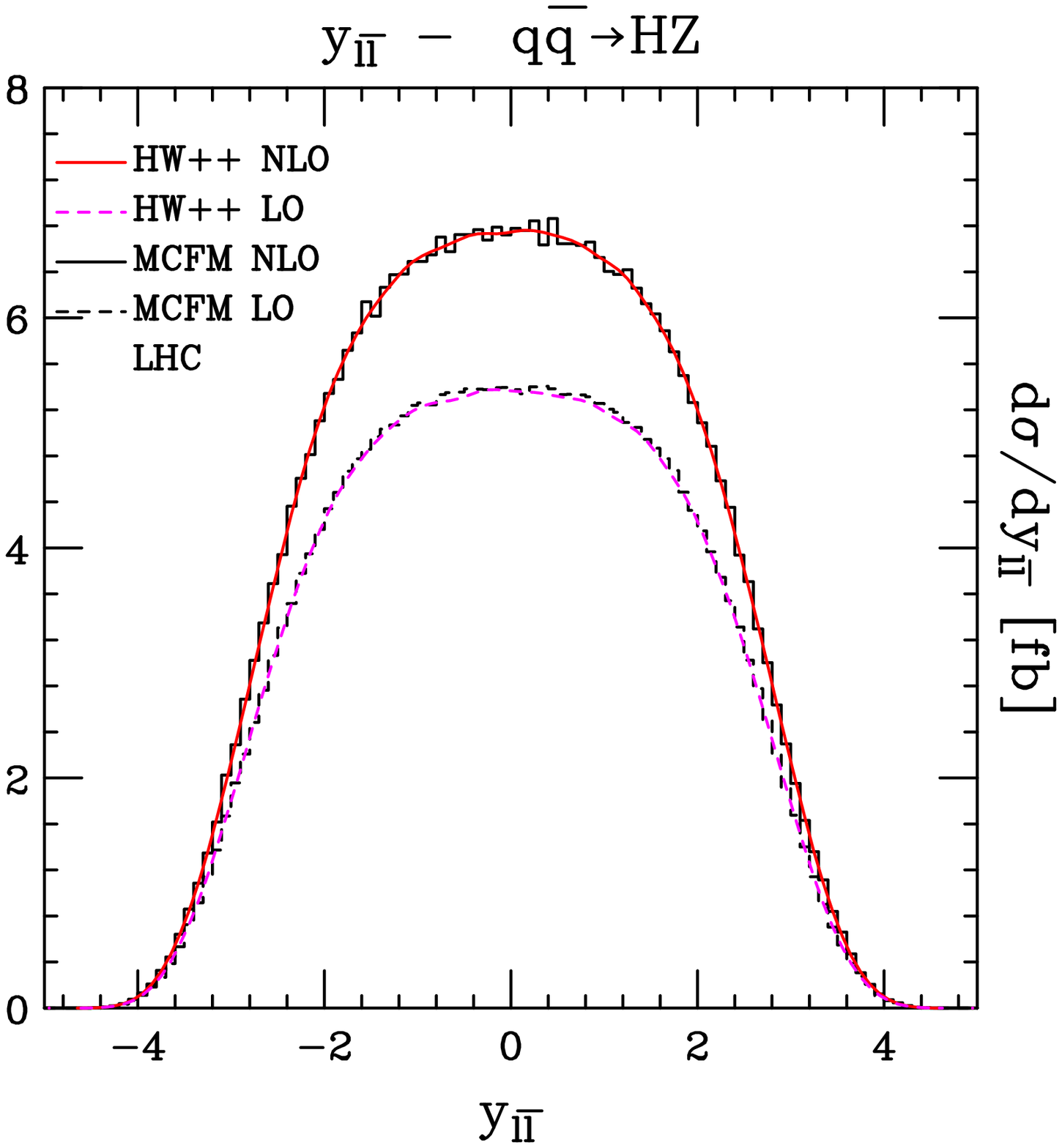}
\par\end{centering}

\caption{The rapidity distributions of the resonant vector boson in the Higgs-strahlung
processes, obtained with MCFM and the \HWPP\
\textsf{POWHEG} implementation. The left-hand plot was obtained considering
the $q\bar{q}\rightarrow HW^{-}$ process while the right-hand plot
relates to the $q\bar{q}\rightarrow HZ$ process. As with the previous
figures, on the left-hand side we show predictions for the Tevatron
and, on the right, predictions for the LHC. This is another fundamental
test of the functioning of the Kleiss trick, since, in using this
trick, the generation of the radiative variables is \emph{completely}
independent of the generation of the decay of the \emph{virtual vector
boson} \emph{viz.} the Higgs boson and the resonant vector boson.}

\label{fig:higgs-strahlung_y_llbar} 
\vspace{8mm}
%\end{figure}
%
%\begin{figure}[H]

\begin{centering}
\includegraphics[width=0.45\textwidth,keepaspectratio,angle=90]{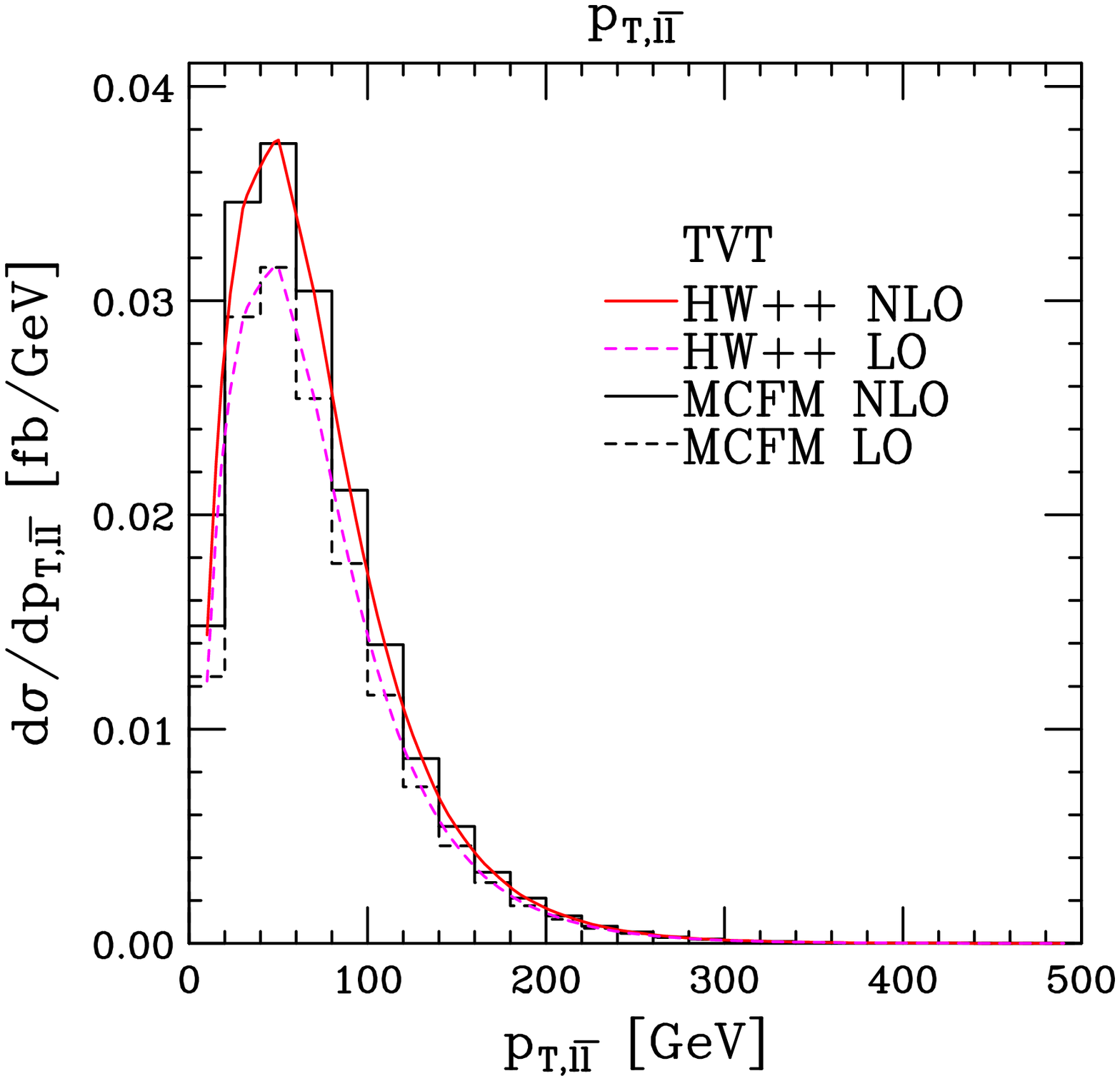}\hfill{}\includegraphics[width=0.45\textwidth,keepaspectratio,angle=90]{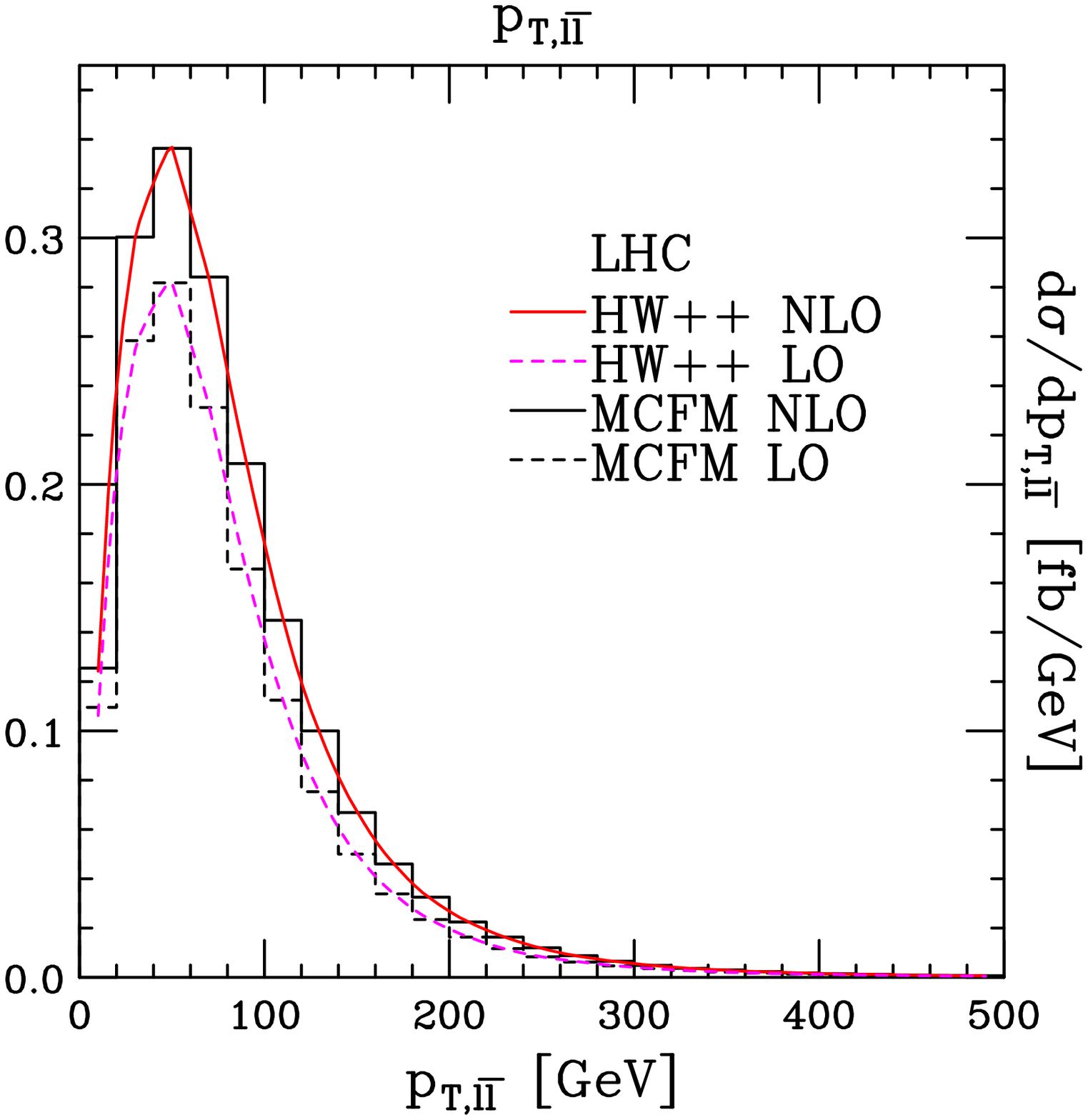}
\par\end{centering}

\caption{The transverse momentum of the $Z$ boson in Higgs-strahlung events
at the Tevatron, compared to MCFM. On the right hand side we show
an analogous comparison for the LHC. This distribution is a further
confirmation that the Kleiss trick is working correctly, since it
is sensitive to the details of the final state and not simply the
production of the initial off-shell vector boson.}

\label{fig:higgs-strahlung_pT_llbar} 
\end{figure}

The variety of Higgs-strahlung processes which can be simulated by
MCFM is limited, unlike \HWPP, hence we opted to carry out the Higgs-strahlung
comparisons using the following processes: $q\bar{q}\rightarrow HW^{+}\rightarrow e^{+}\nu_{e}b\bar{b}$,
$q\bar{q}\rightarrow HW^{-}\rightarrow e^{-}\bar{\nu}_{e}b\bar{b}$,
$q\bar{q}\rightarrow HZ\rightarrow e^{+}e^{-}b\bar{b}$. In Fig.\,\ref{fig:higgs-strahlung_mV_star}
we show the mass of the initial off-shell vector boson, one of the
Born variables for this process, for which there is very good agreement
between our \textsf{POWHEG} result and that of MCFM. In Figures\,\ref{fig:higgs-strahlung_mV_star}-\ref{fig:higgs-strahlung_pT_llbar},
we show a number of distributions with sensitivity to the details
of the decay of the vector boson, specifically, the polar angle of
the lepton produced by the decaying, resonant, vector boson in its
rest frame, the pseudorapidity of the final-state lepton, as well
as the rapidity and transverse momentum of the resonant vector boson
decaying to leptons. In all cases the agreement between our code and
MCFM is very good.

\subsection{Hardest emission generation and showering\label{sub:Hardest-emission-generation-results}}

In this section we focus on the distributions most sensitive to the
generation of the hardest emission (Sect.\,\ref{sub:genhard}) and
any further radiation obtained due to merging with the shower algorithm
(Sects.\,\ref{sec:The-POWHEG-method}, \ref{sub:Truncated-and-vetoed}).
In particular we study the $p_{T}$ spectra of the Higgs boson in
the gluon fusion process and the colourless vector boson plus Higgs
boson system in Higgs-strahlung processes. The distributions of the
rapidity difference between the leading, highest $p_{T}$, jet and
the produced colour neutral systems, $\mathrm{y}_{\mathrm{jet}}-\mathrm{y}_{\mathrm{H}}$
and $\mathrm{y}_{\mathrm{jet}}-\mathrm{y}_{\mathrm{VH}}$, are given
special attention, particularly in view of the differences noted in
previous works on the \textsf{POWHEG} formalism, arising between it
and \textsf{MC@NLO} \cite{Alioli:2008gx,Alioli:2008tz,Frixione:2007nu}.

Since we generate the hardest emission directly in terms of $p_{T}$
it is clear that the $p_{T}$ spectra are a direct test of this part
of the work. The relevance of the $\mathrm{y}_{\mathrm{jet}}-\mathrm{y}_{\mathrm{H}}$
and $\mathrm{y}_{\mathrm{jet}}-\mathrm{y}_{\mathrm{VH}}$ distributions
to these investigations is not immediately obvious. However, the $\mathrm{y}_{\mathrm{jet}}-\mathrm{y}_{\mathrm{H}}$
and $\mathrm{y}_{\mathrm{jet}}-\mathrm{y}_{\mathrm{VH}}$ variables,
for one emission, can be expressed \emph{purely} in terms of the radiative
variables $x$ and $y$, they are in fact both equal to $\mathrm{y}_{k}-\mathrm{y}$
as given by Eq.\,\ref{eq:nlo_1_17}, hence they are also an ideal
probe of the hardest emission generation. In order to gain some physical
insight into what this variable really means, we note that in the
limit that the angle between the radiated parton and colliding beam
partons tends to $90^{\mathrm{o}}$, in the partonic centre-of-mass
frame, Eq.\,\ref{eq:nlo_1_17} approximates to \begin{equation}
\mathrm{y}_{k}-\mathrm{y}=-\frac{2}{1+x}\,\left(\theta-\frac{\pi}{2}\right)\,,\label{eq:yjet-yh_approx_formula}\end{equation}
 where we remind the reader that $\theta$ is the angle between the
emitted parton and the $p_{\oplus}$ parton in that frame. Furthermore,
from Eq.\,\ref{eq:nlo_1_17} one can see that $\mathrm{y}_{k}-\mathrm{y}$
is maximised when the radiated parton is anticollinear to $p_{\oplus}$
and minimised when it is collinear to $p_{\oplus}$. Put simply, the
central region of the $\mathrm{y}_{k}-\mathrm{y}$ distribution is
populated by wide angle radiation while the tails are due to collinear
radiation.

The jets are defined using the longitutinal invariant $k_T$
algorithm~\cite{Catani:1993hr}
with an angular measure \mbox{$\Delta R = \sqrt{\Delta\eta^2+\Delta\phi^2}$},
where $\Delta\eta$ and $\Delta\phi$ are the pseudorapidity and
azimuthal angle differences between two particles respectively,
and the $E$ recomination scheme,
as implemented in the \textsf{KtJet} package~\cite{Butterworth:2002xg}. 

In studying the Higgs-strahlung process we shall compare our results
to three different approaches, namely, the bare angular-ordered parton
shower in \HWPP, the \HWPP\ parton shower including matrix element
corrections and also \textsf{MC@NLO}. In the case of the gluon fusion
process we compare to a further fourth prediction which is given by
modifying the hard component of the matrix element corrections in
\HWPP.

\subsubsection{The dead zone\label{sub:The-dead-zone}}

In order to better understand the predictions from the \HWPP\
shower, with and without matrix element corrections, and also those
of \textsf{MC@NLO}, it will help to understand the phase space available
for the first emission in the shower approach.

\begin{figure}[t]

\begin{centering}
\includegraphics[width=0.45\textwidth,keepaspectratio]{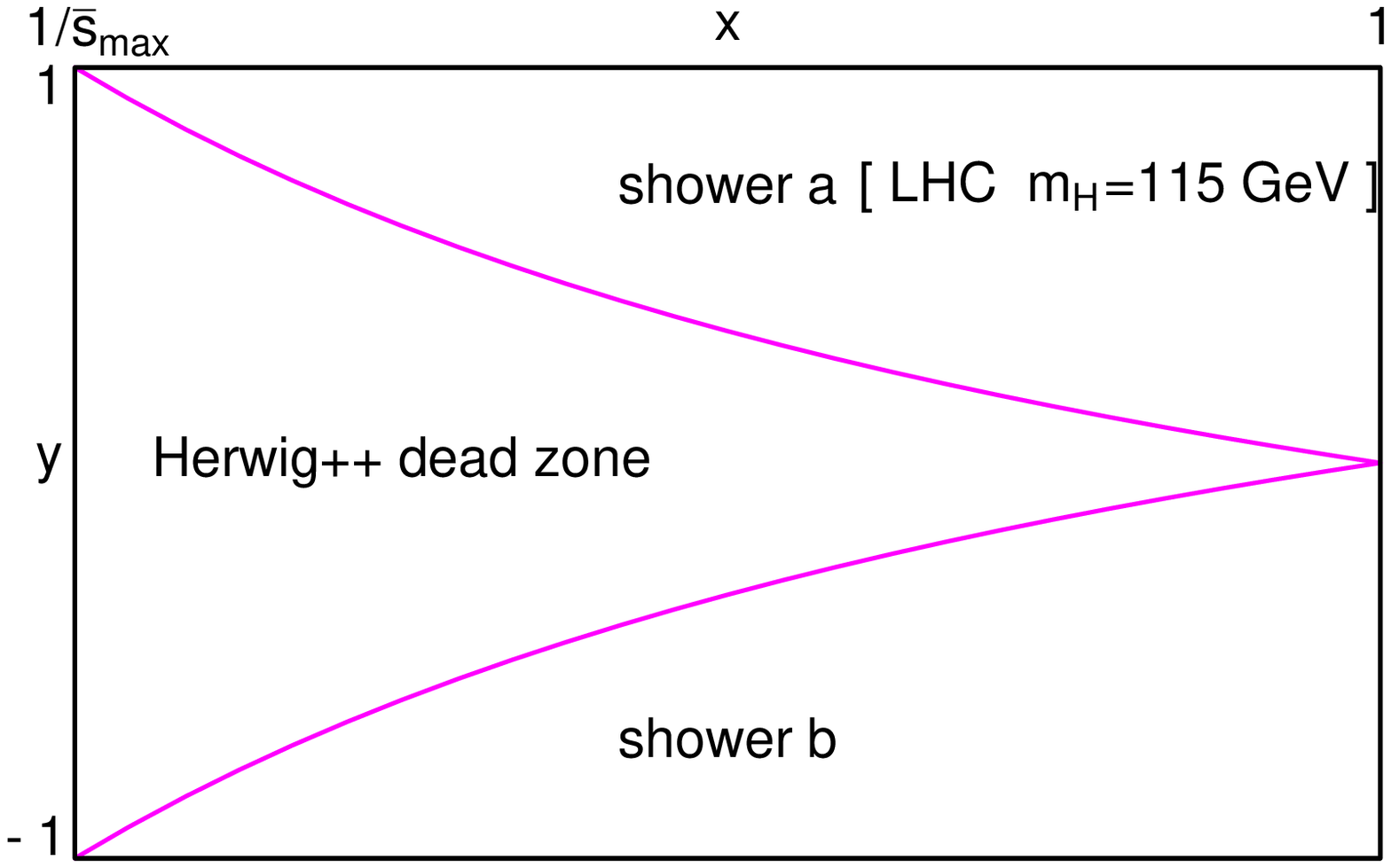}\hfill{}\includegraphics[width=0.45\textwidth,keepaspectratio]{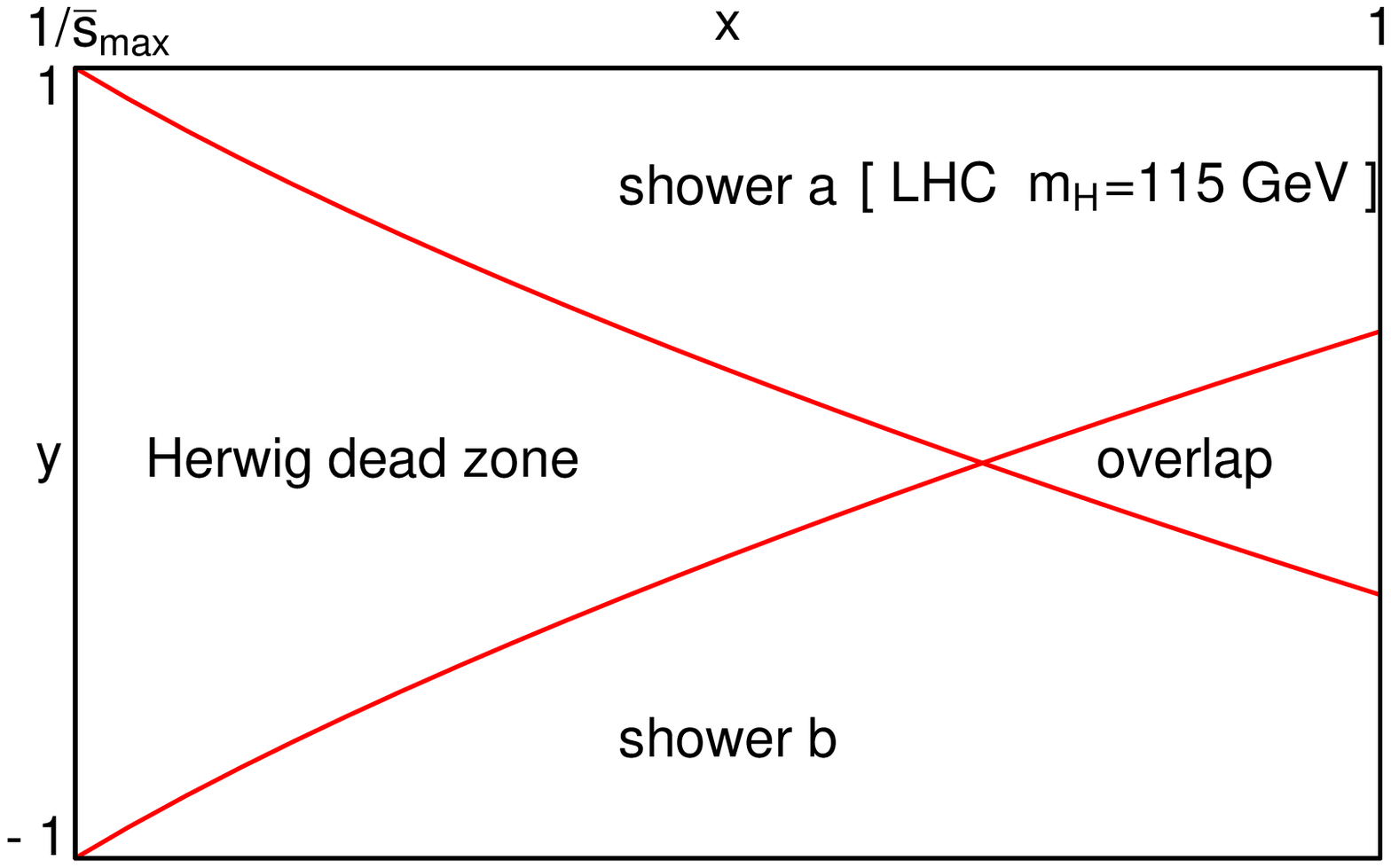}
\par\end{centering}

\caption{\label{fig:phase_space} The \emph{full} radiative phase space is
given by the black rectangle in the $x$, $y$ plane, bounding the
plots on the right. The minimum kinematically allowed value of $x$
is shown as $1/\bar{s}_{max}=p^{2}/s$. The region into which two
colour connected partons $a$ and $b$ cannot emit radiation, according
to the \HWPP\ and \textsf{HERWIG} shower algorithms, is marked `dead
zone' on the left- and right-hand plots respectively. In both algorithms,
radiation into this dead zone is only possible with the help of a
\emph{hard matrix element correction}. Unlike \HWPP, in the case
of the \textsf{HERWIG} algorithm there is a region of phase space
which is double counted by the showering from $a$ and $b$ marked
`overlap'; this double counting is ultimately corrected for by a veto
procedure. These plots correspond specifically to the case of a 115
GeV Higgs boson being produced via the gluon fusion process at LHC
energies, although their form does not change significantly for the
other processes we consider.}

\end{figure}

In general, given a leading-order configuration, one must first specify
starting scales for each parton to evolve down from. The guiding principle
behind the choice of starting scales in the \textsf{HERWIG} and \HWPP\
angular-ordered shower algorithms is as follows: given two colour
connected \emph{shower} \emph{progenitor partons} $a$ and $b$, progenitor
\emph{$a$} cannot emit any \emph{soft} radiation into the hemisphere
defined by the direction of progenitor $b$, in the rest frame of
$a$ and $b$, and \emph{vice-versa}. This avoids any double counting
of the phase space%
\footnote{In practice, in the case of \textsf{HERWIG}, there is a small amount
of overlap in the phase space allotted to each shower progenitor,
although the algorithm later corrects for this by a vetoing procedure. %
}. In other words, the starting scales are fixed by constraining that,
in the limit of soft emissions, the two \emph{jet regions}, of the
first emission phase space (those which either progenitor can emit
into) meet smoothly and do not overlap. If one then plots, in the
full phase space, the contours to which these values of the evolution
variables correspond to, one finds three regions: a jet region into
which progenitor $a$ can emit, a jet region into which progenitor
$b$ can emit and a further region into which neither can emit, the
so-called \emph{dead zone}. We show exactly these phase-space regions for
the \textsf{HERWIG} \cite{Corcella:2000bw,Corcella:1999gs} and \HWPP\ \cite{Gieseke:2003rz,Bahr:2008pv}
algorithms in Fig.\,\ref{fig:phase_space}, taking the case of gluon
fusion at LHC energies as an example.

At this point we wish to remind the reader that in the distributions
of $\mathrm{y}_{\mathrm{jet}}-\mathrm{y}_{\mathrm{H}}$ and $\mathrm{y}_{\mathrm{jet}}-\mathrm{y}_{\mathrm{HV}}$
which follow, the region around zero corresponds to the emission of
the jet at right angles to the colliding partons in the partonic centre
of mass frame (Eq.\,\ref{eq:yjet-yh_approx_formula}). This in turn
corresponds to the region either side of the line $y=0$ in the phase
map shown in Fig.\,\ref{fig:phase_space}, along which the volume
of the dead zone is \emph{maximised}.

\begin{figure}[t]
\begin{centering}
\includegraphics[width=0.45\textwidth,keepaspectratio]{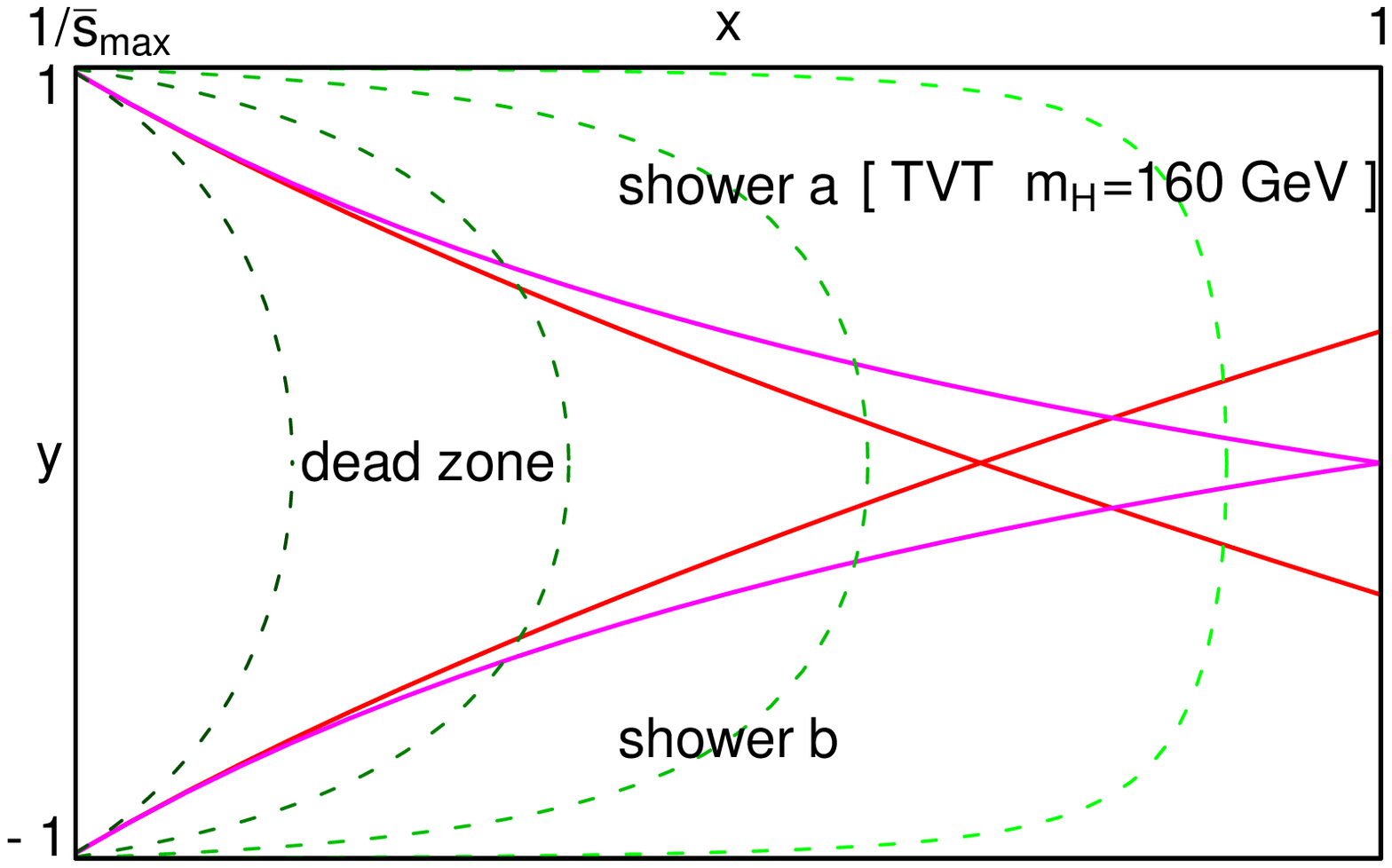}\hfill{}\includegraphics[width=0.45\textwidth,keepaspectratio]{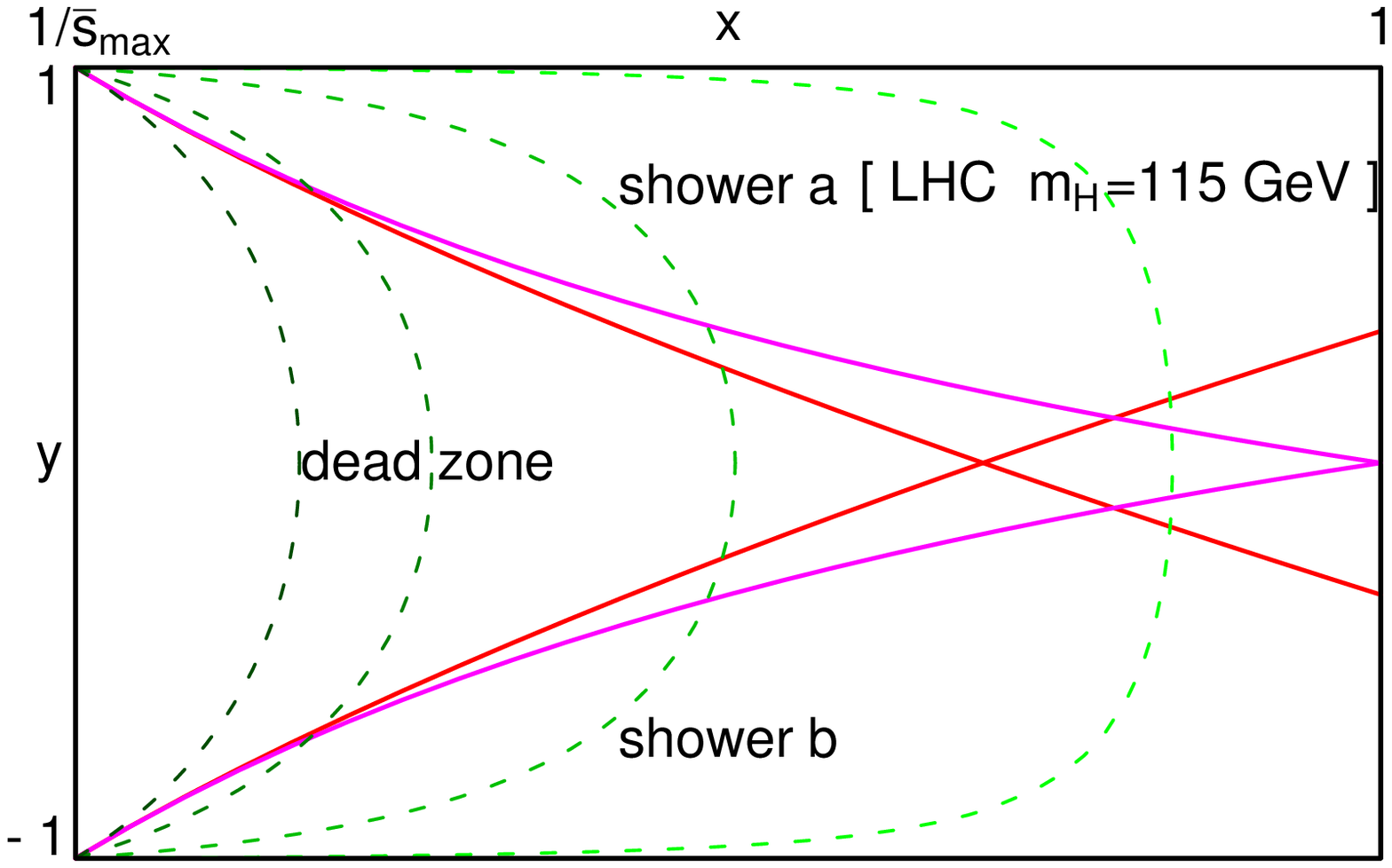}
\par\end{centering}

\begin{centering}
\vspace{5mm}
 \includegraphics[width=0.45\textwidth,keepaspectratio]{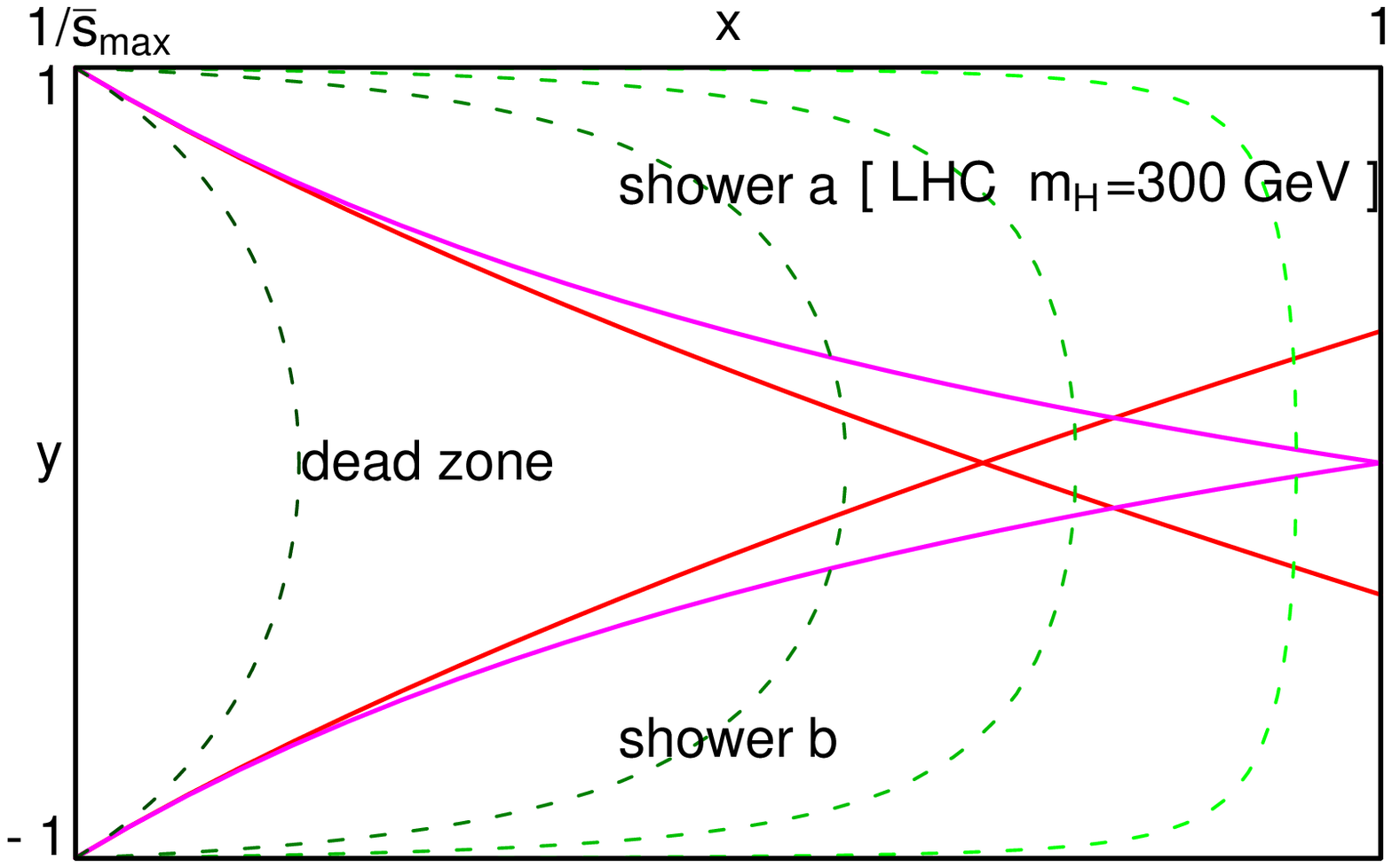}
\par\end{centering}

\caption{\label{fig:phase_space_with_cuts}Radiative phase space in the $x$,
$y$ plane, as in Fig.\,\ref{fig:phase_space} but with contours
corresponding to constant values of $p_{T}$ superimposed in green,
from lightest to darkest (right to left) respectively these are, $p_{T}=$10
GeV, $p_{T}=$40 GeV, $p_{T}=$80 GeV and $p_{T}=m_{H}$. The region
into which the two shower progenitors cannot emit radiation is again
marked `dead zone'; for \textsf{HERWIG} this area is bounded in red
while for \HWPP\ it lies between the magenta lines. The three plots
correspond to the three scenarios under study in the remainder of
the paper: a Higgs boson of mass 160 GeV at the Tevatron, a Higgs
boson of mass of 115 GeV at the LHC, and a Higgs boson of mass 300
GeV also at the LHC. The area with emissions $p_{T}>m_{H}$ is entirely
within the dead zone in all scenarios.}

\end{figure}

In Fig.\,\ref{fig:phase_space_with_cuts} we superimpose, on the
$x$, $y$ phase-space map of the gluon fusion process, four contours
corresponding to constant values of $p_{T}$: 10 GeV, 40 GeV, 80 GeV
and $m_{\mathrm{H}}$. This was done for the three scenarios which
we study using the Monte Carlo predictions, specifically, a Higgs
boson mass of 160 GeV at Tevatron energies (1960 GeV), a Higgs boson
mass of 115 at LHC energies (14 TeV) and finally a Higgs boson mass
of 300 GeV, also at LHC energies. In doing this we see that restricting
the phase space to regions with higher and higher $p_{T}$ leads to
the expected result, namely, that the dead zone begins to fill the
allowed region.%

For the three scenarios we consider, Fig.\,\ref{fig:phase_space_with_cuts}
shows that a cut of $p_{T}>$80 GeV already leads to a great reduction
in the area of phase space populated by the shower, while for $p_{T}>m_{\mathrm{H}}$
the allowed region is fully contained within the dead zone. Similar
plots are obtained for the Higgs-strahlung process by simply replacing
the Higgs boson mass used in the calculation of the gluon fusion phase
space, by the typical mass of the colourless vector boson plus Higgs
boson system (Fig.\,\ref{fig:higgs-strahlung_mV_star} shows this
to be in the range 200-300 GeV). The maps in Figs.\,\ref{fig:phase_space}
and \ref{fig:phase_space_with_cuts} are key to a good understanding
of the results in Sects.\,\ref{sub:Gluon-fusion-plots}, \ref{sub:Higgs-strahlung-plots}.

As noted earlier, the dead zone of phase space, into which the shower
cannot emit, can be filled with the aid of a hard matrix element correction.
This involves populating that region according to the single real
emission matrix element squared \cite{Seymour:1994df,Bahr:2008pv}.
In principle this is a simple procedure, since the dead zone does
not run into any singular regions of phase space%
\footnote{Although the `throat' of the dead zone in \HWPP\ touches the soft
boundary, it does so only in a vanishing region of the dead zone phase
space, which is anyway cut off by the gluon mass regulator used in
the shower.%
}: given an underlying \emph{N}-body configuration one selects whether
an emission into the dead zone occurs according to the conditional
probability,\begin{equation}
\mathcal{P}_{\mathrm{dead}}^{\mathrm{HW}}\left(\Phi_{B}\right)=1-\exp\left[-\int_{\mathrm{dead}}\mathrm{d}\Phi_{R}\,\frac{\widehat{R}\left(\Phi_{B},\Phi_{R}\right)}{B\left(\Phi_{B}\right)}\,\right]\,,\label{eq:me_corr_dead_zone_prob}\end{equation}
 and, if an emission is to be generated, it is distributed according
to the single real emission matrix element squared including PDFs
($R\left(\Phi_{B},\Phi_{R}\right)$). Neglecting terms beyond NLO
accuracy, $\mathcal{P}_{\mathrm{dead}}^{\mathrm{HW}}\left(\Phi_{B}\right)$,
integrated over the Born variables, gives the fraction of the NLO
cross section which the dead zone would contribute.

In all cases the hard matrix element correction is accompanied by
a soft matrix element correction, which corrects the distribution
of the hardest emission in the parton shower regions so that it is
also given by $R\left(\Phi_{B},\Phi_{R}\right)$. The combination
of the soft and hard matrix element corrections ensures that, to $\mathcal{O}\left(\alpha_{\mathrm{S}}\right)$,
the distribution of real radiation is exactly matched either side
of the dead zone boundaries, \emph{i.e.} any sensitivity on the position
of the dead zone boundary will be at the level of $\mathcal{O}\left(\alpha_{\mathrm{S}}^{2}\right)$
terms, which should not be logarithmically enhanced since the boundary
is predominantly in the high $p_{T}$ region.

The same exact matching at $\mathcal{O}\left(\alpha_{\mathrm{S}}\right)$
is true in the \textsf{MC@NLO} program, which feeds events to the
\textsf{HERWIG} parton shower. In this case the generation of the
first emission is done according to the full NLO differential cross
section with additional, resummed, higher order corrections entering
in the shower regions of the phase space. Sensitivity to the dead
zone boundary is present in the NLO calculation through the shower
subtraction terms, which are required to avoid double counting of
the NLO contributions through the subsequent showering with \textsf{HERWIG}.
Ultimately, in \textsf{MC@NLO}, one has that the emission rate in
the shower regions is given by the resummed rate in the shower's Sudakov
form factor, corrected at $\mathcal{O}\left(\alpha_{\mathrm{S}}\right)$,
while in the dead zone it is related to the fraction which that area
contributes to the total NLO cross section. This is much like the
case of the matrix element correction procedure and, as with that
method, any mis-match between the shower region and the dead zone
should therefore again relate to unenhanced $\mathcal{O}\left(\alpha_{\mathrm{S}}^{2}\right)$
terms. We stress that, unlike the matrix element correction method,
the \textsf{MC@NLO} program includes also exact NLO virtual corrections
to the process in the event generation process, giving full NLO accuracy.
However, as we now wish to focus on the shapes of the distributions
sensitive to the effects of real radiation, neglecting the overall
normalisation, differences arising from virtual corrections only enter
the results through terms beyond NLO accuracy. This last point is
addressed further at the beginning of Sect.\,\ref{sub:Gluon-fusion-plots}. 

As we have already described in Sects.\,\ref{sec:The-POWHEG-method}
and \ref{sec:Implementation}, the \textsf{POWHEG} method generates
the hardest emission completely independently of the detailed workings
of the shower, it may be considered as being a $p_{T}$ ordered shower
in its own right, albeit for a single emission. Unlike the other methods,
\textsf{POWHEG} therefore has no dependence whatsoever on the dead
zone boundary and the way in which it generates the hardest emission
generation is the same throughout the whole phase space.

\subsubsection{Gluon fusion\label{sub:Gluon-fusion-plots}}

In this subsection we compare our \textsf{POWHEG} simulation against
predictions from the bare angular-ordered parton shower in \HWPP,
the \HWPP\ parton shower including matrix element corrections and
\textsf{MC@NLO}. We also compare against a further prediction from
the matrix element correction (MEC) procedure, in which we decrease
the rate of emission into the dead zone by unenhanced terms of $\mathcal{O}\left(\alpha_{S}^{2}\right)$,
by changing the denominator in $\mathcal{P}_{\mathrm{dead}}^{\mathrm{HW}}\left(\Phi_{B}\right)$
from $B\left(\Phi_{B}\right)$ to $\overline{B}\left(\Phi_{B}\right)$:
\begin{equation}
\mathcal{P}_{\mathrm{dead}}^{\mathrm{HW}}\left(\Phi_{B}\right)\rightarrow\mathcal{P}_{\mathrm{dead}}^{\mathrm{NLO}}\left(\Phi_{B}\right)=\int_{\mathrm{dead}}\mathrm{d}\Phi_{R}\mbox{ }\frac{\widehat{R}\left(\Phi_{B},\Phi_{R}\right)}{\overline{B}\left(\Phi_{B}\right)}\,.\label{eq:me_corr_mod_dead_zone_prob}\end{equation}
 Whereas integrating $\mathcal{P}_{\mathrm{dead}}^{\mathrm{HW}}\left(\Phi_{B}\right)$
over the Born variables gives the fraction of the NLO cross section
contributed by the dead zone neglecting higher order terms, performing
the same integral with $\mathcal{P}_{\mathrm{dead}}^{\mathrm{NLO}}\left(\Phi_{B}\right)$
gives this fraction \emph{exactly}%
\footnote{For this reason we choose to distinguish the modified probability
by the superscript NLO.%
}. Although it should be obvious to the reader that this is technically
only an alteration at the level of terms beyond NLO accuracy, we do
not wish to give the impression that it is \emph{a priori} a small
change, at least not for the gluon fusion process (recall Fig.\,\ref{fig:ggh_born variables}).
We expect that this alteration should mean that the modified MEC predictions
should reproduce well the rate at which \textsf{MC@NLO} emits radiation
into the dead zone.

\begin{figure}[H]
\begin{centering}
\includegraphics[width=0.38\textwidth,height=0.48\textwidth,angle=90]{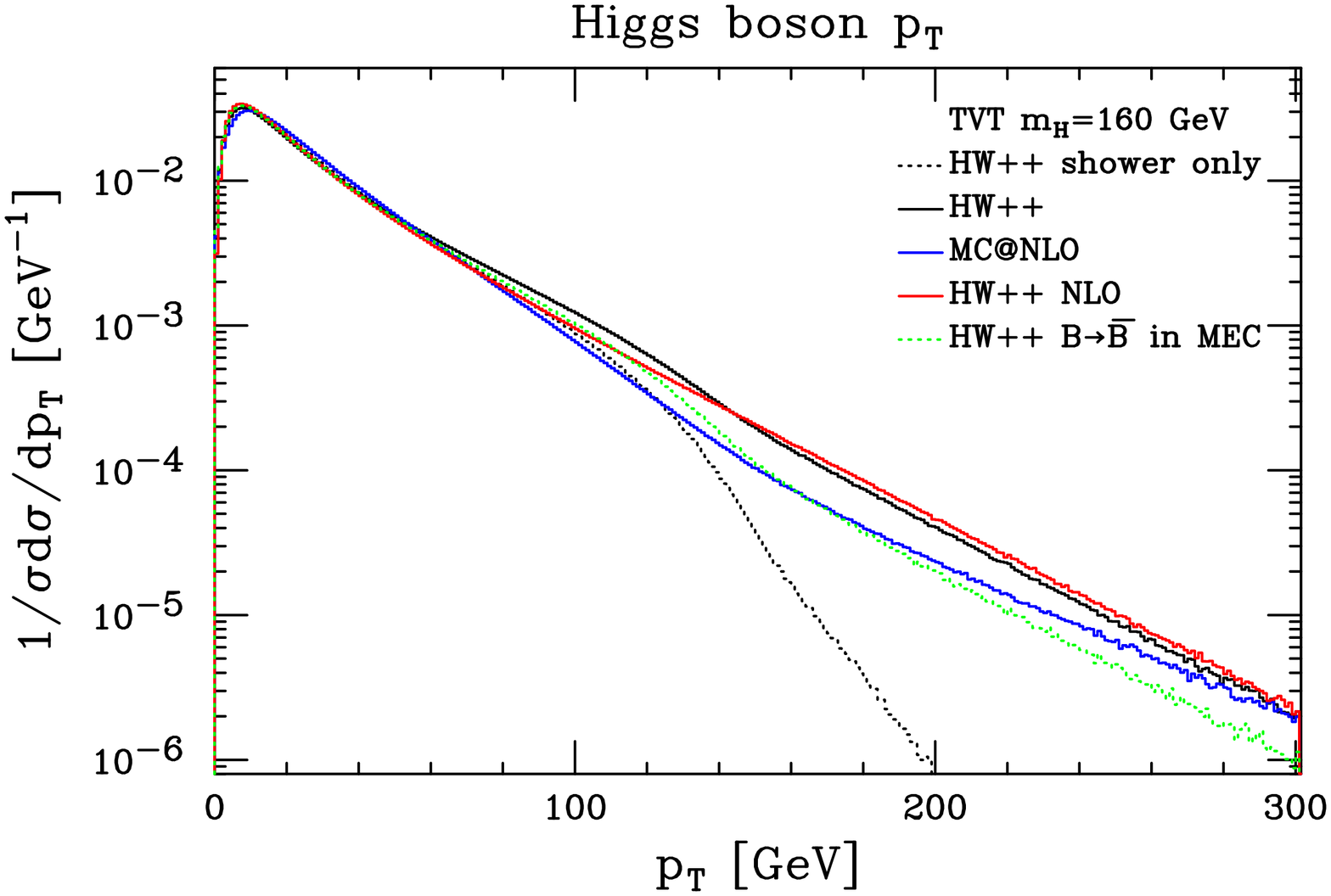}\hfill{}\includegraphics[width=0.38\textwidth,height=0.48\textwidth,angle=90]{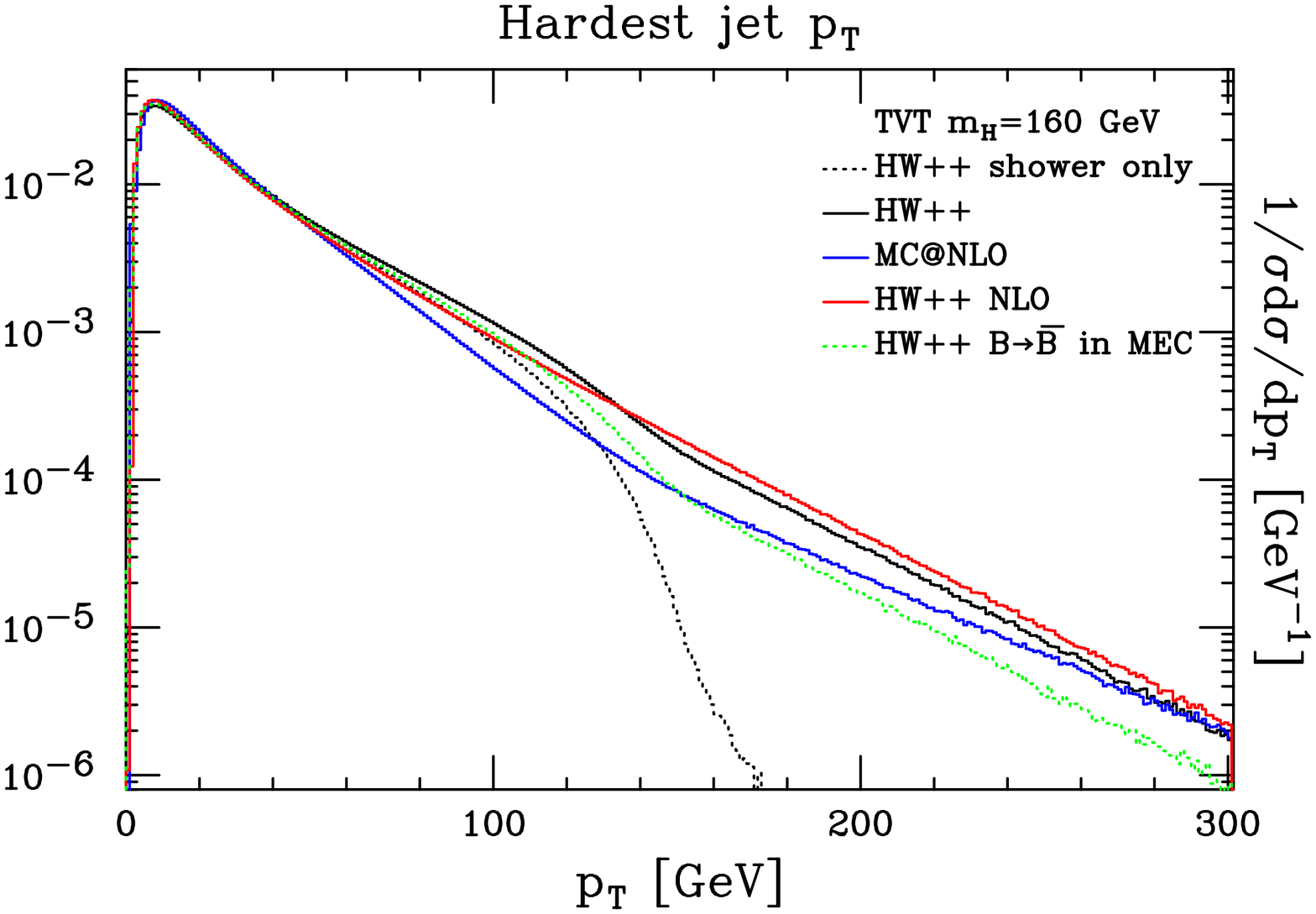}
\par\end{centering}

\vspace{10mm}

\begin{centering}
\includegraphics[width=0.38\textwidth,height=0.48\textwidth,angle=90]{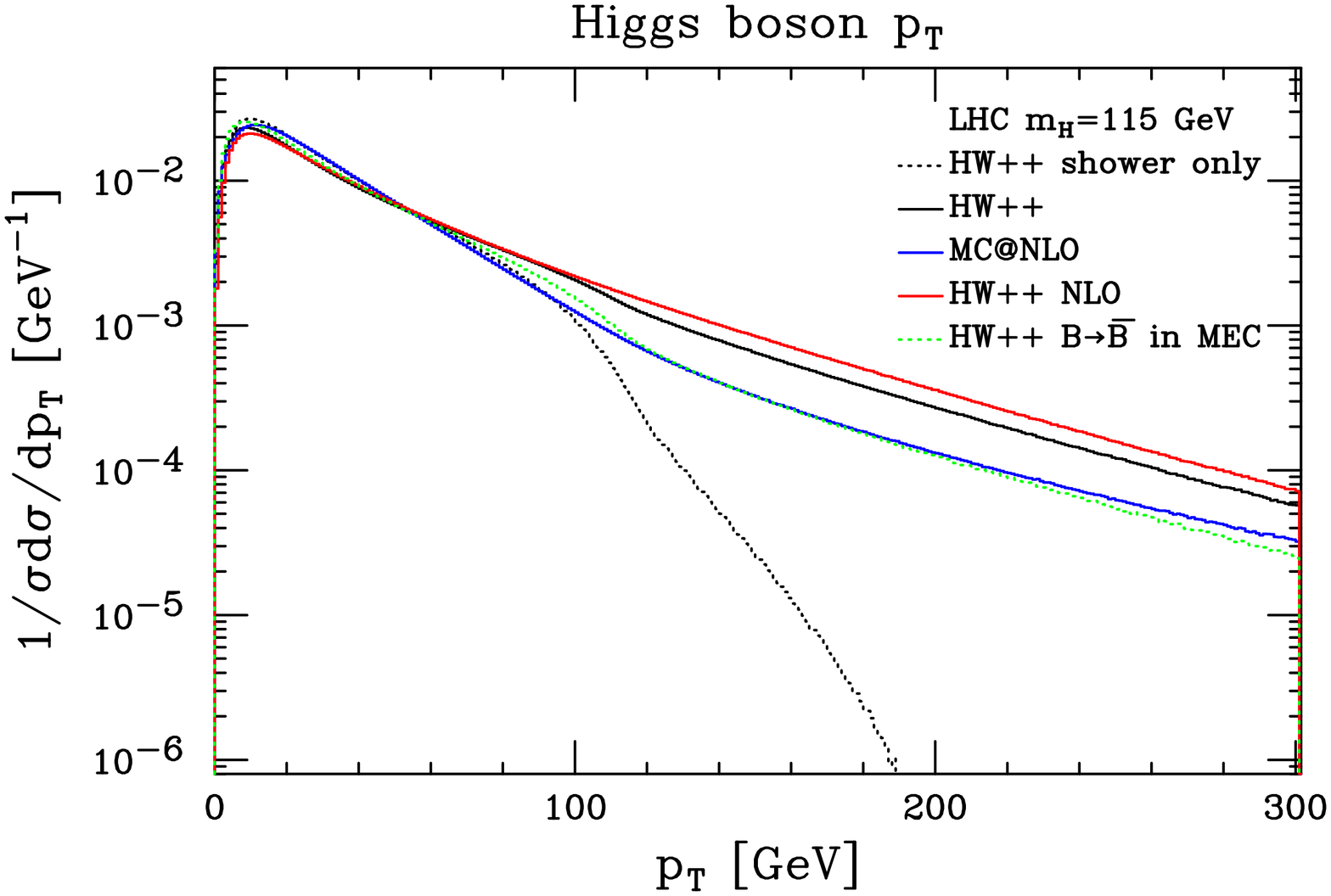}\hfill{}\includegraphics[width=0.38\textwidth,height=0.48\textwidth,angle=90]{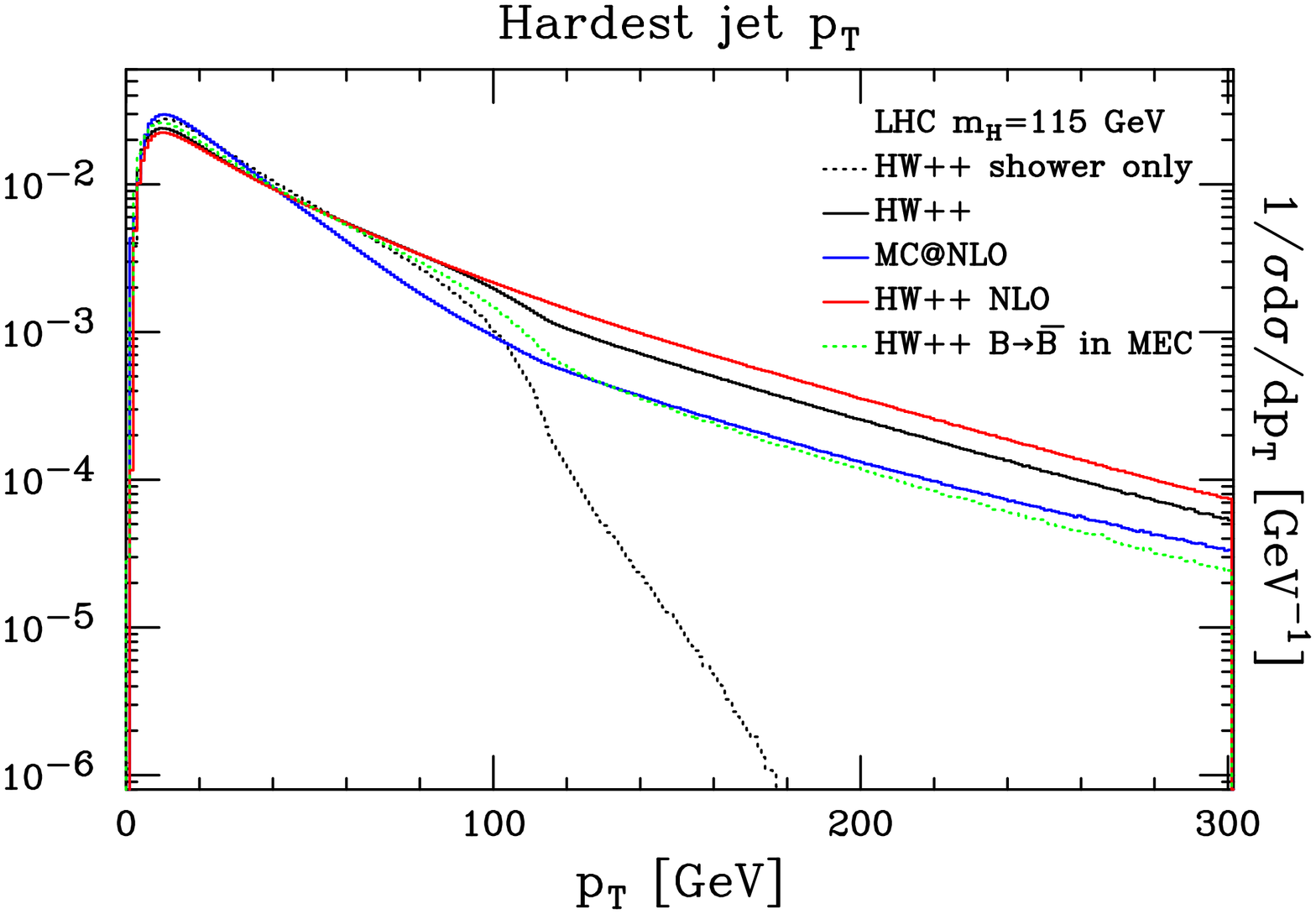}
\par\end{centering}

\vspace{10mm}

\begin{centering}
\includegraphics[width=0.38\textwidth,height=0.48\textwidth,angle=90]{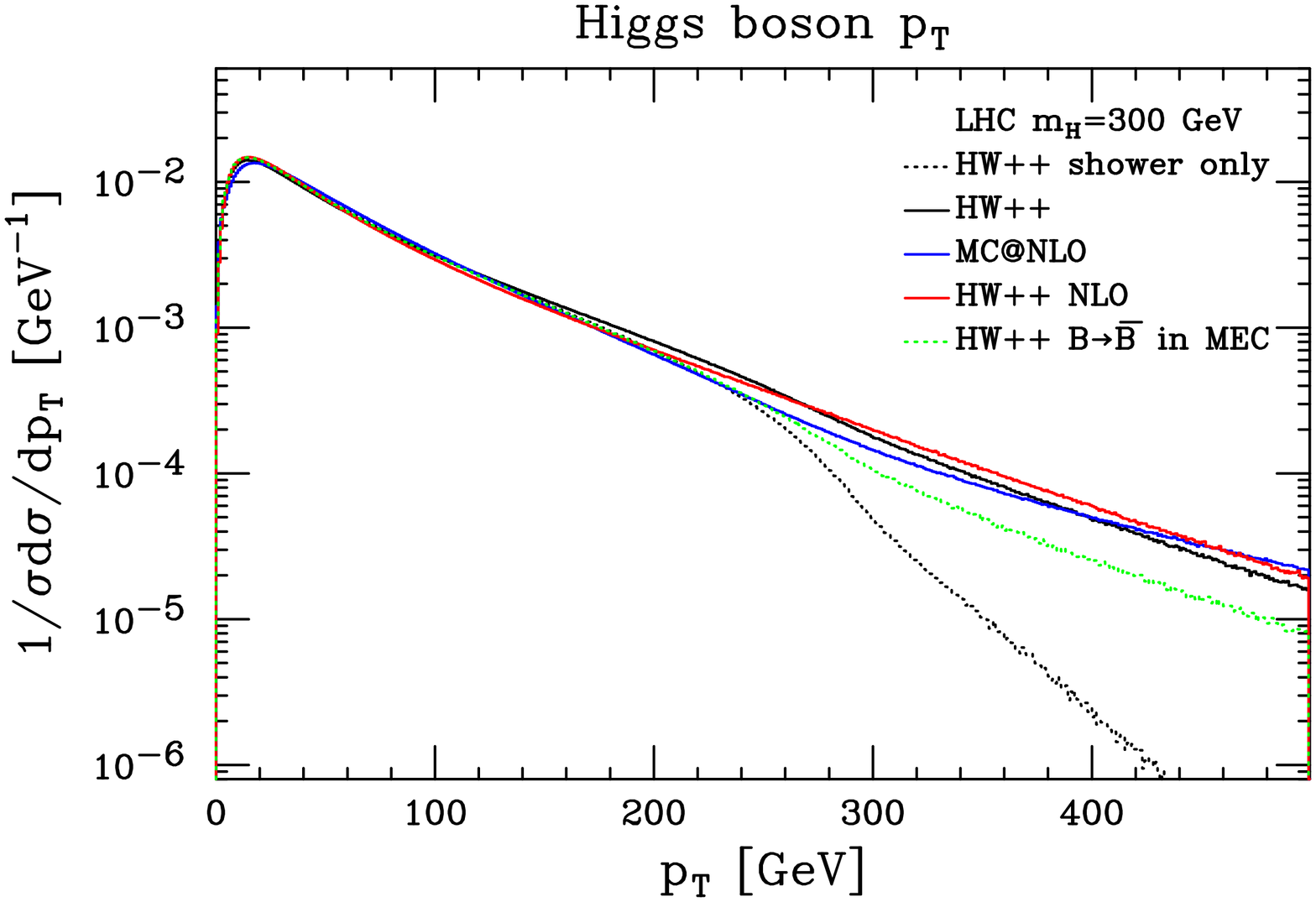}\hfill{}\includegraphics[width=0.38\textwidth,height=0.48\textwidth,angle=90]{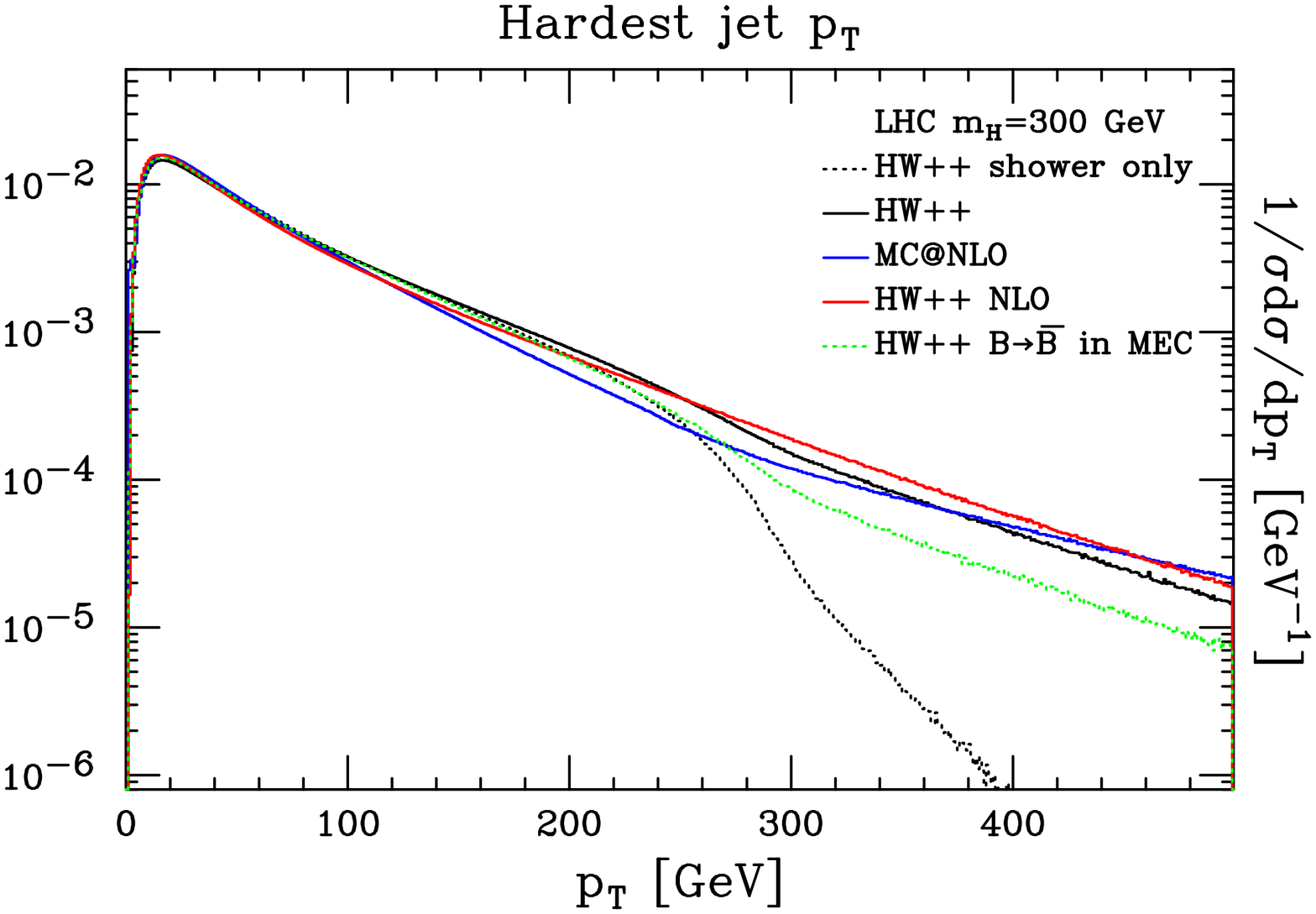}
\par\end{centering}

\caption{\label{fig:ggh_pT_spectra} Transverse momentum spectra for the Higgs
boson and the leading, highest $p_{T}$, jet, obtained using \HWPP\ with
matrix element corrections (black), \HWPP\ without matrix element
corrections, \emph{i.e.} the uncorrected parton shower (dotted), \textsf{MC@NLO}
(blue) and our \textsf{POWHEG} simulation inside \HWPP\ (red). The
green curve is obtained by modifying the hard component of the matrix
element corrections, to decrease the amount of radiation produced
in the associated high $p_{T}$, wide-angle, dead-zone, by terms beyond
next-to-leading-order accuracy. This modification is discussed further
in Sect.\,\ref{sub:Discussion}.}

\end{figure}

In figure \ref{fig:ggh_pT_spectra} we show the $p_{T}$ spectrum
of the Higgs boson and also that of the hardest jet. One can see that
the $p_{T}$ spectra at the Tevatron are less hard than those obtained
at the LHC, as one would expect given the greater centre-of-mass energies
of the latter. One also expects that, at the LHC, the larger mass
of a 300 GeV Higgs boson would automatically give rise to it having
a harder spectrum than that of a 115 GeV Higgs boson, this is indeed
the case for all five Monte Carlo predictions.

In each of our three phenomenological scenarios we see that the behaviour
of the different simulations is more-or-less the same with respect
to one another. The effect of the radiation dead zone is very clear
as a sharp \emph{knee} in the spectra from the uncorrected \HWPP\
parton shower (black dotted lines), as the transverse momentum approaches
the Higgs boson mass. Only the effects of multiple emission mean that
this is not an abrupt cut off (Fig.\,\ref{fig:phase_space_with_cuts}).
As predicted, turning on the MECs and hence filling the dead zone
in \HWPP, leads to a spectrum in much better agreement with all of
the other methods (black lines). A very slight kink is still visible
in this default MEC prediction, which we already forecast in Sect.\,\ref{sub:The-dead-zone}
as a mis-match, at $\mathcal{O}\left(\alpha_{\mathrm{S}}^{2}\right)$,
across the dead zone boundary.

The modified MEC (green dotted lines) shows a similar trend with respect
to the uncorrected parton shower prediction, although it comes as
no surprise that it has a softer spectrum than the regular MEC, simply
because the rate of emission into the dead zone is reduced by an amount
approximately given by the NLO K-factor \emph{cf.} Eqs.\,\ref{eq:me_corr_dead_zone_prob},
\ref{eq:me_corr_mod_dead_zone_prob}. From our earlier investigations
concerning the dead zone one should expect that the difference in
the emission rates in Eqs.\,\ref{eq:me_corr_dead_zone_prob} and
\ref{eq:me_corr_mod_dead_zone_prob} directly manifests itself as
the same relative difference in the upper values of the $p_{T}$ spectra
of the Higgs boson and the leading jet. This is indeed seen to be
the case in Fig.\,\ref{fig:ggh_pT_spectra}, where the unmodified
MEC prediction is around a factor of two higher than the modified
one on the right-hand side of each plot.

In all cases the \HWPP\ NLO \textsf{POWHEG} prediction (red lines)
is seen to be in very close agreement with the predictions obtained
using the normal MEC procedure (black lines). As we shall discuss
more in Sect.\,\ref{sub:Discussion}, we attribute this to the fact
that the emission rates for the MEC method and the \textsf{POWHEG}
implementation can be expected to converge to the same value at high
$p_{T}$.

The \textsf{MC@NLO} $p_{T}$ spectra are in general somewhat softer
than those of the MEC procedure and \textsf{POWHEG} NLO prediction,
they tend to lie between the prediction obtained using the modified
MEC (green dotted lines) and the \HWPP\ \textsf{POWHEG}/default
MEC lines. Good agreement between the \textsf{MC@NLO} and modified
MEC predictions can be seen for the case of a 115 GeV Higgs boson
at the LHC, similarly, the agreement in the case of the Tevatron is
good below 200 GeV, however, for a 300 GeV Higgs boson at the LHC
the agreement is not as good as hoped beyond $p_{T}\simeq m_{\mathrm{H}}$.
We suggest that these discrepancies arise from presence of an improvement
introduced between versions 3.2 and 3.3 of \textsf{MC@NLO}, with the
aim of improving the description of the Higgs boson \textsf{$p_{T}$}
spectrum \cite{Frixione:2006gn}. 

\begin{figure}[H]
\begin{centering}
\includegraphics[width=0.38\textwidth,height=0.48\textwidth,angle=90]{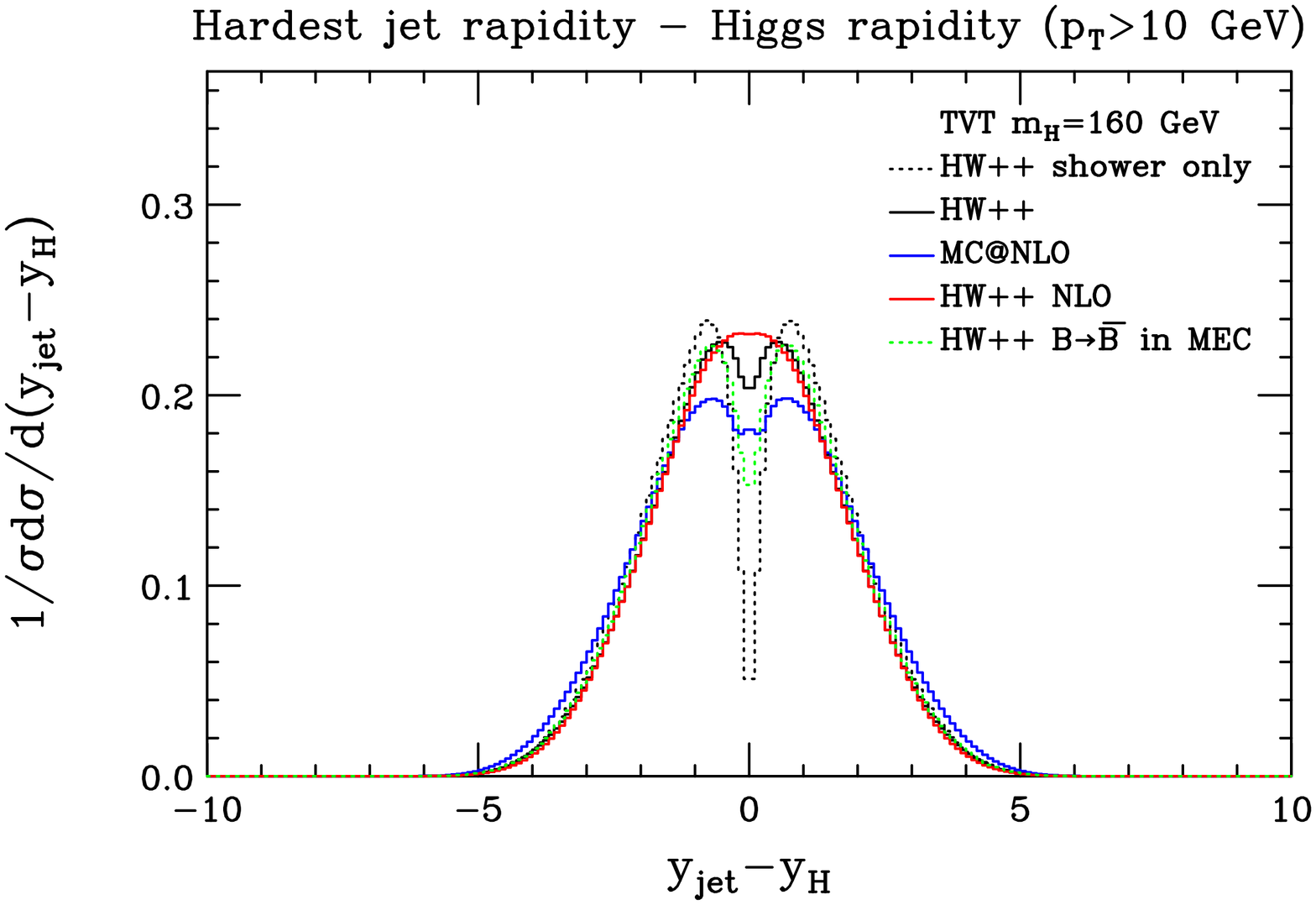}\hfill{}\includegraphics[width=0.38\textwidth,height=0.48\textwidth,angle=90]{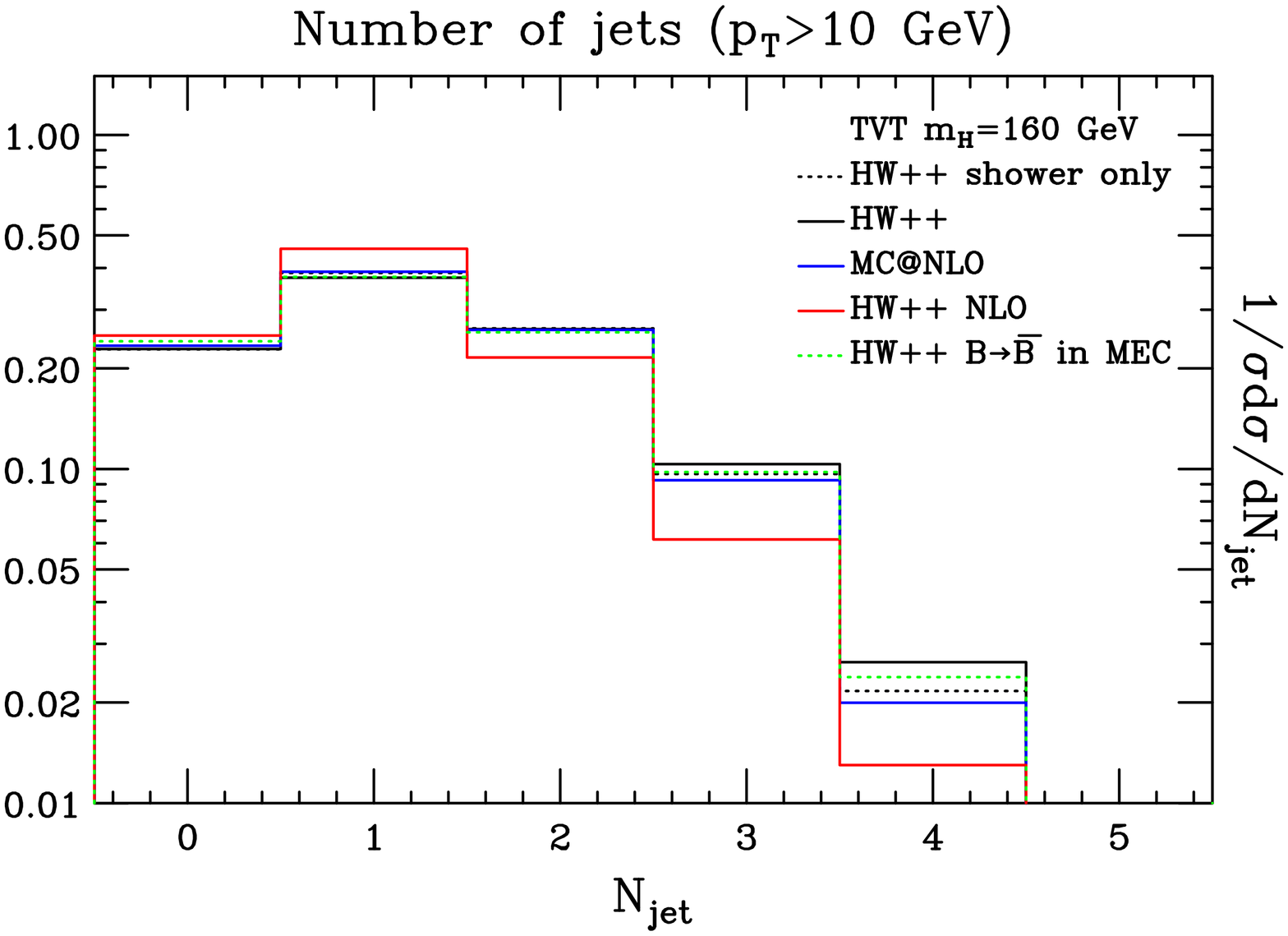}
\par\end{centering}

\vspace{7mm}

\begin{centering}
\includegraphics[width=0.38\textwidth,height=0.48\textwidth,angle=90]{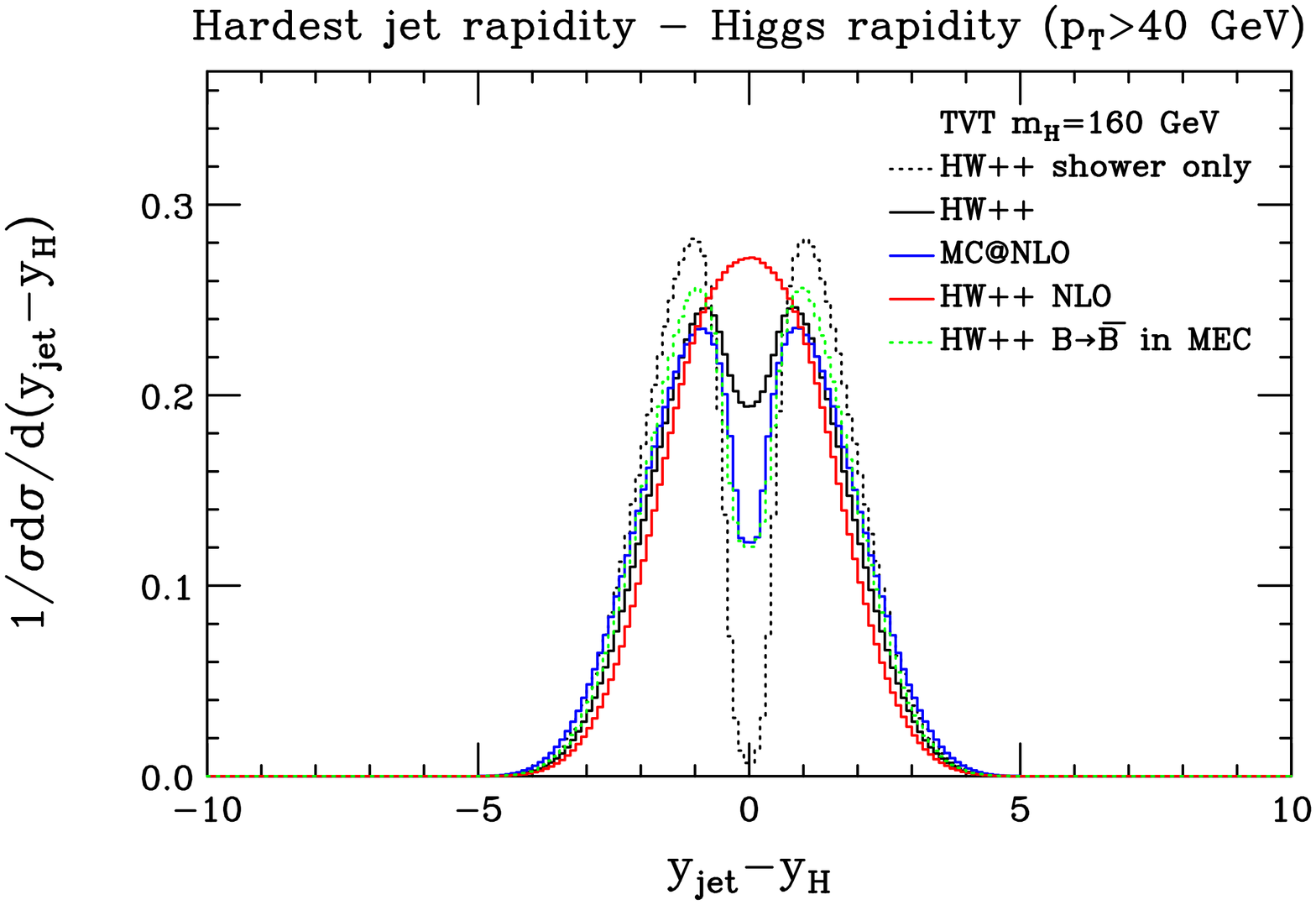}\hfill{}\includegraphics[width=0.38\textwidth,height=0.48\textwidth,angle=90]{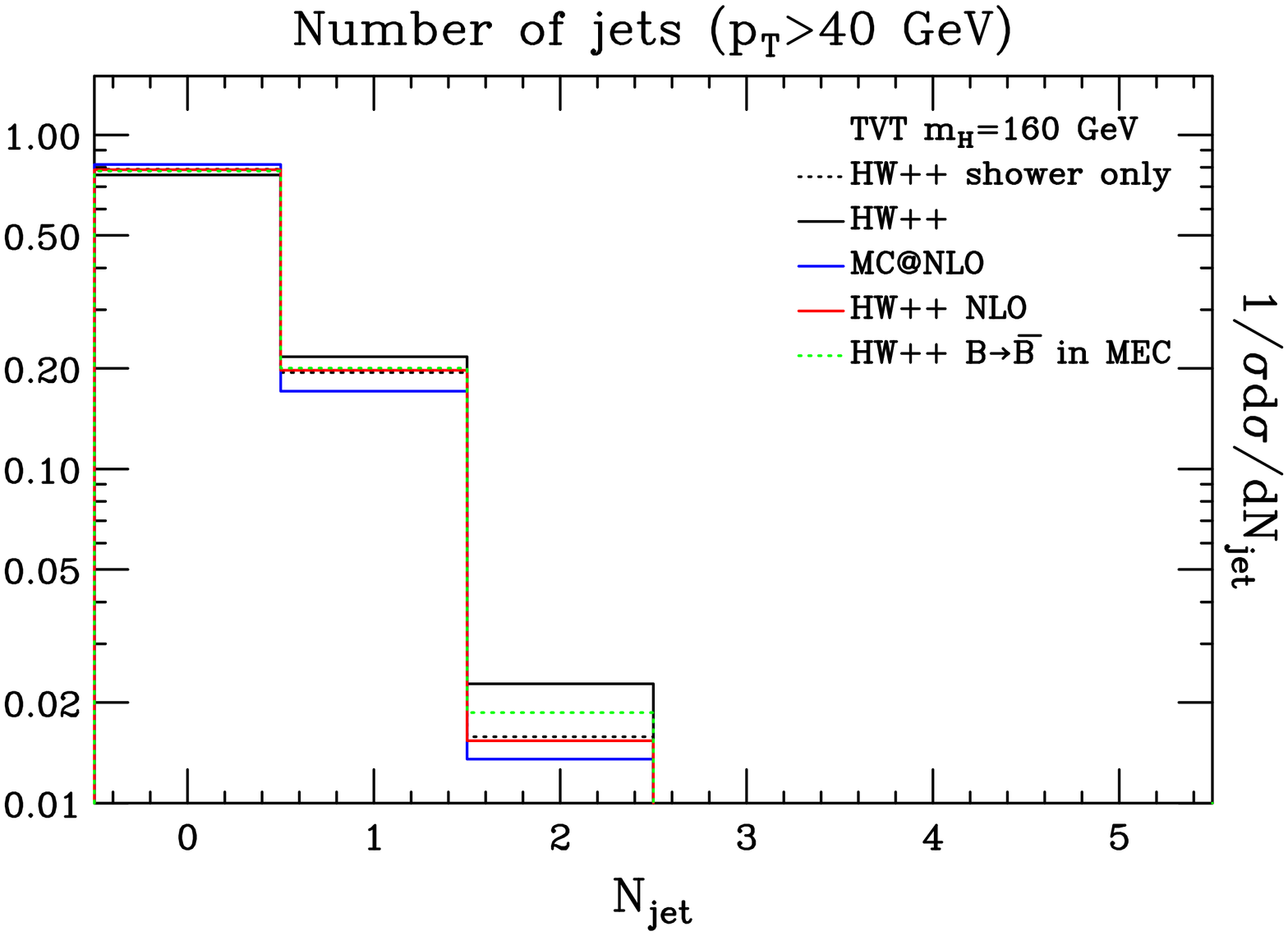}
\par\end{centering}

\vspace{7mm}

\begin{centering}
\includegraphics[width=0.38\textwidth,height=0.48\textwidth,angle=90]{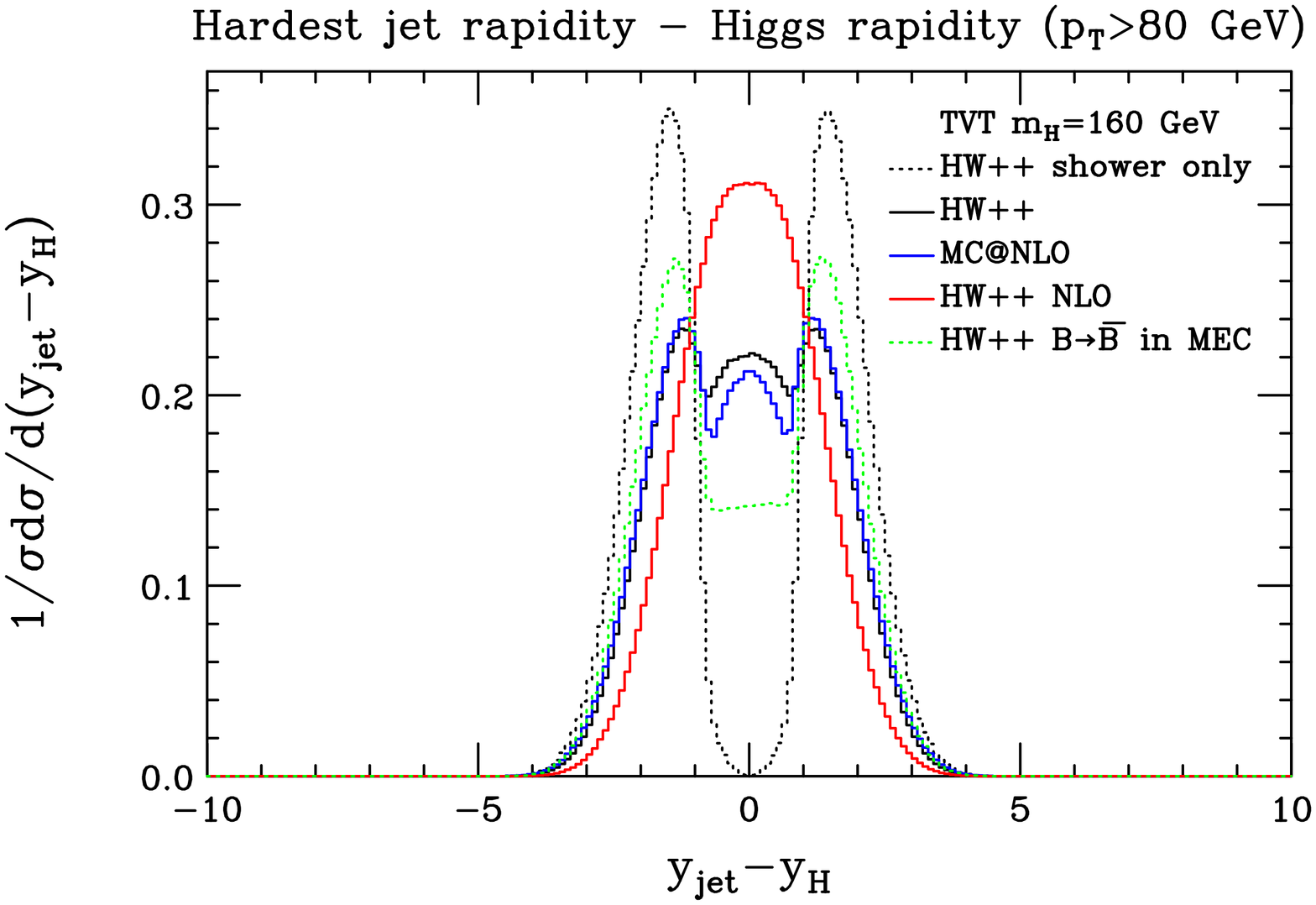}\hfill{}\includegraphics[width=0.38\textwidth,height=0.48\textwidth,angle=90]{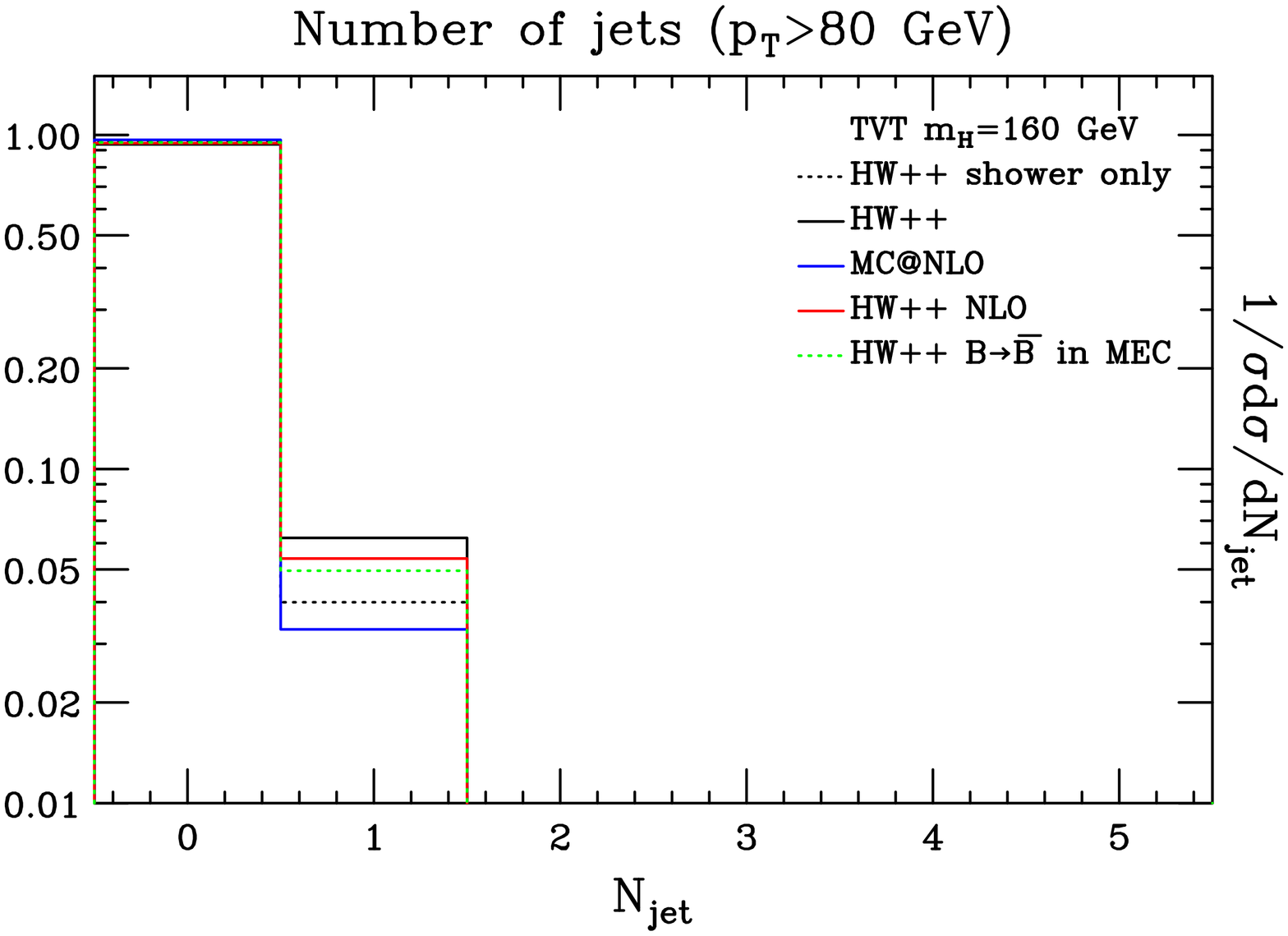}
\par\end{centering}

\caption{\label{fig:ggh_tvt_yj-yh_njets} In the left hand column we show distributions
of the difference in rapidity between the leading jet and a Higgs
boson of mass 160 GeV at the Tevatron, for increasing cuts on the
$p_{T}$ of the leading jet. For a single emission, in the region
close to zero, this variable is proportional to the angle between
the emitted parton and the transverse direction in the partonic centre-of-mass
frame (Eq.\,\ref{eq:yjet-yh_approx_formula}). In the right hand
column we show the corresponding jet multiplicity distributions. The
black and dotted lines show the predictions obtained using \HWPP\
with and without matrix element corrections, respectively. The blue
line shows the prediction from \textsf{MC@NLO} and the red line is
that of our \textsf{POWHEG} simulation in \HWPP. The green line is
obtained using a modified version of the hard matrix element correction,
effectively decreasing the amount of radiation that this method produces
in the high $p_{T}$, dead zone, by terms beyond NLO accuracy (see
Sect.\,\ref{sub:Discussion}).}

\end{figure}

\begin{figure}[H]
\begin{centering}
\includegraphics[width=0.38\textwidth,height=0.48\textwidth,angle=90]{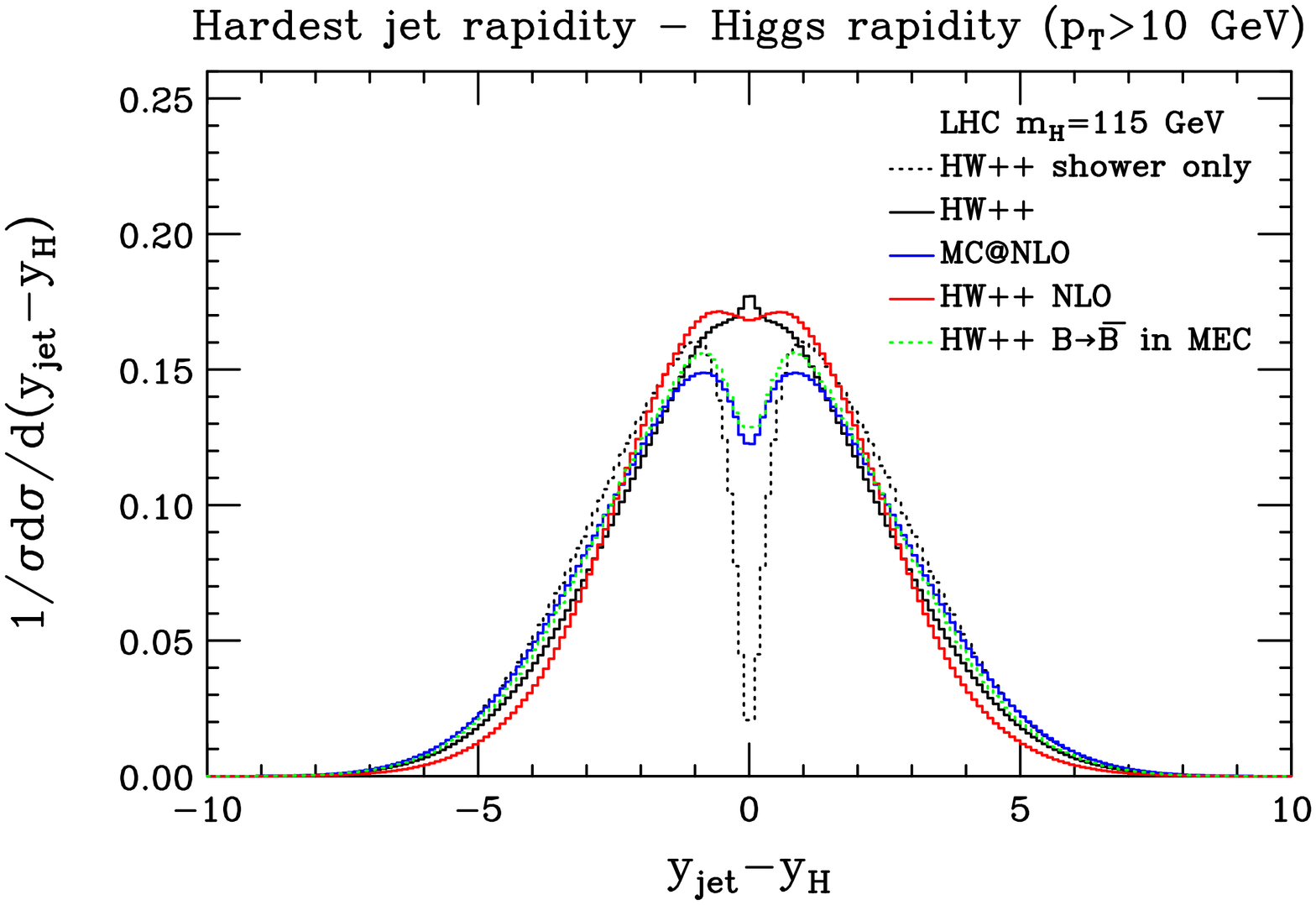}\hfill{}\includegraphics[width=0.38\textwidth,height=0.48\textwidth,angle=90]{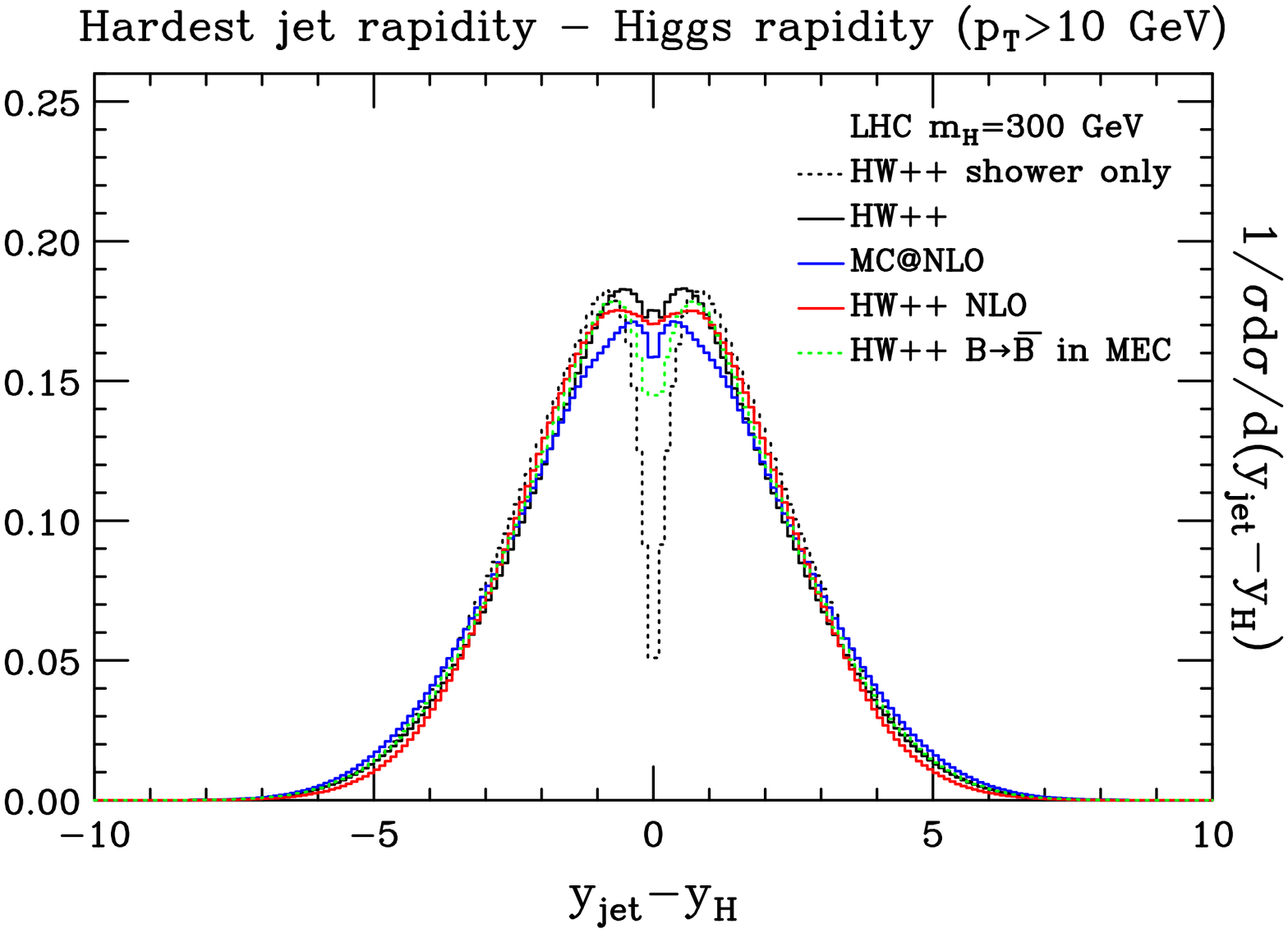}
\par\end{centering}

\vspace{10mm}

\begin{centering}
\includegraphics[width=0.38\textwidth,height=0.48\textwidth,angle=90]{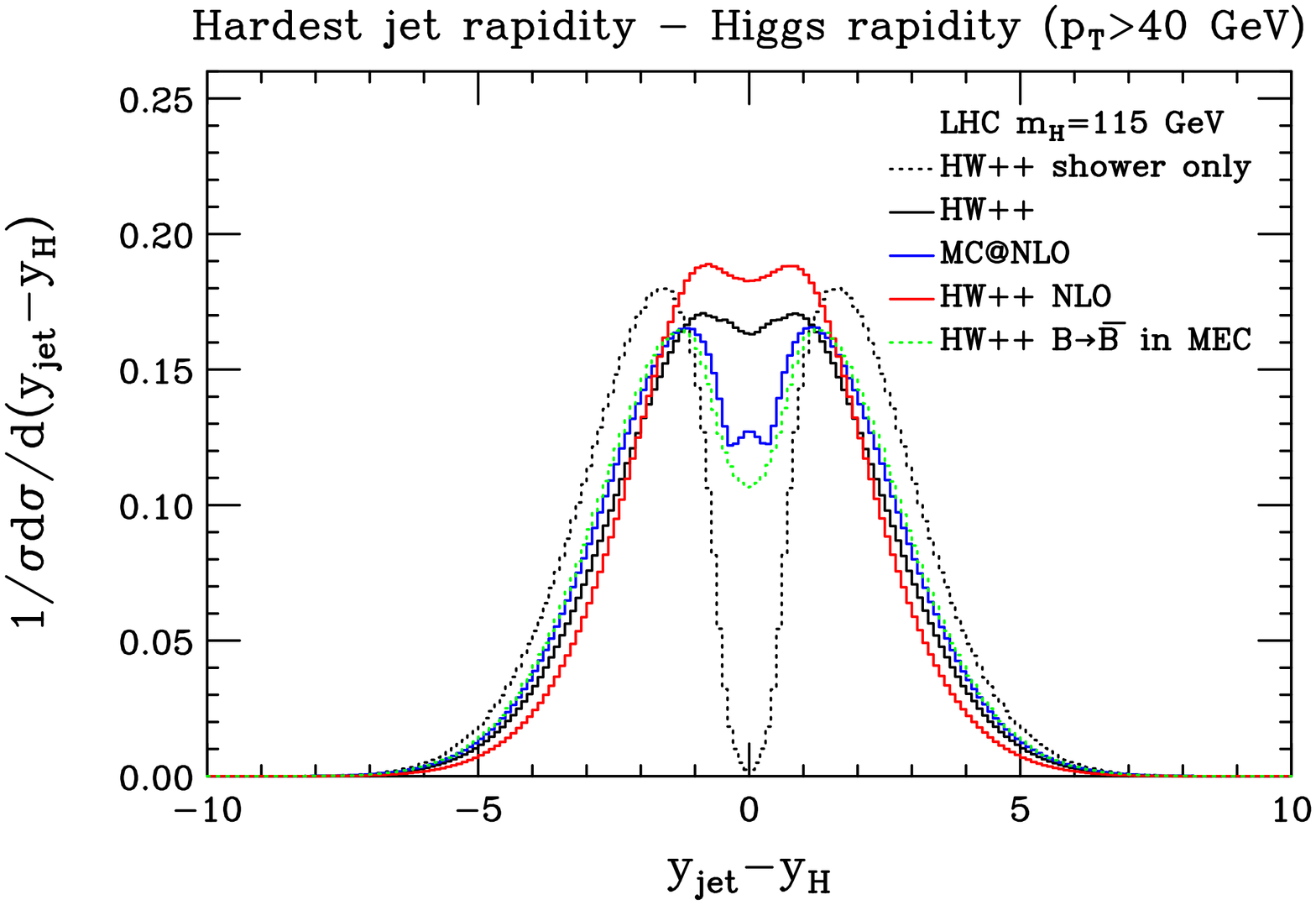}\hfill{}\includegraphics[width=0.38\textwidth,height=0.48\textwidth,angle=90]{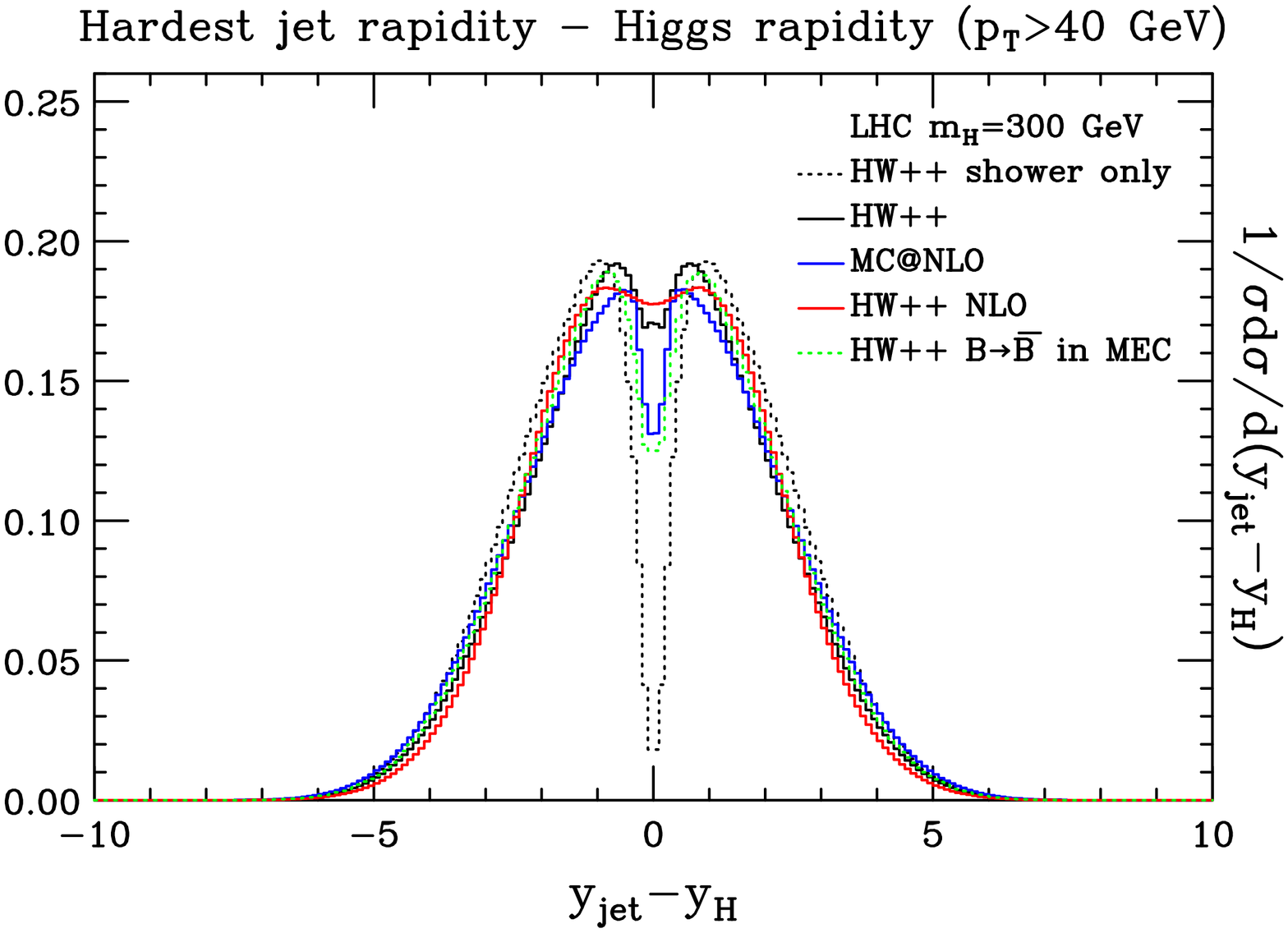}
\par\end{centering}

\vspace{10mm}

\begin{centering}
\includegraphics[width=0.38\textwidth,height=0.48\textwidth,angle=90]{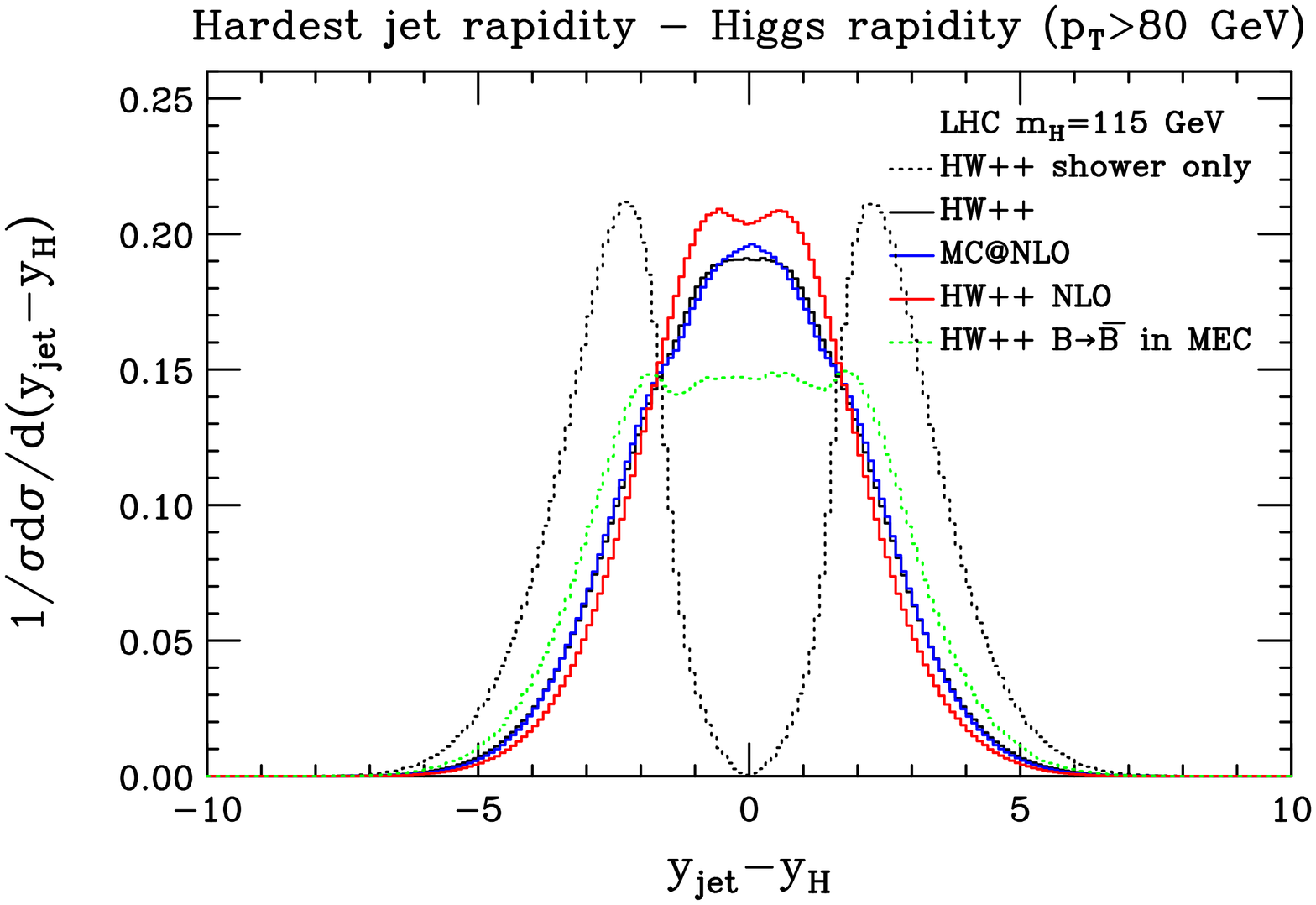}\hfill{}\includegraphics[width=0.38\textwidth,height=0.48\textwidth,angle=90]{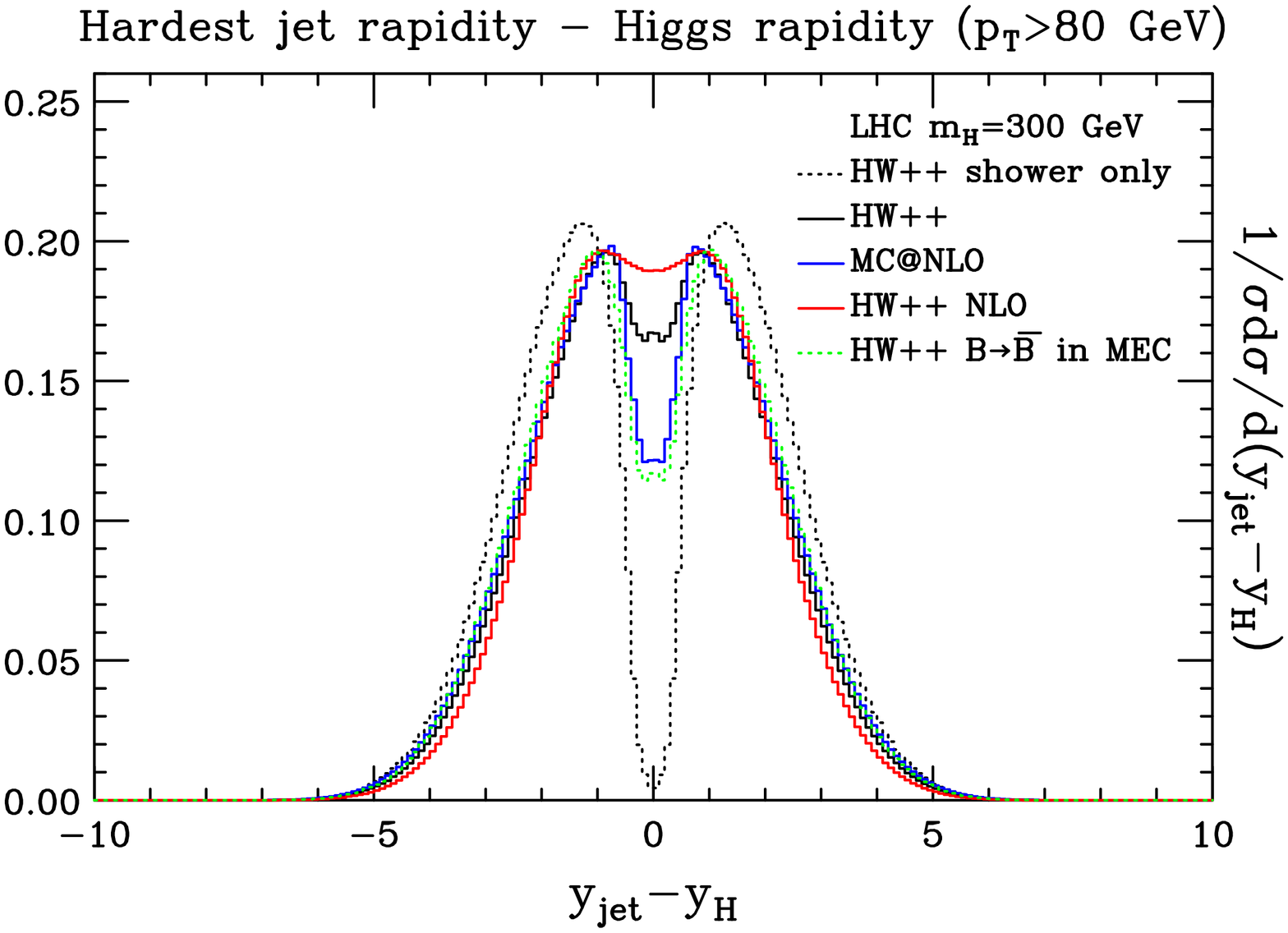}
\par\end{centering}

\caption{\label{fig:ggh_lhc_yj-yh} Distributions of the difference in rapidity
between the leading jet and the Higgs boson in the gluon fusion process
at the LHC, for increasing cuts on the $p_{T}$ of the leading jet.
The series of plots on the left hand side are obtained for a Higgs
boson of mass 115 GeV, while those on the right correspond to a Higgs
boson of mass 300 GeV. The colour assignment of the various predictions
is described inset, it is the same as for earlier Tevatron predictions
in Fig\,\ref{fig:ggh_tvt_yj-yh_njets}. }

\end{figure}

In figures \ref{fig:ggh_tvt_yj-yh_njets} and \ref{fig:ggh_lhc_yj-yh}
we show the distribution of the rapidity difference between the leading
jet and the Higgs boson, for increasingly hard cuts on the $p_{T}$
of the jet. Starting again with the bare parton shower prediction
(black dotted lines), for each scenario and each value of the $p_{T}$
cut on the leading jet, the structure is broadly the same, the distribution
rises from the tails at either side of the plot into a hump before
falling again into a very deep dip in the centre, at $\mathrm{y_{jet}}-\mathrm{y}_{\mathrm{H}}=0$.

These \emph{volcano} formations are a direct manifestation of the
radiation dead zone; the volume of the dead zone is maximised along
the direction corresponding to wide angle radiation in the partonic
centre-of-mass frame, $y\approx0$, $\theta\approx\frac{\pi}{2}$
(Sect.\,\ref{sub:The-dead-zone}, Figs.\,\ref{fig:phase_space},
\ref{fig:phase_space_with_cuts}) and the $\mathrm{y_{\mathrm{jet}}-}\mathrm{y}_{\mathrm{H}}$
variable is proportional to $\theta-\frac{\pi}{2}$ (Sect.\,\ref{sub:Hardest-emission-generation-results},
Eq.\,\ref{eq:yjet-yh_approx_formula}). The only thing which may
then be slightly surprising is that the distribution does not in fact
go exactly to zero at $\mathrm{y_{jet}}-\mathrm{y}_{\mathrm{H}}=0$,
this is simply due to the effects of multiple emissions (the phase
space in Figs.\,\ref{fig:phase_space} and \ref{fig:phase_space_with_cuts}
is only exact for the case of a single emission).

Looking in more detail at the structure of the uncorrected \HWPP\
predictions, we see that the throat in each of volcano distributions
widens as the $p_{T}$ cut on the hardest jet is increased. This broadening
occurs because the dead zone accounts for
\emph{proportionally} more and more of the smaller/less wide angle
regions of the phase space
as the $p_{T}$ increases. This trend of the broadening throat with
the hardening of the $p_{T}$ cut is therefore visible in each of
the three scenarios we consider and it will come as no surprise that
the same occurs for the Higgs-strahlung process. %

The hard matrix element correction begins to fill in the central throat
region by emitting wide angle radiation into the dead zone. Nevertheless,
both the modified and unmodified MEC results carry some residual sensitivity
to the dead zone boundary, as evidenced by their inheritance of dips
in the central region, dips which follow the same trend as that set
by the uncorrected parton shower, to broaden and deepen as the $p_{T}$
cut on the hardest jet is increased. We attribute this dip behaviour
as being almost entirely due to the mis-match in the emission rate
across the dead zone boundary, which occurs for terms beyond NLO accuracy,
as discussed in Sect.\,\ref{sub:The-dead-zone}. This conclusion
may not appear to hold for the 115 GeV Higgs boson when the $p_{T}$
cut on the jet reaches 80 GeV. However, in this plot, the absence
of a dip actually reinforces our assertion, as can be seen by consulting
the 80 GeV $p_{T}$ contour in the corresponding phase-space map;
this contour shows that the allowed region for emissions is almost
identically in the dead zone, so any matching across the dead zone
boundary and, conversely, any mis-matching, is extremely limited in
this special case.

Having understood how the dead zone is manifest in the rapidity difference
plots, the differences observed between the modified and unmodified
MEC methods are rather unremarkable: since the emission rate of the
modified MEC into the dead zone is reduced by around a factor of two
with respect to the unmodified correction, the former emits more wide
angle radiation and so populates the central region of the $\mathrm{y}_{\mathrm{jet}}-\mathrm{y}_{\mathrm{H}}$
to a greater extent.

Although the \textsf{HERWIG} program has a slightly different phase-space
coverage for its parton showers (Figs.\,\ref{fig:phase_space}, \ref{fig:phase_space_with_cuts}),
they are apparently not so different, particularly away from the soft
region. Considering the region of the phase space allowed by just
a 10 GeV $p_{T}$ cut (Fig.\ref{fig:phase_space_with_cuts}) it is
already clear that the coverage by \textsf{HERWIG} and \HWPP\
is really very similar and that it is basically identical by the time
the cut reaches 80 GeV. The predictions of the uncorrected \textsf{HERWIG}
shower, which showers the positive and negative unit weight events
fed to it from \textsf{MC@NLO}, will then be very similar to those
shown here for \HWPP, in particular, the volcano structures in the
$\mathrm{y}_{\mathrm{jet}}-\mathrm{y}_{\mathrm{H}}$ distributions.
This being the case, it is understandable that the \textsf{MC@NLO}
distributions (blue lines) also exhibit dips in the central region,
and that the behaviour of these dips with the varying $p_{T}$ cut
follows that of the MEC method; we also note that a number of these
results are markedly similar to those obtained using the modified
MEC.

Before discussing the \textsf{POWHEG} results we reiterate that this
method is wholly independent of the details of the partitioning of
phase space in the shower algorithms, it generates the NLO emission
effectively as a self contained $p_{T}$ ordered parton shower, albeit
with NLO accuracy, and thereby circumvents such issues. Hence, the
appearance of a slight dip in the distributions of $\mathrm{y}_{\mathrm{jet}}-\mathrm{y}_{\mathrm{H}}$
cannot be explained in the same way as those seen in the predictions
of the \HWPP\ shower, with or without the MECs, nor those of \textsf{MC@NLO}.
The central dip appearing in the \textsf{POWHEG} results alters the
height of these distributions, in all cases, by less than 5\%, it
is therefore characteristic of the uncertainties typical of \emph{NNLO}
calculations, and is many times smaller than those seen in the other
predictions. Given the smallness of the effect, its absence in the
Tevatron distributions, and also the fact that it exhibits no discernible
response to the changing $p_{T}$ cut, we cannot comment on its origin.

Finally we note that only the \textsf{POWHEG} predictions clearly
exhibit the expected physical behaviour on increasing the $p_{T}$
cut: that the $\mathrm{y}_{\mathrm{jet}}-\mathrm{y}_{\mathrm{H}}$
distribution should become \emph{squeezed} toward the central region,
as the phase space available for small angle emissions, which populate
the tails, becomes reduced relative to the phase space available for
wide angle emissions. This trend is somewhat obscured in the distributions
predicted by the other methods.

At this point we do not wish to give the impression that the dips
exhibited by the MEC and \textsf{MC@NLO} predictions are incorrect.
It is our contention, however, that these predictions can be consistently
explained by ascribing the origin of the dips to a mis-match at $\mathcal{O}\left(\alpha_{\mathrm{S}}^{2}\right)$
in the emission rates either side of the dead zone boundary. Although
the dead zone boundary is an unphysical partition in the phase space,
we stress that the mis-match involves terms beyond the stated accuracy
of either method. Looking ahead, beyond NLO accuracy, it should then
be the case that the MEC and \textsf{MC@NLO} approaches fail to approximate
any higher order terms, while the \textsf{POWHEG} method, being independent
of any kind of artificial phase space partitioning, should fare much
better.

In figure \ref{fig:ggh_tvt_yj-yh_njets} we also show jet multiplicity
distributions associated to each of the Tevatron $\mathrm{y}_{\mathrm{jet}}-\mathrm{y}_{\mathrm{H}}$
plots. As expected, these plots show, in all cases, that the jet multiplicity
distribution decreases rapidly as the $p_{T}$ cut on the leading
jet is increased, with only $\sim$5\% of events containing a jet
once the cut reaches 80 GeV. In the case of the soft $p_{T}$ cut
($p_{T}$>10~GeV) one can see that the \textsf{POWHEG} approach predicts
events with lower multiplicity than the other methods (which broadly
agree with one another).

\subsubsection{Higgs-strahlung\label{sub:Higgs-strahlung-plots}}

\begin{figure}[H]

\begin{centering}
\includegraphics[width=0.38\textwidth,height=0.48\textwidth,angle=90]{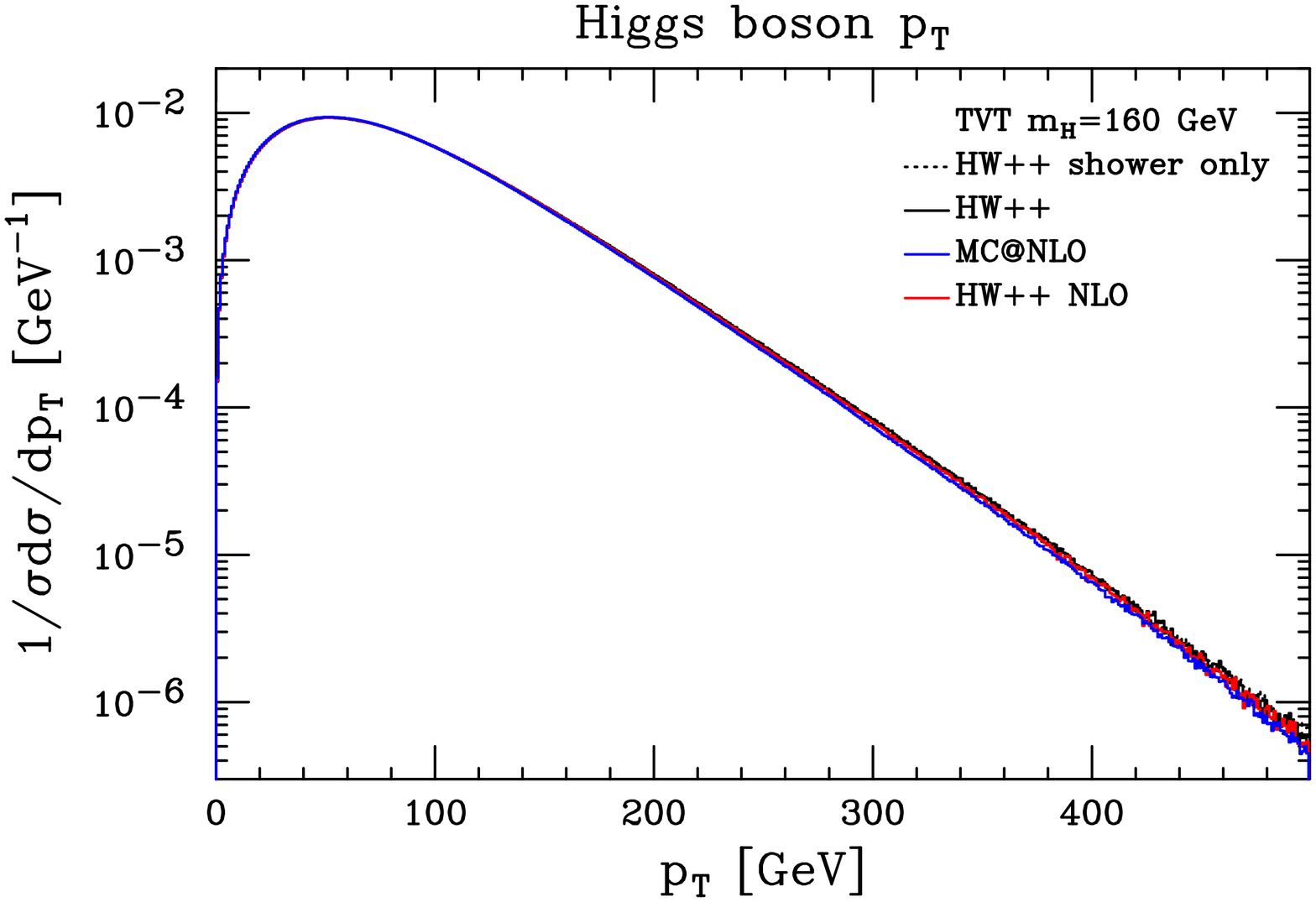}\hfill{}\includegraphics[width=0.38\textwidth,height=0.48\textwidth,angle=90]{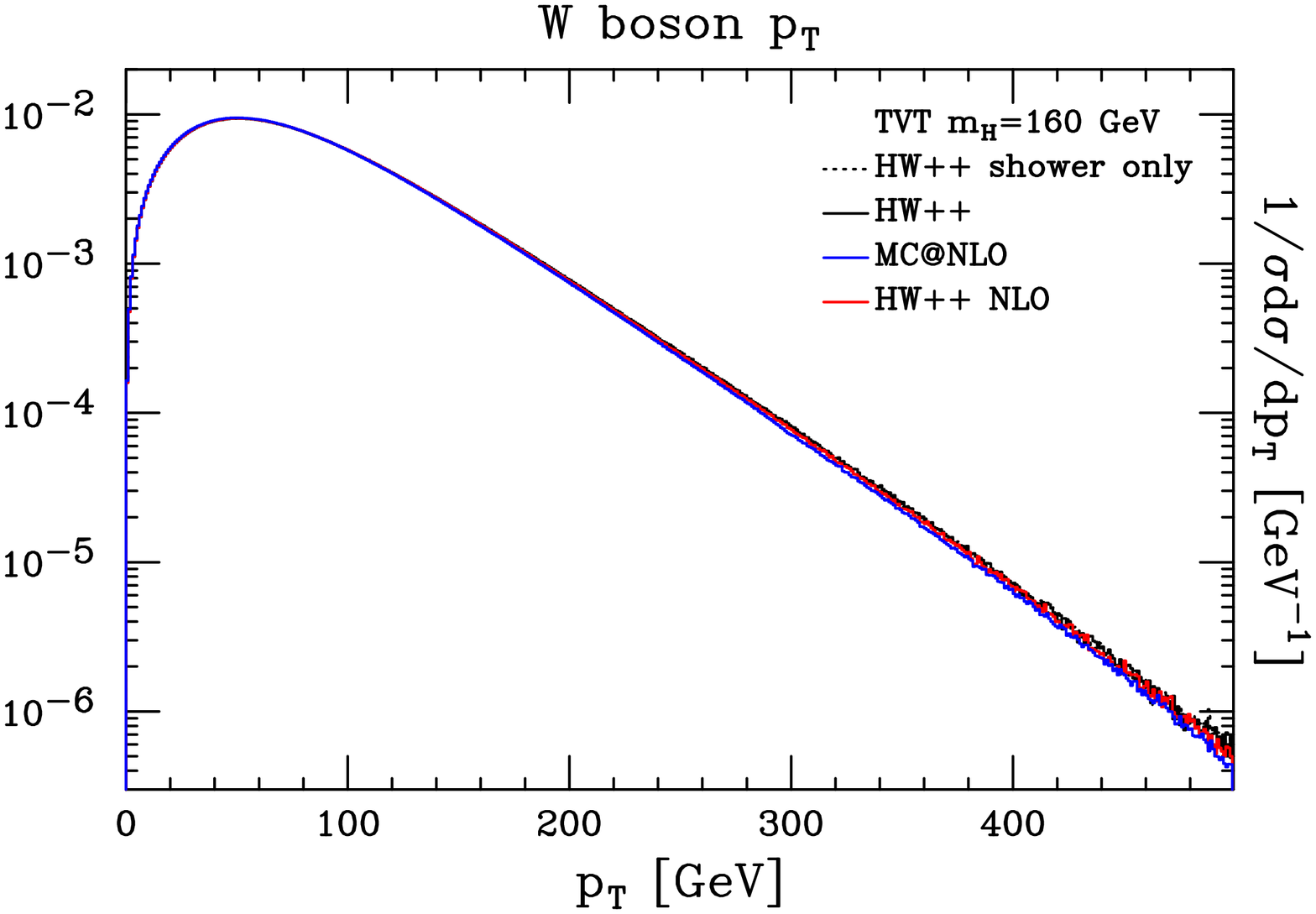}
\par\end{centering}

\vspace{10mm}

\begin{centering}
\includegraphics[width=0.38\textwidth,height=0.48\textwidth,angle=90]{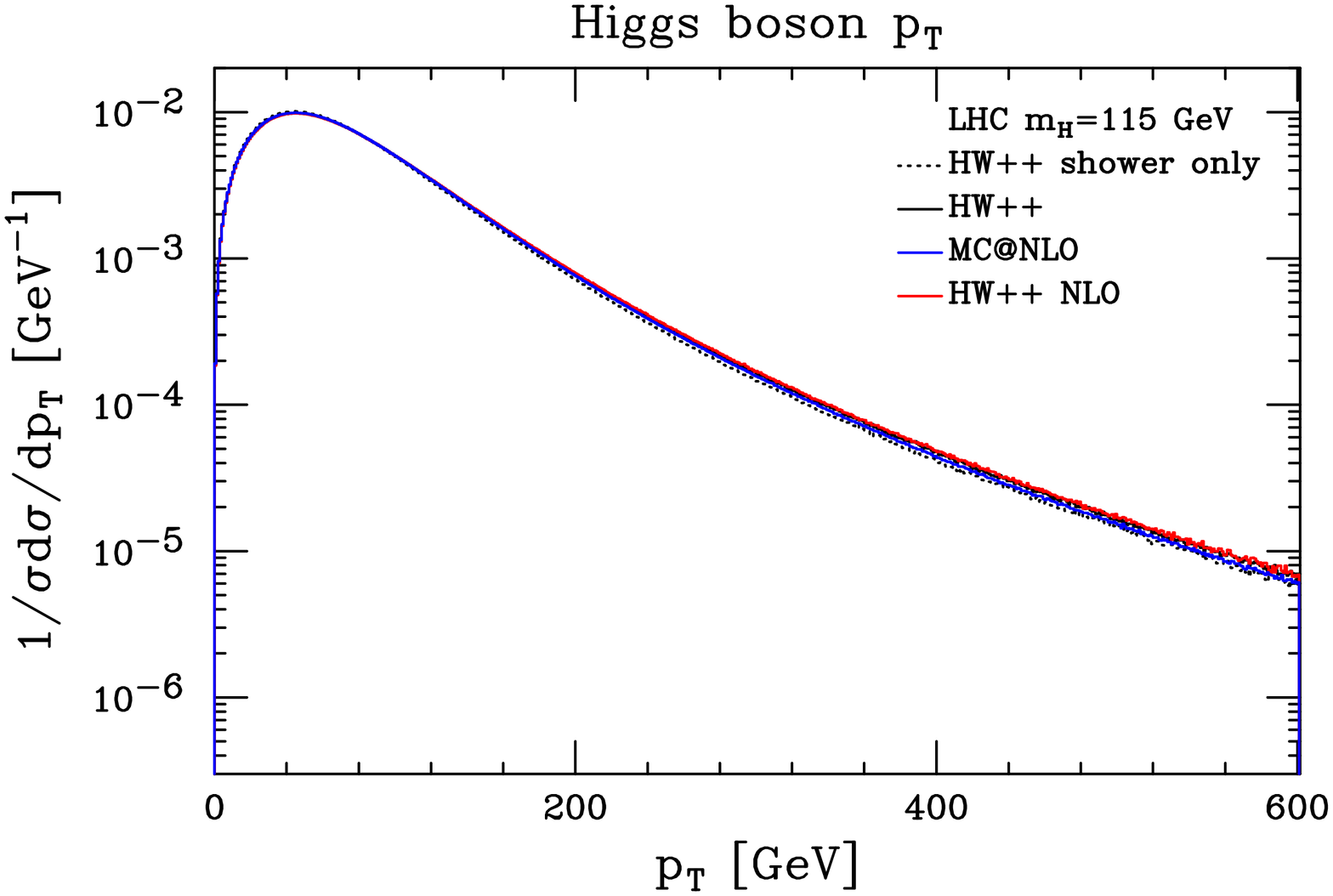}\hfill{}\includegraphics[width=0.38\textwidth,height=0.48\textwidth,angle=90]{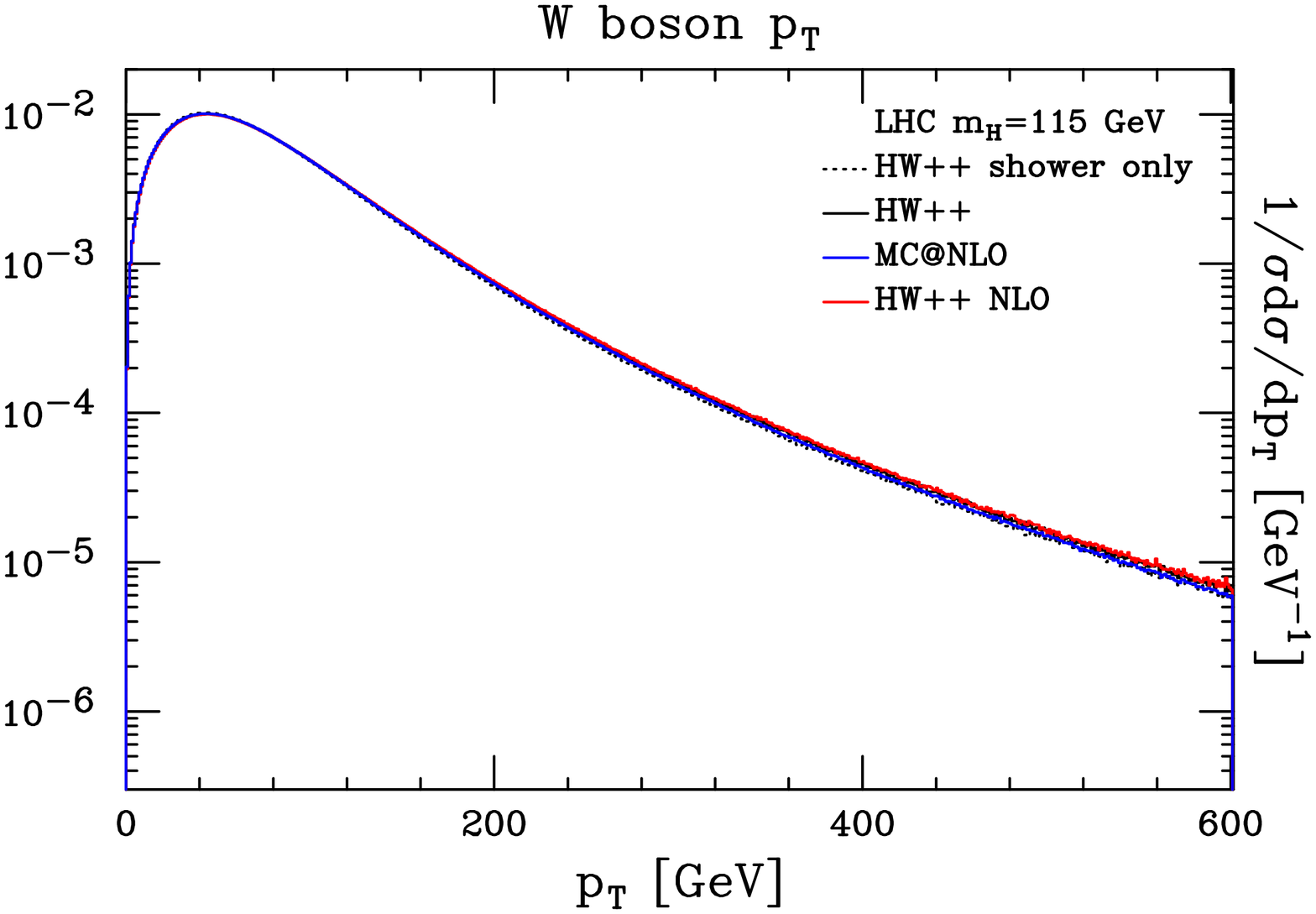}
\par\end{centering}

\vspace{10mm}

\begin{centering}
\includegraphics[width=0.38\textwidth,height=0.48\textwidth,angle=90]{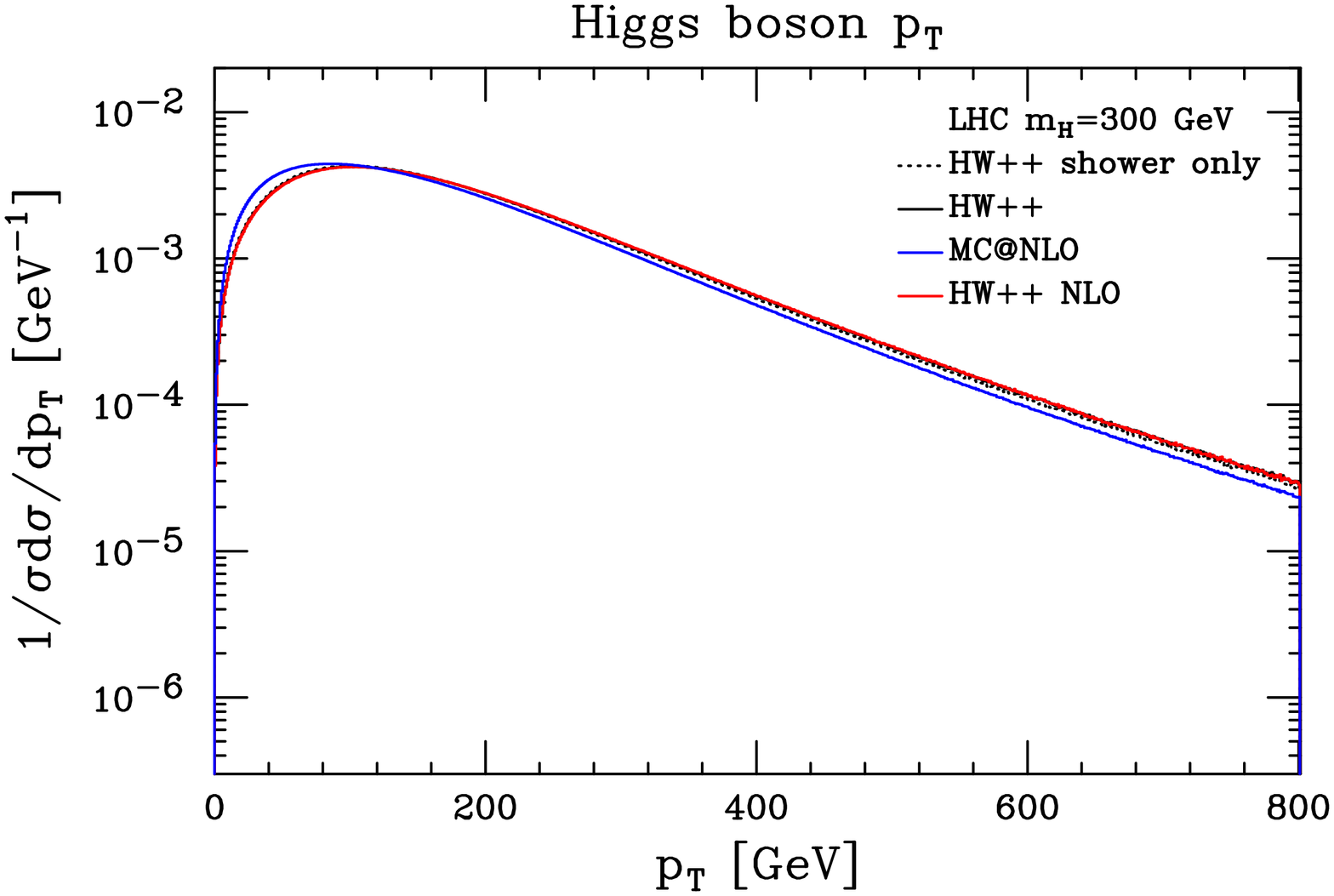}\hfill{}\includegraphics[width=0.38\textwidth,height=0.48\textwidth,angle=90]{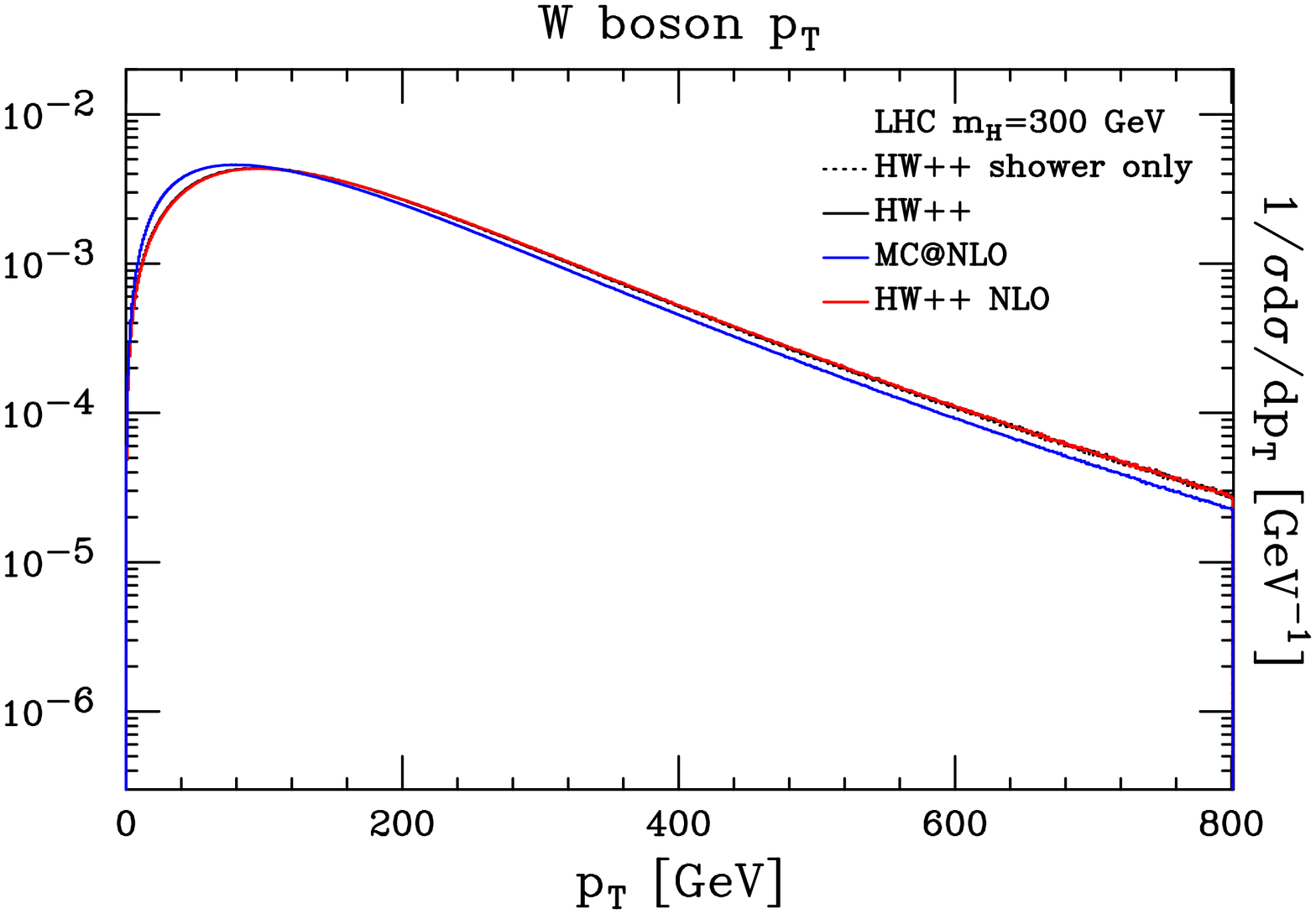}
\par\end{centering}

\caption{\label{fig:higgs_and_W_pT_spectra} Transverse momentum spectra for
the Higgs boson (left) and the $W^{\pm}$ boson in $q\bar{q}\rightarrow HW^{\pm}$
events obtained using \HWPP\ with and without matrix element corrections
(black and black dots respectively), \textsf{MC@NLO} (blue) and our
\textsf{POWHEG} simulation inside \HWPP\ (red). The first uppermost
predictions, for the Tevatron, are obtained assuming a Higgs boson
mass of 160 GeV. The following four plots are analogous projections
for the LHC for Higgs boson masses of 115 GeV and 300 GeV.}

\end{figure}

In figure \ref{fig:higgs_and_W_pT_spectra} we compare the transverse
momentum spectra of the Higgs boson and $W$ boson assuming a 160
GeV Higgs boson mass at the Tevatron and Higgs boson masses of 115
GeV and 300 GeV at the LHC. In each case we compare the results obtained
using the uncorrected \HWPP\ parton shower, the parton shower with
MECs, \textsf{MC@NLO} and our \textsf{POWHEG} implementation. All
four approaches agree remarkably well. The fact that the uncorrected,
leading-order, parton shower prediction agrees so well with the other
methods, which include at least the NLO real emission corrections,
indicates that these distributions are rather insensitive to the emission
of additional radiation, therefore one should not expect to see differences
among the more sophisticated approaches.

\begin{figure}[H]
\begin{centering}
\includegraphics[width=0.38\textwidth,height=0.48\textwidth,angle=90]{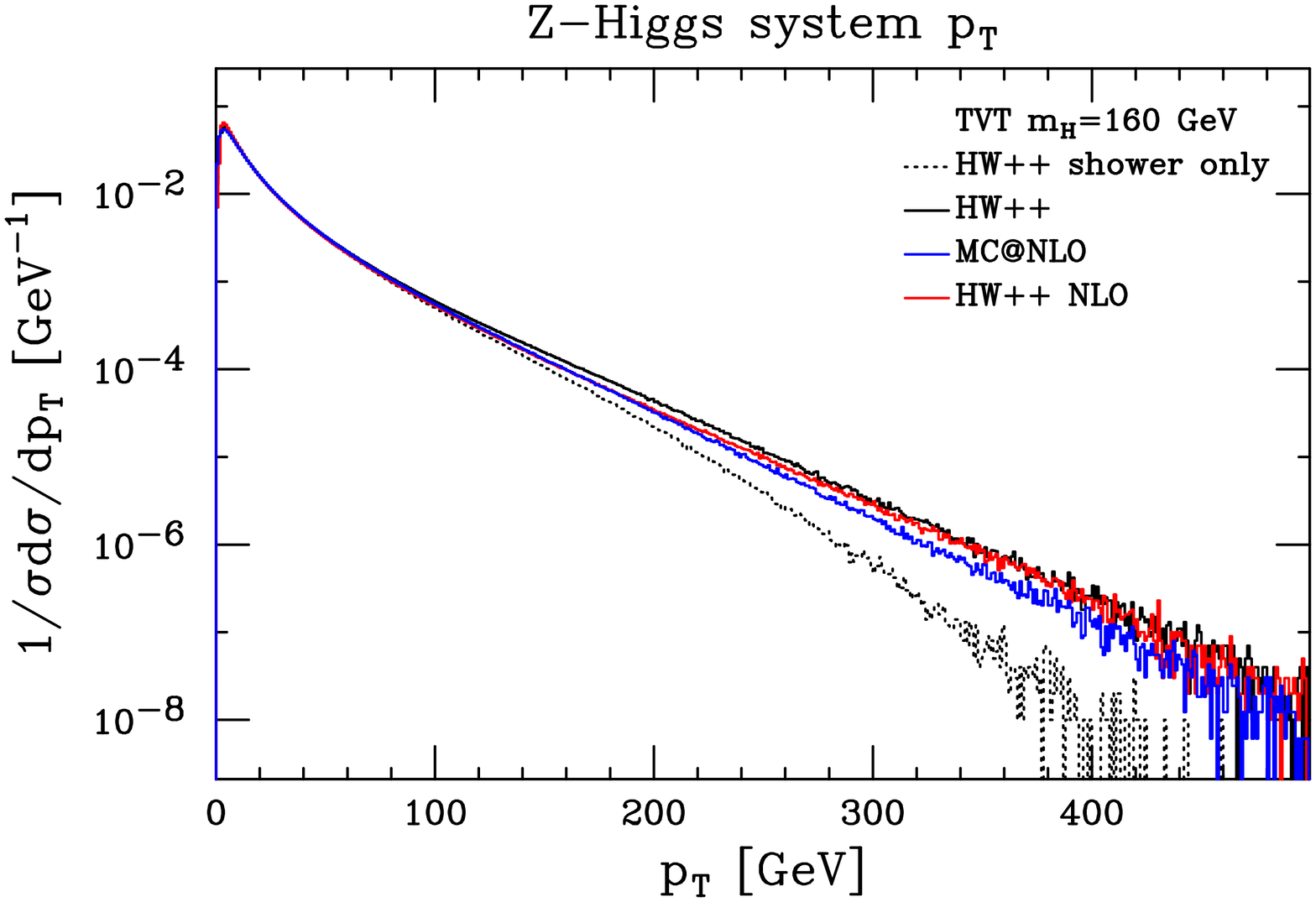}\hfill{}\includegraphics[width=0.38\textwidth,height=0.48\textwidth,angle=90]{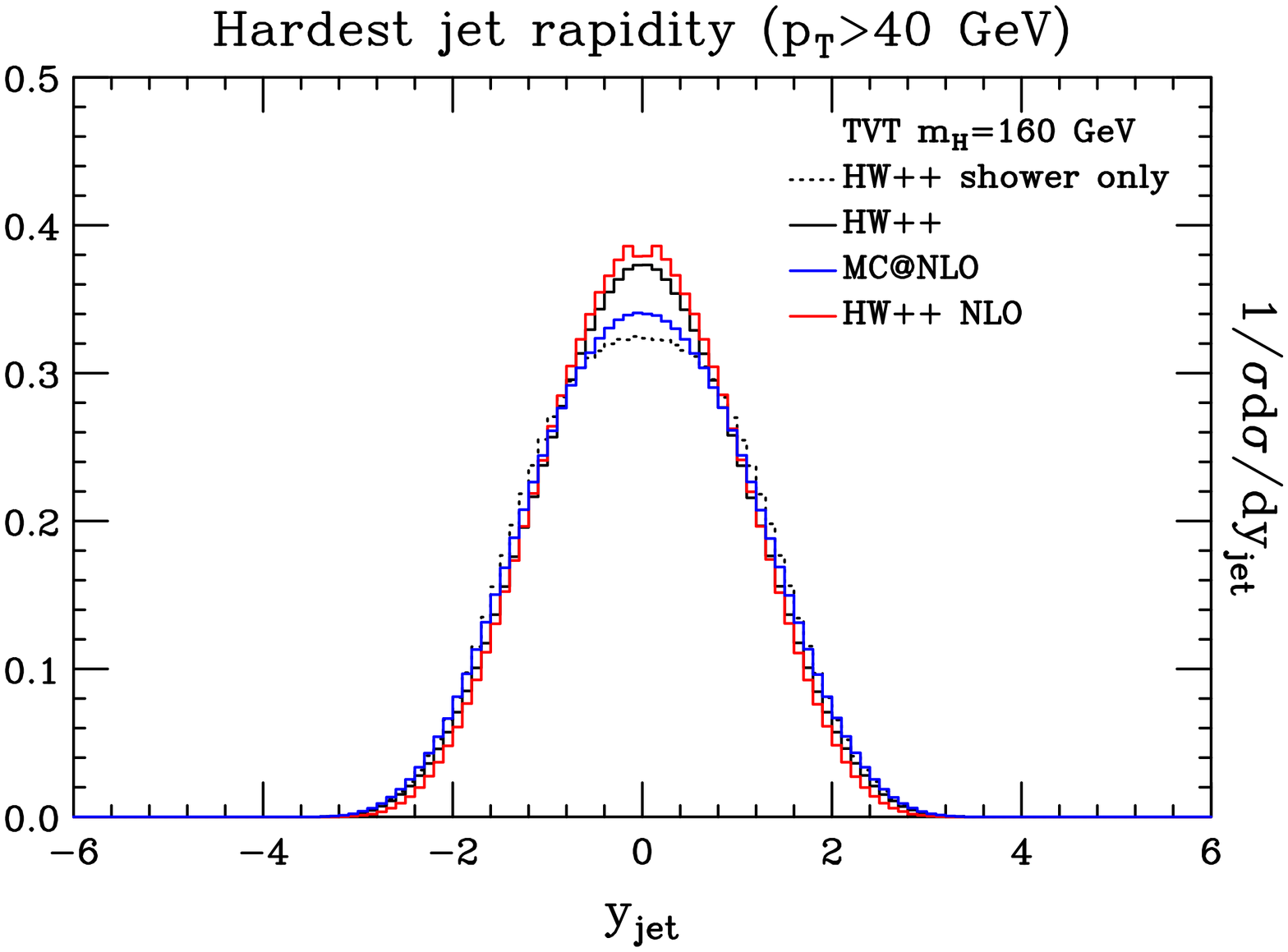}
\par\end{centering}

\vspace{10mm}

\begin{centering}
\includegraphics[width=0.38\textwidth,height=0.48\textwidth,angle=90]{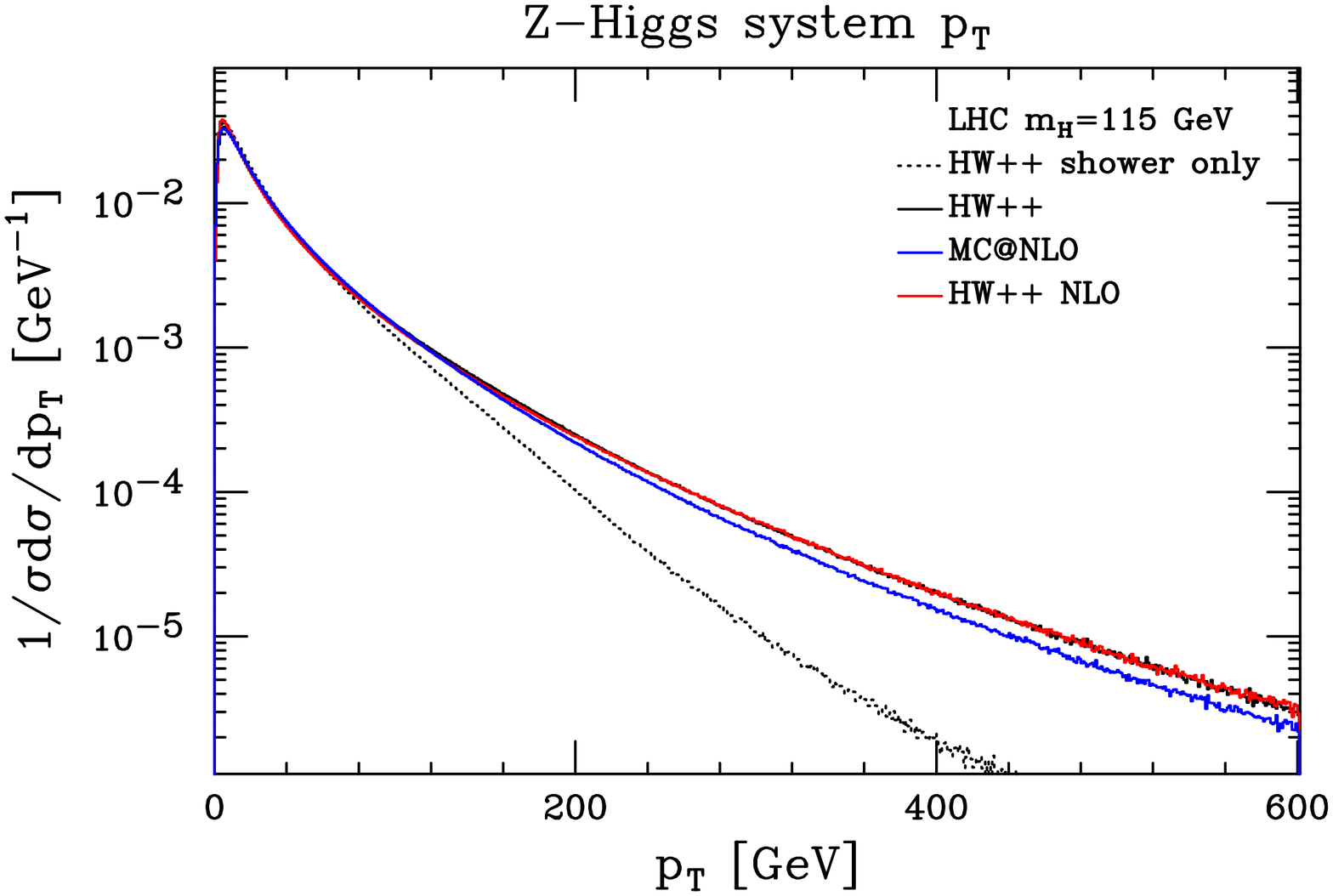}\hfill{}\includegraphics[width=0.38\textwidth,height=0.48\textwidth,angle=90]{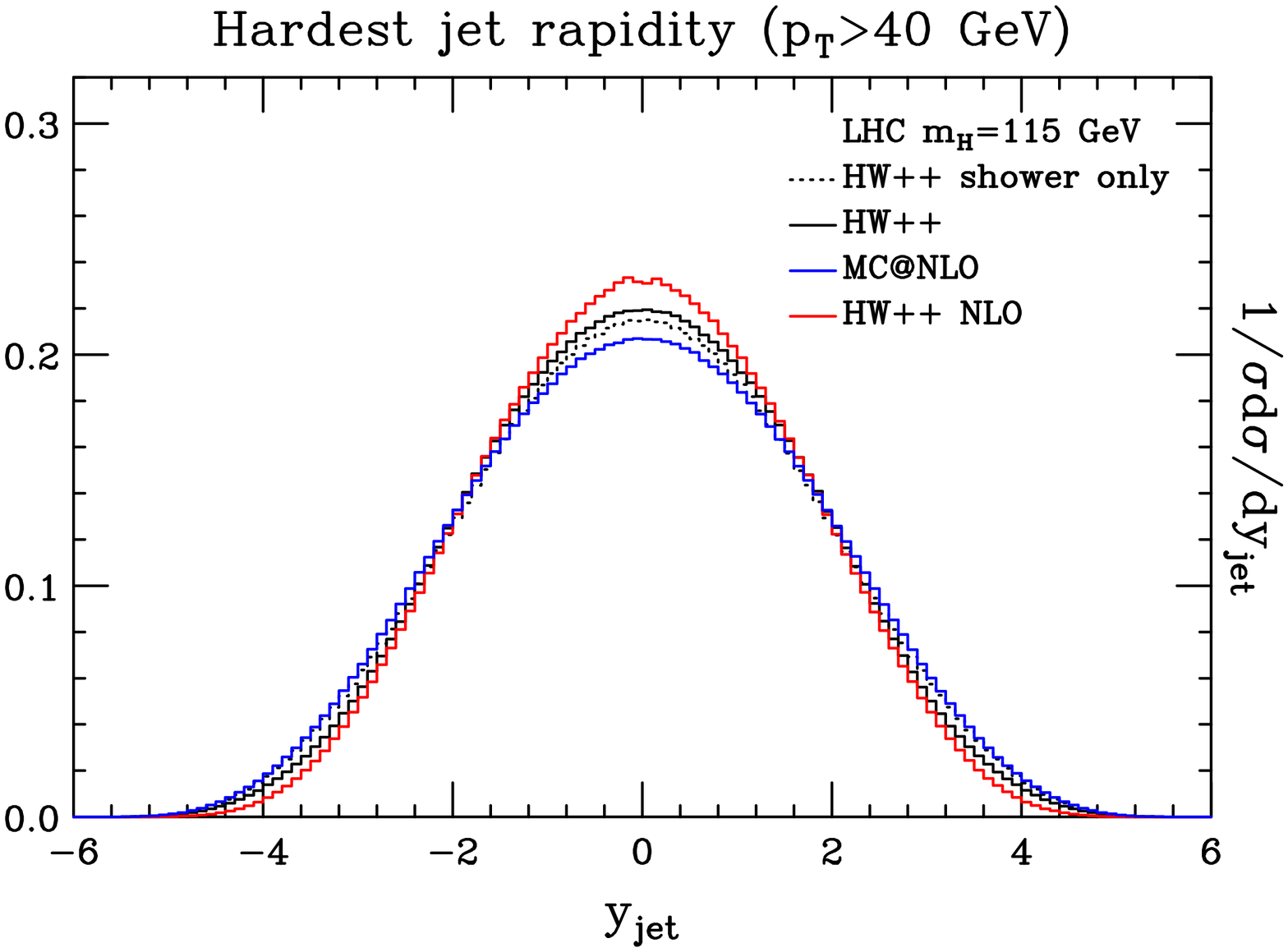}
\par\end{centering}

\vspace{10mm}

\begin{centering}
\includegraphics[width=0.38\textwidth,height=0.48\textwidth,angle=90]{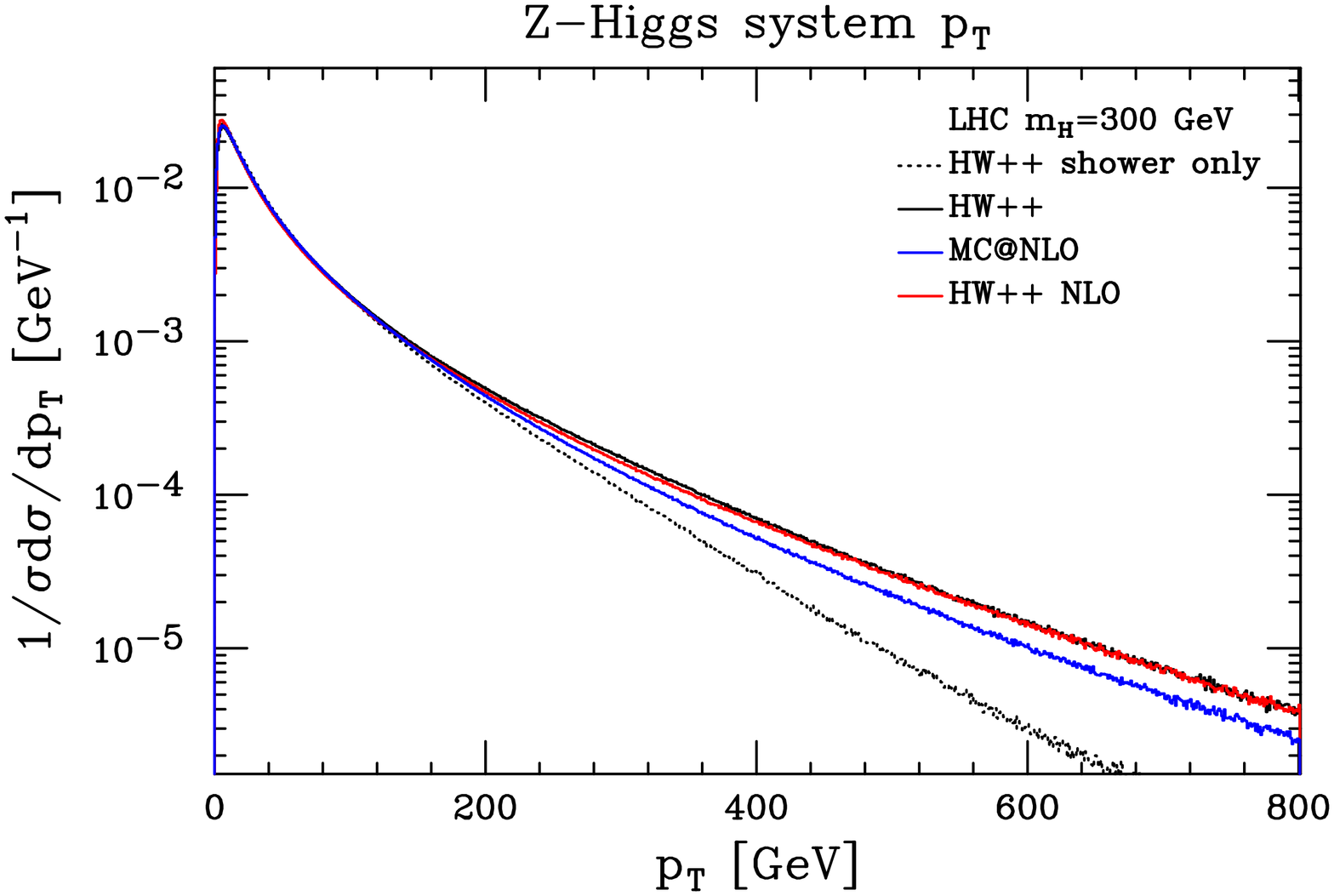}\hfill{}\includegraphics[width=0.38\textwidth,height=0.48\textwidth,angle=90]{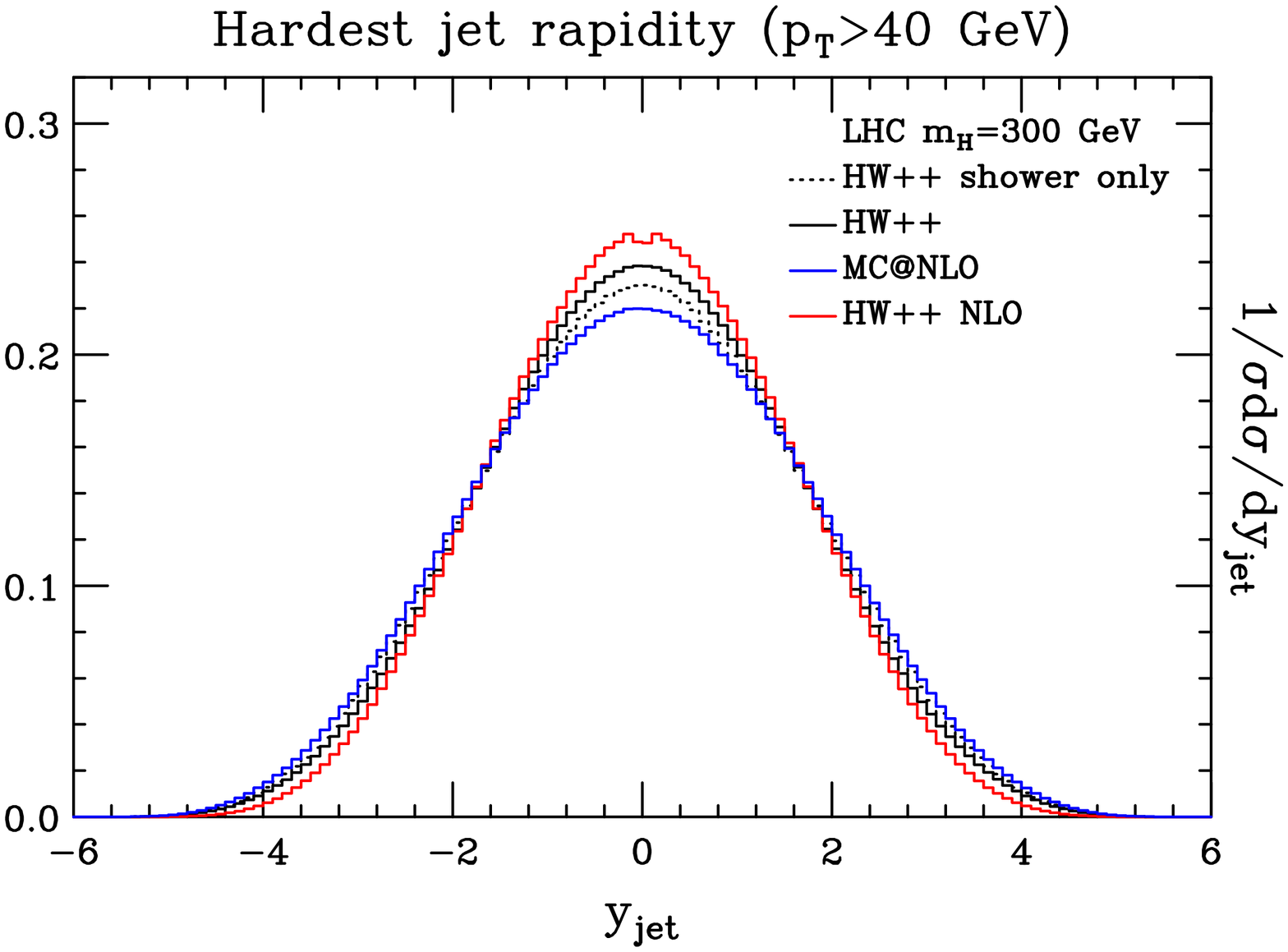}
\par\end{centering}

\caption{\label{fig:ZH_pt_and_jet_rapidity} On the left we compare predictions
for the transverse momentum spectrum of the colourless final-state
system comprised of the $Z$ and Higgs boson (left) in $q\bar{q}\rightarrow ZH$
events. From top to bottom, respectively, these results are obtained
for Tevatron energies with a Higgs boson mass of 160 GeV, LHC energies
with a Higgs boson mass of 115 GeV and LHC energies with a Higgs boson
mass of 300 GeV. On the right hand side we show the corresponding
distributions for the rapidity of the leading, highest $p_{T}$, jet.
The colour assignment for the different approaches used is shown inset,
and is the same as that in Fig.\,\ref{fig:higgs_and_W_pT_spectra}.}

\end{figure}

On the right of figure \ref{fig:ZH_pt_and_jet_rapidity} we show the
rapidity of the leading jet in $q\bar{q}\rightarrow ZH$ events. These
rapidity distributions are concentrated more in the central region
in the case of the Tevatron than the LHC. This behaviour can be inferred
from the fact that there is more phase space available for extra radiation
at LHC energies. The same line of reasoning also explains why the
rapidity distribution in the case of the 300 GeV Higgs boson at the
LHC is also more central than that of the 115 GeV Higgs boson. There
is a tendency in all of the plots for the \textsf{MC@NLO} distributions
to contain more events in the tails, conversely, the \textsf{POWHEG}
results show more jets produced in the central region. The predictions
from the uncorrected parton shower and the parton shower with a MEC
lie between those of \textsf{MC@NLO} and \textsf{POWHEG}, with the
former being slightly closer to \textsf{MC@NLO} and the latter closer
to \textsf{POWHEG}.

The plots on the left of figure \ref{fig:ZH_pt_and_jet_rapidity}
are a more interesting test of our methods, they show the transverse
momentum of the colourless \emph{Z}-Higgs boson system and also the
rapidity of the hardest jet in the same $q\bar{q}\rightarrow ZH$
events. The $p_{T}$ of the vector boson plus Higgs boson system is
generated directly by the Higgs-strahlung \textsf{POWHEG} simulation.
As in the case of the gluon fusion process we see that the \textsf{POWHEG}
results and those of the parton shower including matrix element corrections
are essentially the same, and that both are harder than the corresponding
\textsf{MC@NLO} prediction. However, the degree by which the latter
predictions are above those of \textsf{MC@NLO} is significantly reduced
with respect to that seen in the gluon fusion case. We again attribute
this to the relative differences in the rate of emission into the
dead zone, in the MEC method this occurs with probability $\mathcal{P}_{\mathrm{dead}}^{\mathrm{HW}}\left(\Phi_{B}\right)$
(Eq.\,\ref{eq:me_corr_dead_zone_prob}) while in \textsf{MC@NLO}
the analogous probability will be essentially given by the fraction
which the dead zone contributes to the total NLO cross section \emph{i.e.}
$\mathcal{P}_{\mathrm{dead}}^{\mathrm{NLO}}\left(\Phi_{B}\right)$
(Eq.\,\ref{eq:me_corr_mod_dead_zone_prob}). Since the denominators
in these emission probabilities differ by an amount of order the NLO
K-factor the differences arising in the gluon fusion case should be
large whereas they should be small in the Higgs-strahlung case. It
is difficult to see it clearly at the upper end of these Higgs-strahlung
$p_{T}$ spectra but the \textsf{MC@NLO} result is below that of \textsf{POWHEG}
and the MEC methods by roughly 30\%, which is compatible with the
enhancement due to the NLO corrections seen in the comparisons with
MCFM in \emph{e.g.} Fig.\,\ref{fig:higgs-strahlung_mV_star}.

\begin{figure}[H]
\begin{centering}
\includegraphics[width=0.38\textwidth,height=0.48\textwidth,angle=90]{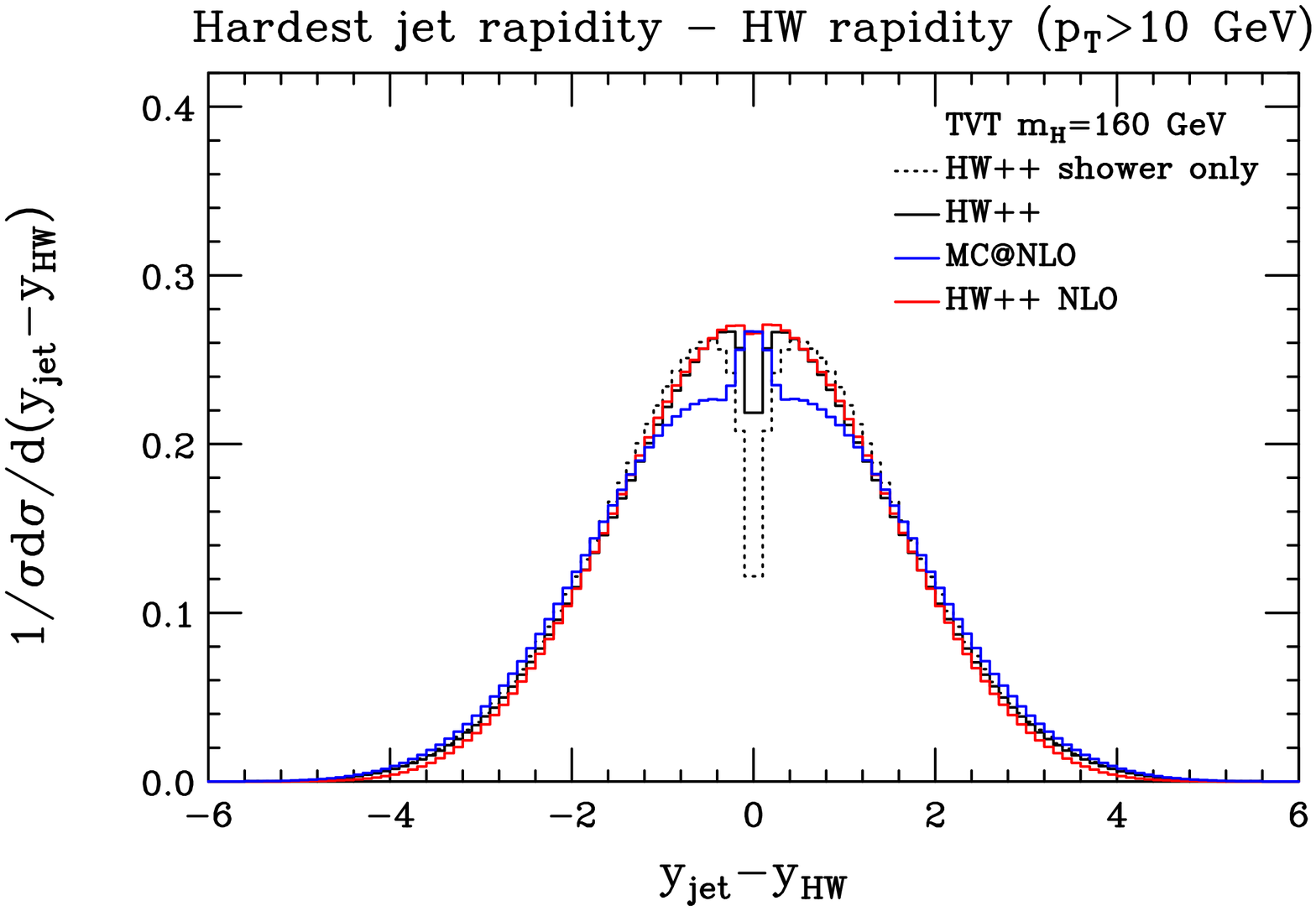}\hfill{}\includegraphics[width=0.38\textwidth,height=0.48\textwidth,angle=90]{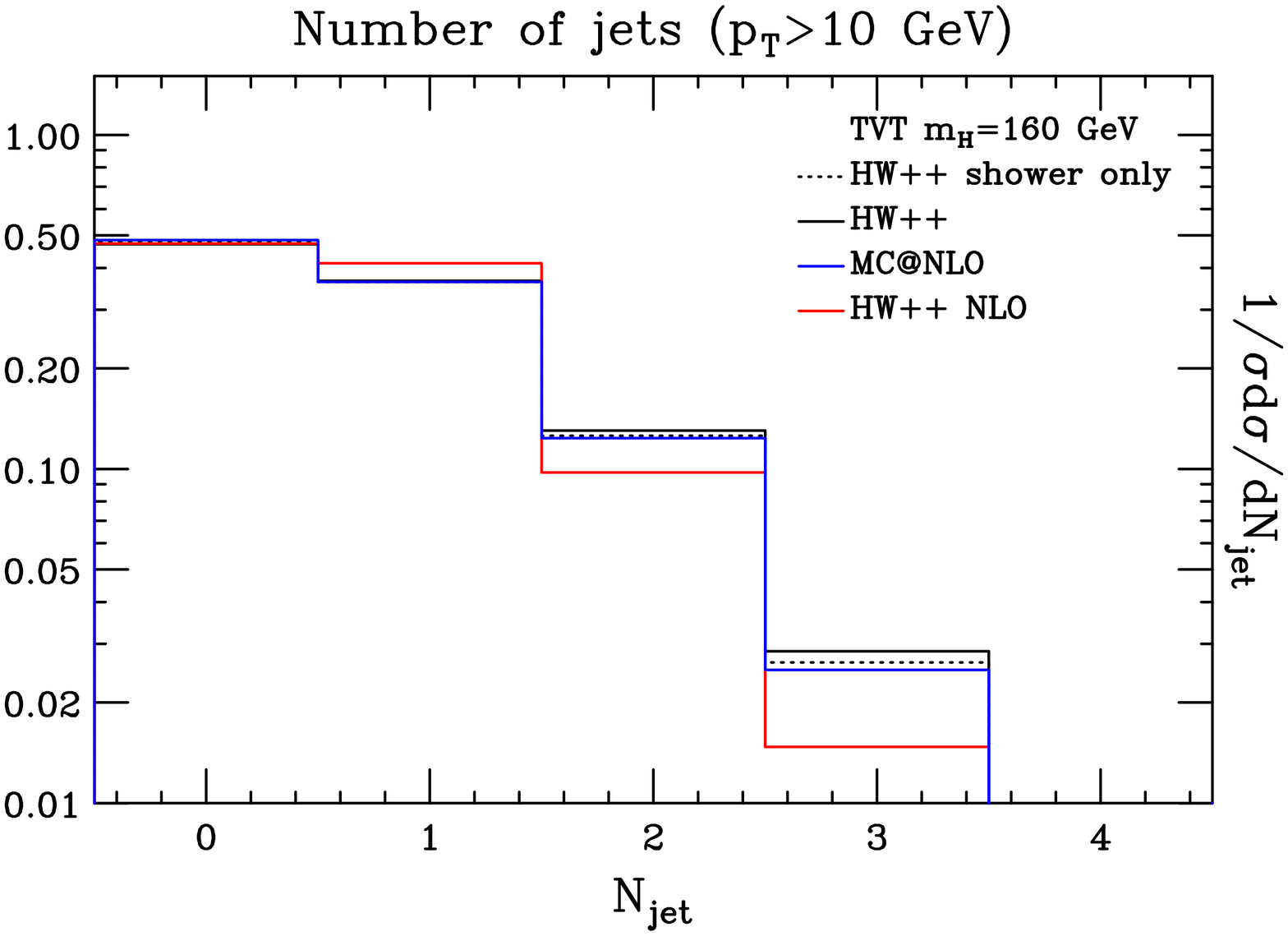}
\par\end{centering}

\vspace{10mm}

\begin{centering}
\includegraphics[width=0.38\textwidth,height=0.48\textwidth,angle=90]{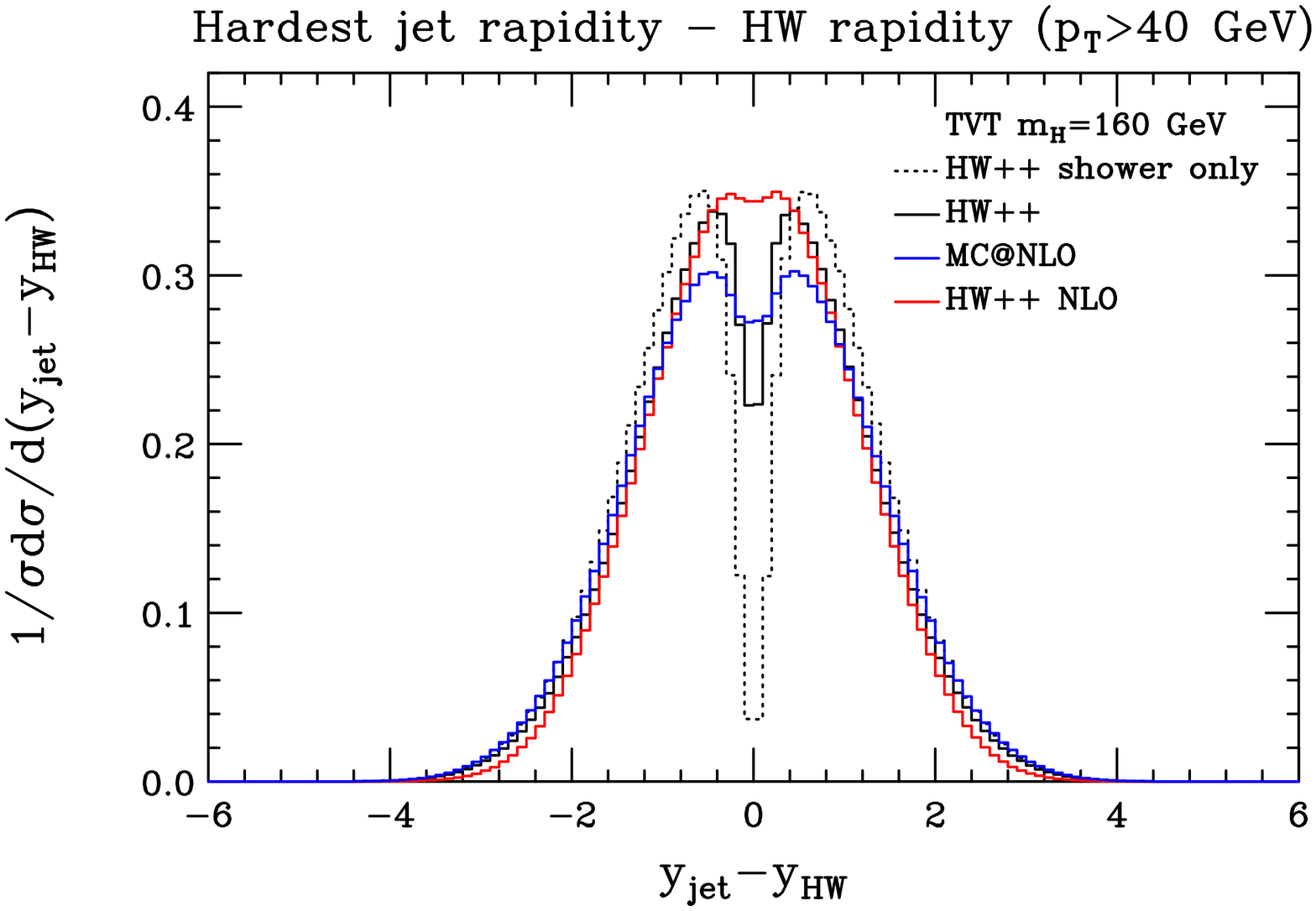}\hfill{}\includegraphics[width=0.38\textwidth,height=0.48\textwidth,angle=90]{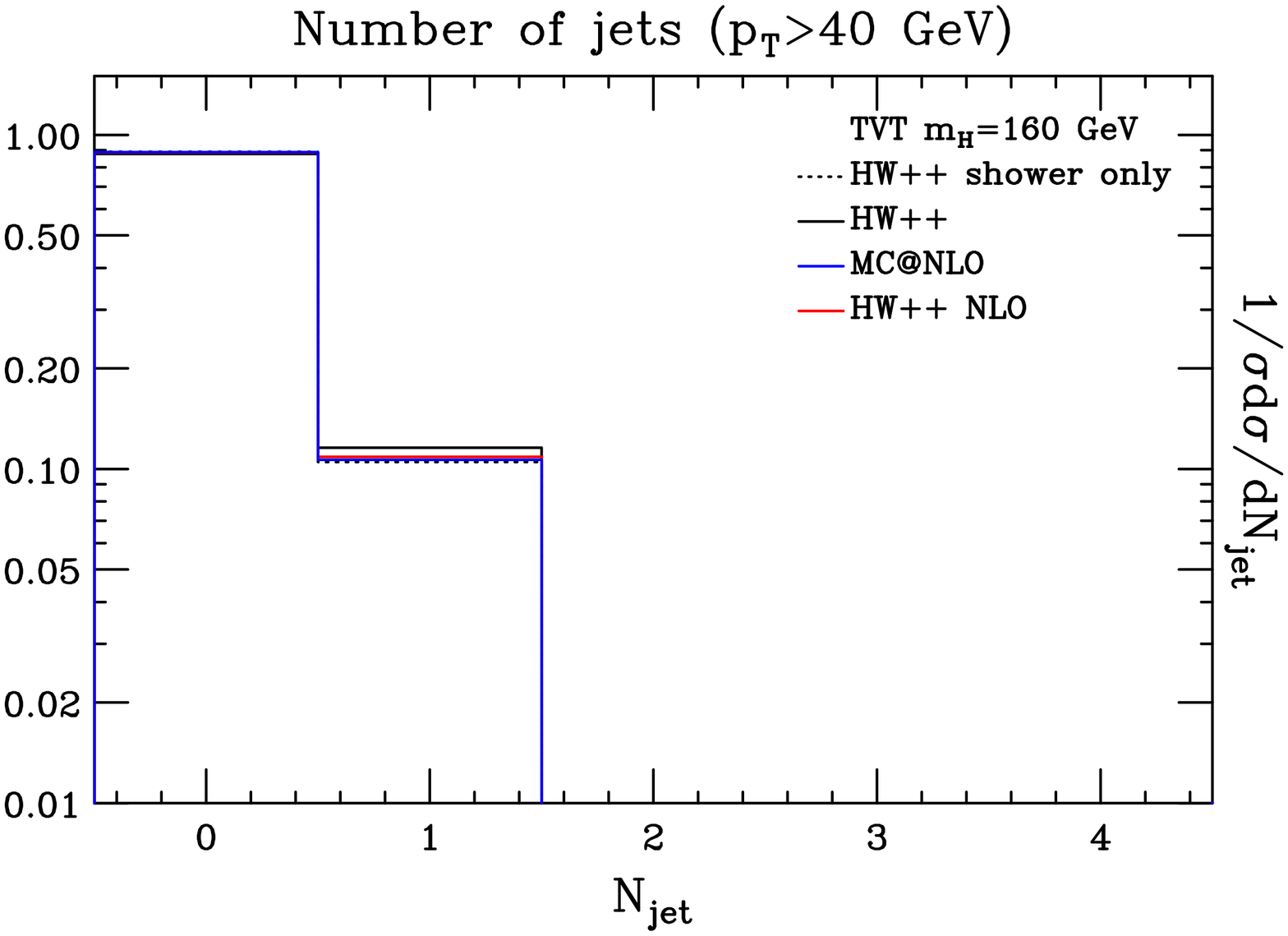}
\par\end{centering}

\vspace{10mm}

\begin{centering}
\includegraphics[width=0.38\textwidth,height=0.48\textwidth,angle=90]{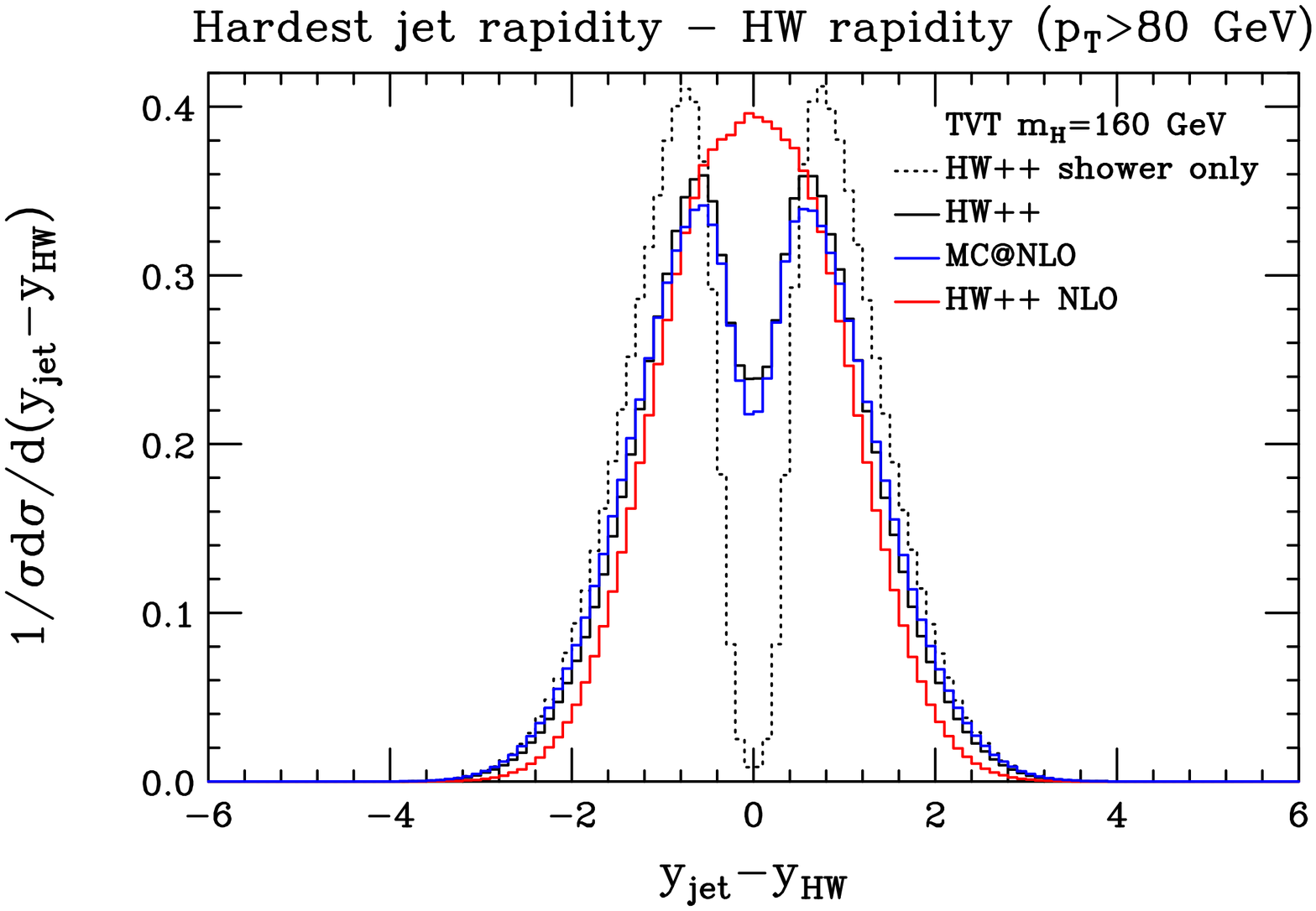}\hfill{}\includegraphics[width=0.38\textwidth,height=0.48\textwidth,angle=90]{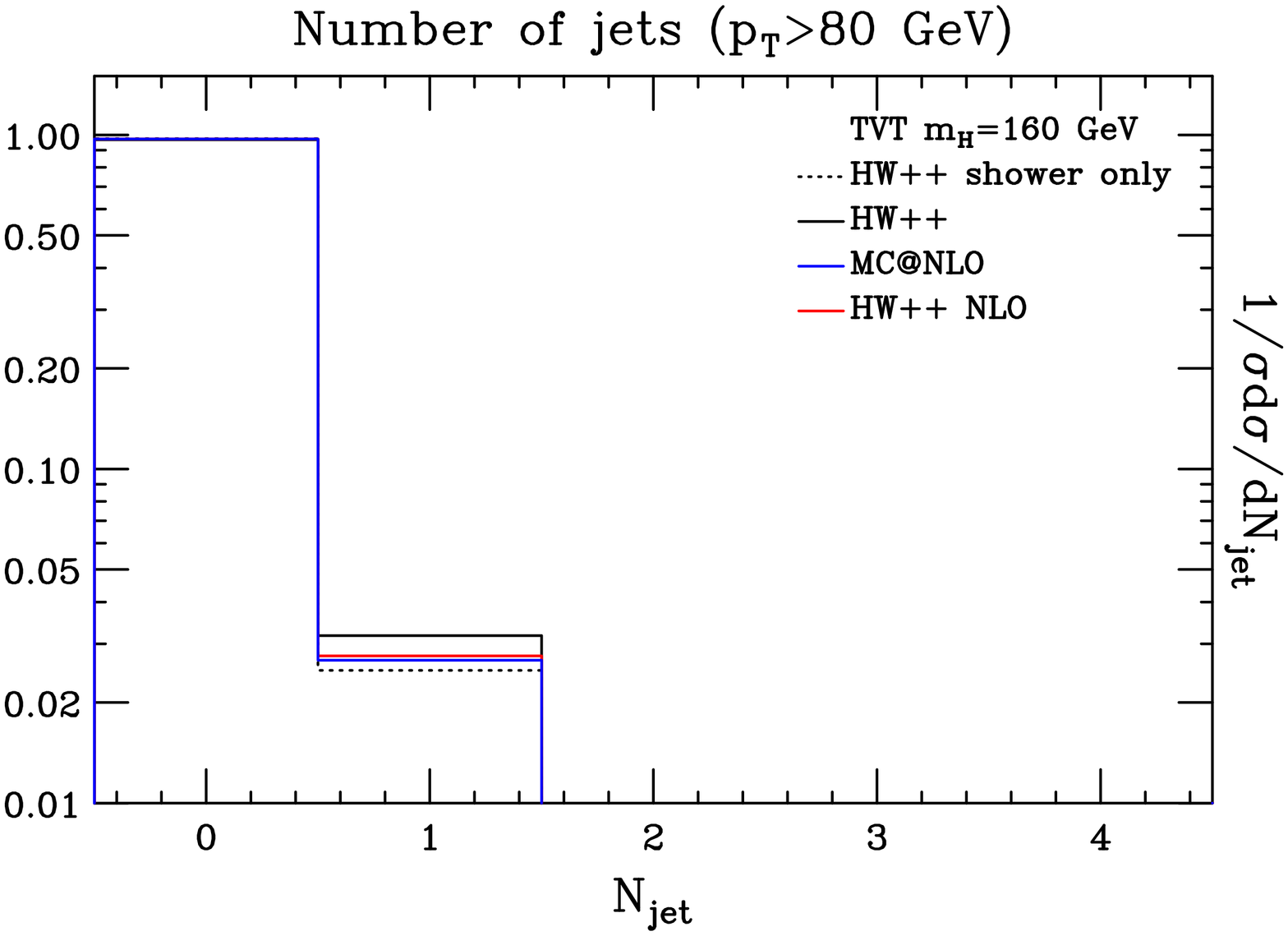}
\par\end{centering}

\caption{\label{fig:WH_tvt_yj-yh_njets} In the left-hand column are distributions
of the difference in rapidity between the leading jet and the colourless
$W-$Higgs boson system, in $q\bar{q}\rightarrow HW^{\pm}$ events
at the Tevatron, for increasing cuts on the $p_{T}$ of the jet. For
a single emission, the central value $\mathrm{y}_{\mathrm{jet}}-\mathrm{y}_{\mathrm{HW}}=0$
corresponds to a configuration where the $W$-Higgs boson and the
colliding partons travel at right-angles in the partonic centre-of-mass
frame (Eqs.\,\ref{eq:nlo_1_17}, \ref{eq:yjet-yh_approx_formula}).
In the plots on the right-hand side we show the corresponding jet
multiplicity distributions. The black and dotted lines show the predictions
obtained using \HWPP\ with and without matrix element corrections
respectively. The blue line shows the predictions of \textsf{MC@NLO}
and the red line corresponds to our \textsf{POWHEG} simulation inside
\HWPP. }

\end{figure}

\begin{figure}[H]

\begin{centering}
\includegraphics[width=0.38\textwidth,height=0.48\textwidth,angle=90]{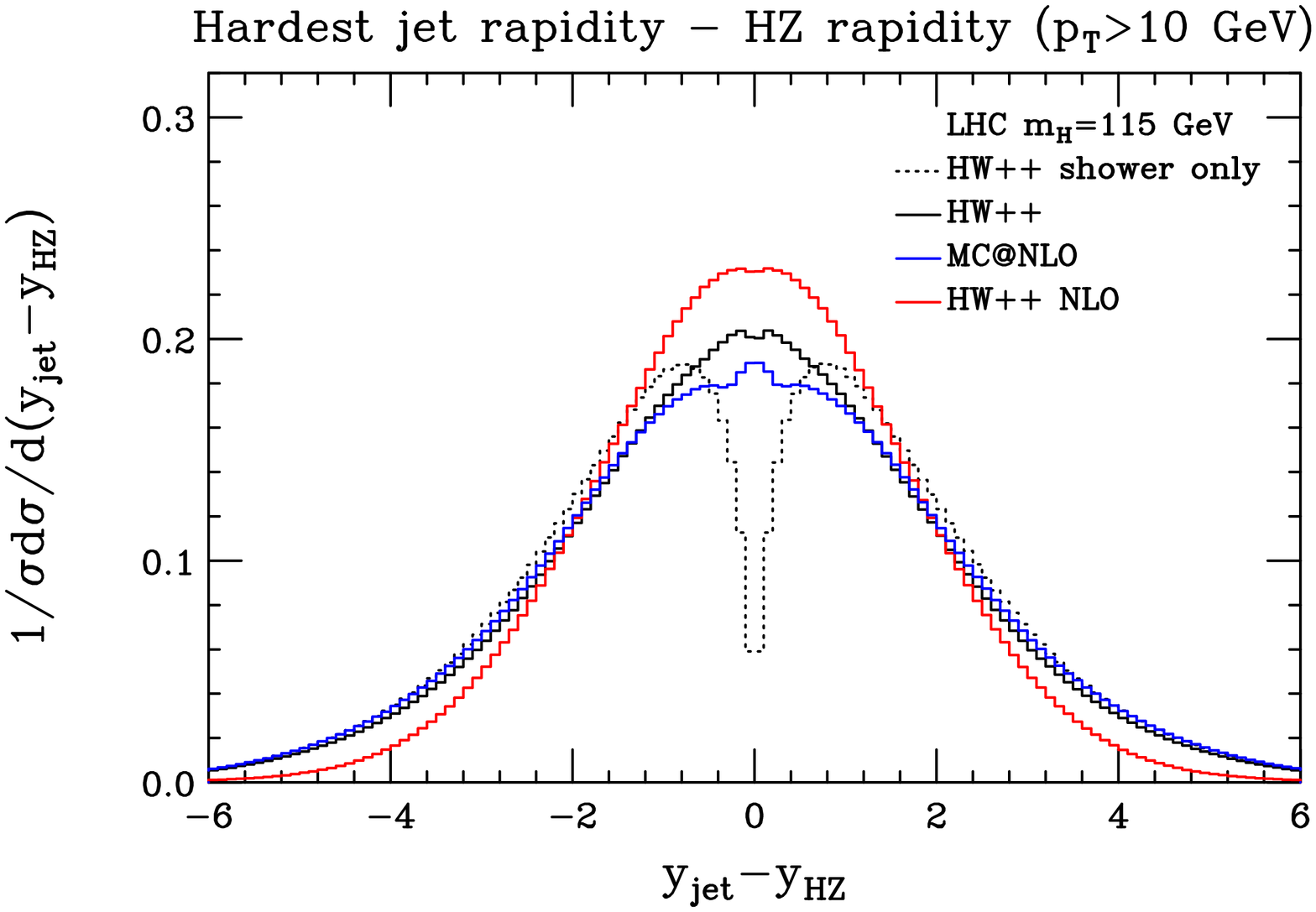}\hfill{}\includegraphics[width=0.38\textwidth,height=0.48\textwidth,angle=90]{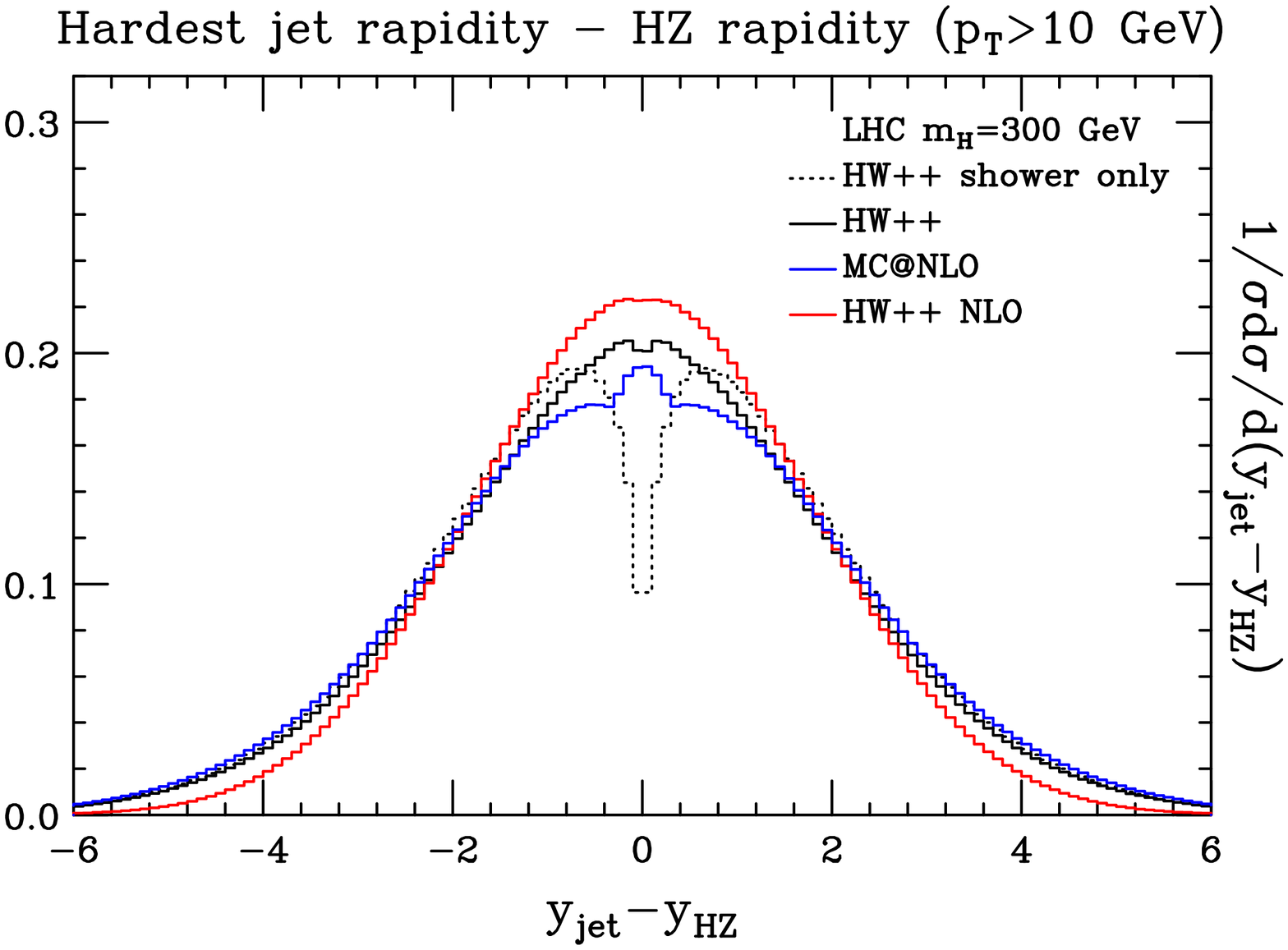}
\par\end{centering}

\vspace{10mm}

\begin{centering}
\includegraphics[width=0.38\textwidth,height=0.48\textwidth,angle=90]{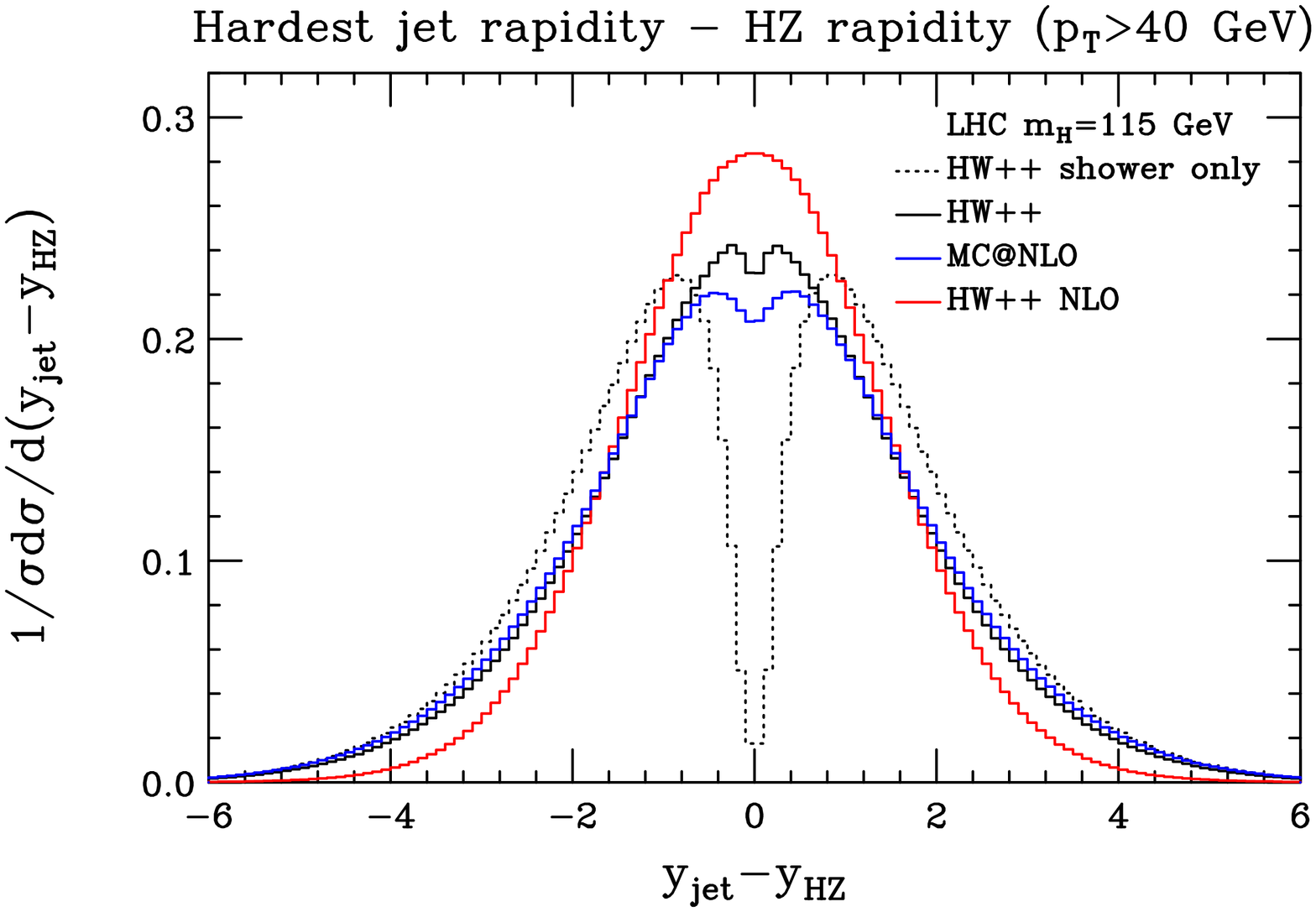}\hfill{}\includegraphics[width=0.38\textwidth,height=0.48\textwidth,angle=90]{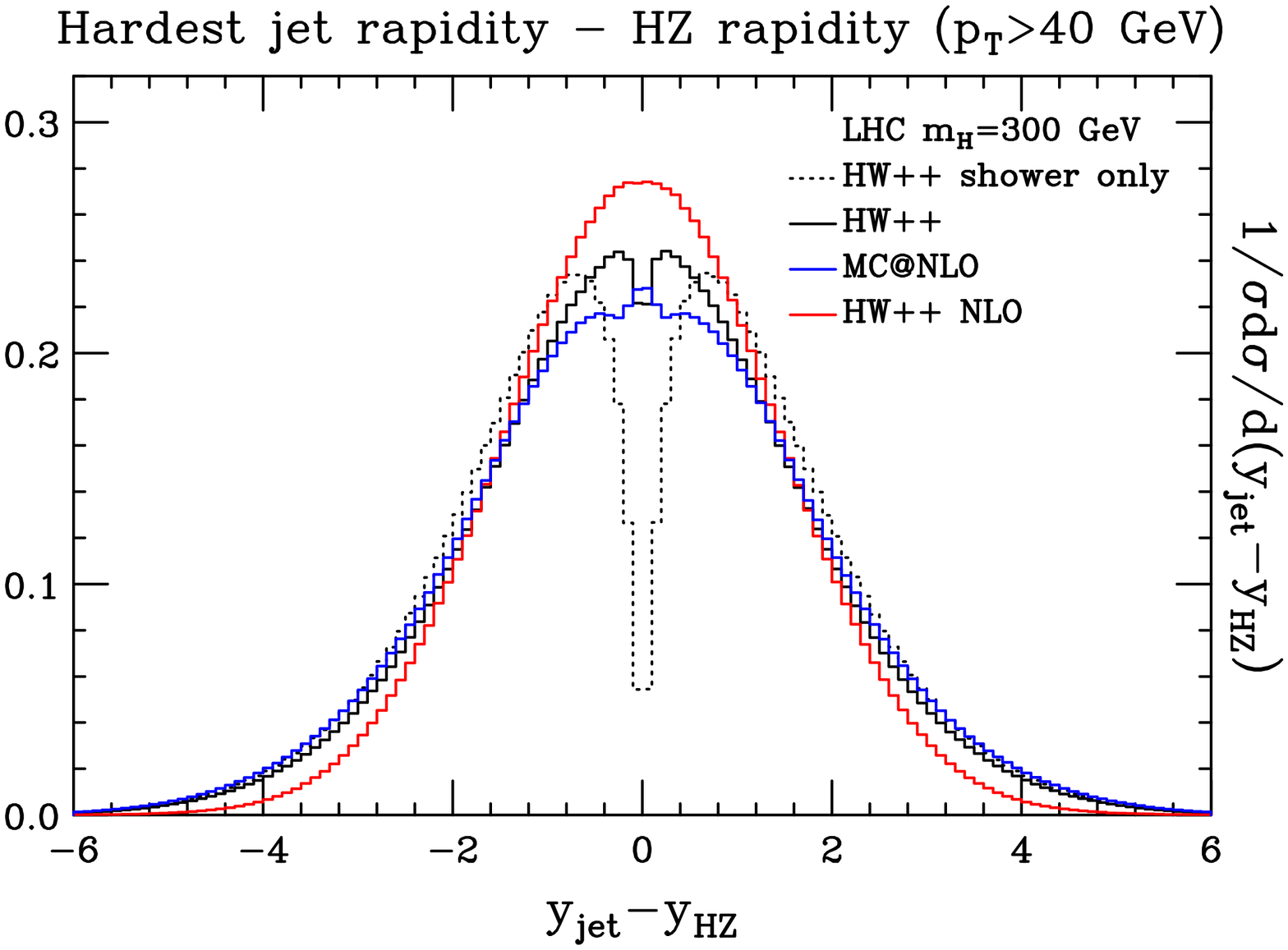}
\par\end{centering}

\vspace{10mm}

\begin{centering}
\includegraphics[width=0.38\textwidth,height=0.48\textwidth,angle=90]{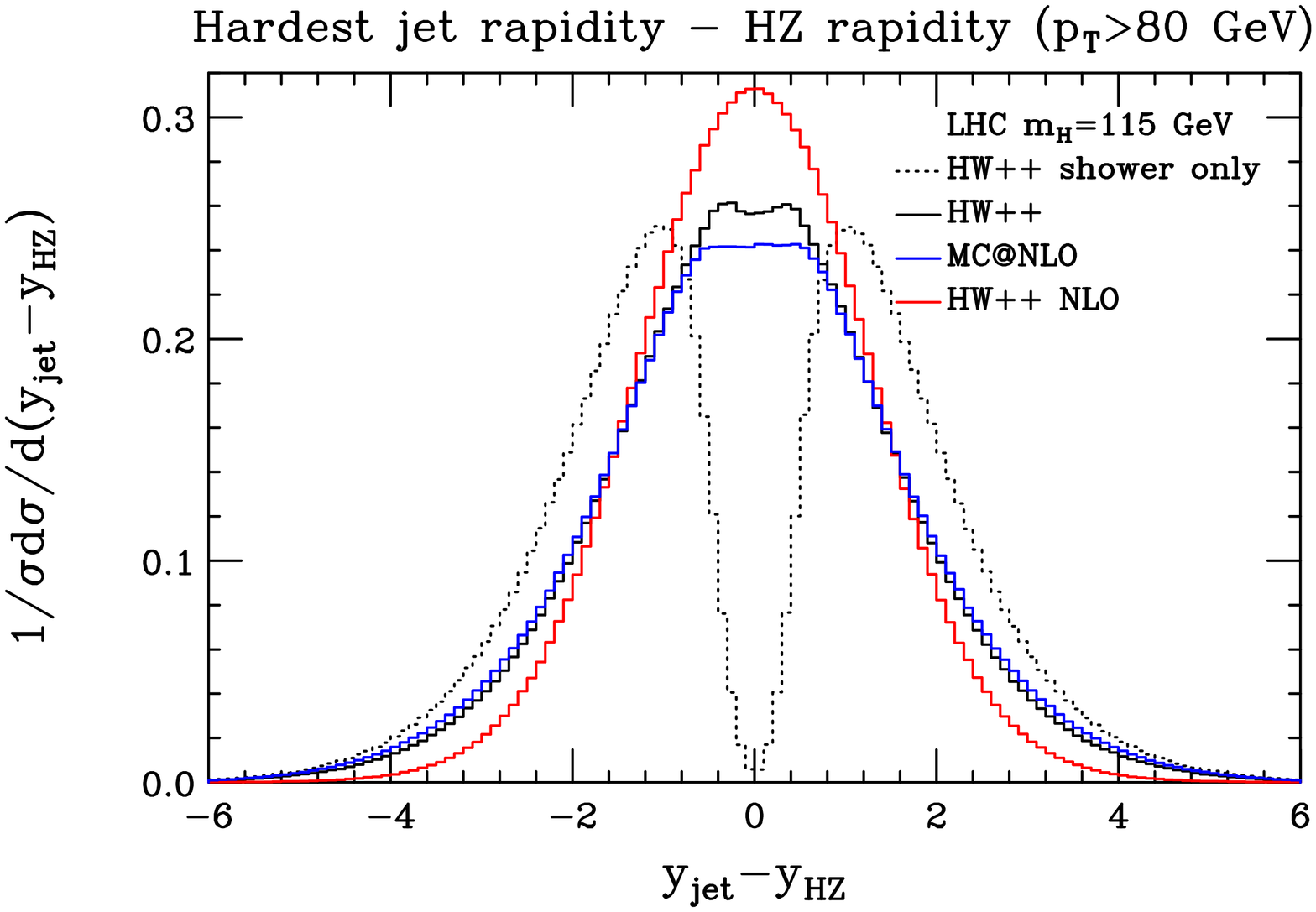}\hfill{}\includegraphics[width=0.38\textwidth,height=0.48\textwidth,angle=90]{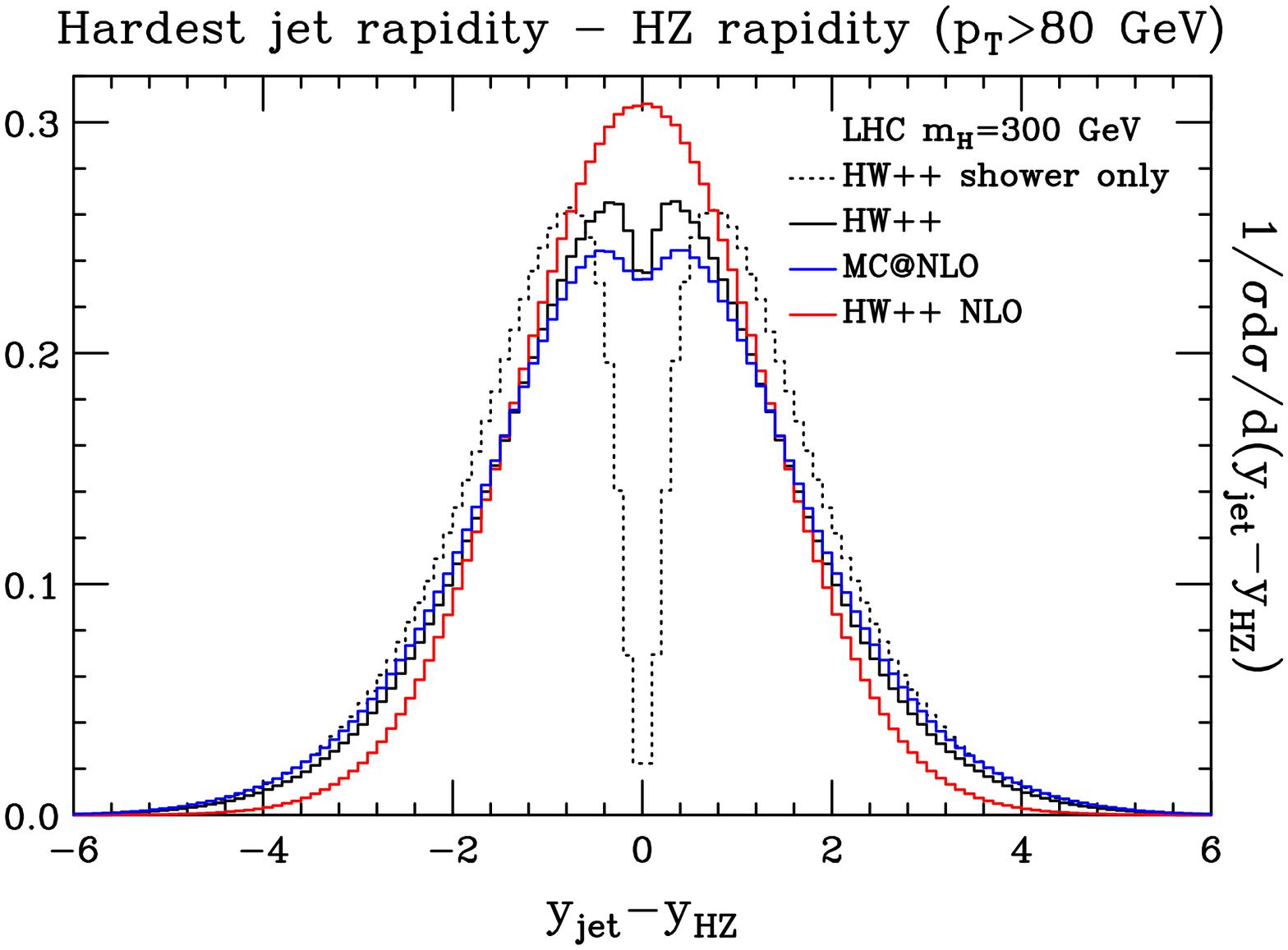}
\par\end{centering}

\caption{\label{fig:ZH_lhc_yj-yh_njets} Distributions of the difference in
rapidity between the leading jet and the Higgs boson in the $q\bar{q}\rightarrow ZH$
process at the LHC, for increasing cuts on the $p_{T}$ of the leading
jet. The series of plots on the left hand side are obtained for a
Higgs boson of mass 115 GeV, while those on the right correspond to
a Higgs boson of mass 300 GeV. The colour assignment of the various
predictions is described inset, it is the same as for earlier Tevatron
predictions for $q\bar{q}\rightarrow HW^{\pm}$ in Fig\,\ref{fig:WH_tvt_yj-yh_njets}.}

\end{figure}

In figures \ref{fig:WH_tvt_yj-yh_njets} and \ref{fig:ZH_lhc_yj-yh_njets}
we show distributions of the rapidity differences 
\mbox{$\mathrm{y}_{\mathrm{jet}}-\mathrm{y}_{\mathrm{ZH}}$}
and \mbox{$\mathrm{y}_{\mathrm{jet}}-\mathrm{y}_{\mathrm{WH}}$}, for the
$q\bar{q}\rightarrow WH$ and $q\bar{q}\rightarrow ZH$ processes,
at the Tevatron and LHC respectively. These plots display the same
features and trends as seen in the gluon fusion case. As before, the
uncorrected parton shower predictions give rise to volcano shaped
distributions due to the dead zone in the phase space and increasing
the $p_{T}$ cut on the leading jet again has the effect of broadening
the throat, reflecting the fact that the dead zone occupies an increasingly
large fraction of the allowed phase space.

The MEC and \textsf{MC@NLO} methods emit radiation into the dead zone,
filling the throat region in the $\mathrm{y}_{\mathrm{jet}}-\mathrm{y}_{\mathrm{WH}}$
and $\mathrm{y}_{\mathrm{jet}}-\mathrm{y}_{\mathrm{ZH}}$ plots, however,
they still exhibit a clear sensitivity to the dead zone boundary which
is manifest as irregularities in the central regions of the distributions.
For the case that the $p_{T}$ cut on the leading jet is soft, $p_{T}>$10
GeV, the MEC predictions show a tiny, sharp dip around the centre,
while those of \textsf{MC@NLO} show the formation of a small tower
in the same place. This is plainly a mis-match of $\mathcal{O}\left(\alpha_{\mathrm{S}}^{2}\right)$
terms across the phase-space partition. As the $p_{T}$ veto increases
the dips in the MEC predictions tend to deepen and in the case of
\textsf{MC@NLO}, the small towers turn to small dips. In all cases
the \textsf{POWHEG} distributions are smooth, exhibiting no irregular
\emph{volcanic} features, as expected, furthermore they are more concentrated
in the central region than all of the other predictions, indicating
a tendency to emit proportionally more wide angle radiation.

We point out that in all of the MEC and \textsf{MC@NLO} predictions
the residual effects of the phase-space dead zone are felt much less
strongly in the case of Higgs-strahlung than in gluon fusion. These
behaviours are less marked for the same reason that the \textsf{MC@NLO}
and \textsf{POWHEG} $p_{T}$ spectra agree better for Higgs-strahlung
than for gluon fusion: the fact that the NLO corrections are substantially
smaller for Higgs-strahlung means $\overline{B}\left(\Phi_{B}\right)$
is less different to $B\left(\Phi_{B}\right)$ and so the $\mathcal{O}\left(\alpha_{\mathrm{S}}^{2}\right)$
differences in the rates at which each method populates the dead zone
are greatly reduced, \emph{cf.} Eqs.\,\ref{eq:me_corr_dead_zone_prob},
\ref{eq:me_corr_mod_dead_zone_prob}.

\subsection{Emission rates in the dead zone\label{sub:Discussion}}

In this subsection we present an heuristic discussion of the rates
with which each approach emits into the high $p_{T}$ region. In particular
we consider the area of phase space corresponding to transverse momenta
$p_{T}>m_{n}$, where $m_{n}$ is the mass of the colourless final-state
system. This region is completely contained within the dead zone and
the contribution which it makes to the cross section is not logarithmically
enhanced (Sect.\,\ref{sub:The-dead-zone}). 

From Eq.\,\ref{eq:me_corr_dead_zone_prob} it follows that the probability
for \HWPP\  to generate an emission with $p_{T}>m_{n}$ is \begin{equation}
\mathcal{P}_{m_{n}}^{\mathrm{HW}}\left(\Phi_{B}\right)=\int_{m_{n}}\mathrm{d}\Phi_{R_{1}}\,\frac{\widehat{R}_{1}\left(\Phi_{B},\Phi_{R_{1}}\right)}{B\left(\Phi_{B}\right)}\,,\label{eq:hard_ME_prob}\end{equation}
where the omitted higher-order terms in the exponential series are
negligible, not simply because they carry higher powers of the coupling
constant but also, significantly, because the contribution to the
total cross section from this region is in general very small. In
Eq.\,\ref{eq:hard_ME_prob} $\widehat{R}_{1}\left(\Phi_{B},\Phi_{R_{1}}\right)$
is, as before, the real single-emission matrix element squared, including
flux factors and PDFs, and $\Phi_{B}$ and $\Phi_{R_{1}}$ are the
Born and radiative variables parametrising the two-body phase space.
The lower limit on the integral signifies that it extends over all
available phase space above $p_{T}>m_{n}$. Our adding a subscript
`1' to each $R$, has no significance here but it will be useful later,
in discussing NNLO corrections. 

The fraction of \textsf{MC@NLO} events with emissions in this region
is given by the corresponding fraction of the NLO cross section\begin{equation}
\mathcal{P}_{m_{n}}^{\mathrm{NLO}}\left(\Phi_{B}\right)=\int_{m_{n}}\mathrm{d}\Phi_{R_{1}}\mbox{ }\frac{\widehat{R}_{1}\left(\Phi_{B},\Phi_{R_{1}}\right)}{\overline{B}\left(\Phi_{B}\right)}\,.\label{eq:MC@NLO_prob}\end{equation}
For what follows it will be useful to note that if we expand the denominator
in Eq.\,\ref{eq:MC@NLO_prob}, neglecting terms $\mathcal{O}\left(\alpha_{S}^{3}\right)$
and above, we have\begin{equation}
\mathcal{P}_{m_{n}}^{\mathrm{NLO}}\left(\Phi_{B}\right)=\int_{m_{n}}\mathrm{d}\Phi_{R_{1}}\mbox{ }\frac{\widehat{R}_{1}\left(\Phi_{B},\Phi_{R_{1}}\right)}{B\left(\Phi_{B}\right)}\left(2-\frac{\overline{B}\left(\Phi_{B}\right)}{B\left(\Phi_{B}\right)}\right)\,.\label{eq:MC@NLO_prob_alphaS^2}\end{equation}

In \textsf{POWHEG} the Sudakov form factor, $\Delta_{\hat{R}}\left(p_{T}\right)$
in Eq\,\ref{eq:powheg_3}, is the probability that no radiation is
emitted from the leading-order partons in evolving from the maximum
kinematically allowed transverse momentum, down to $p_{T}$. Hence
a \textsf{POWHEG} simulation will emit radiation into the same high
$p_{T}$ region with probability $\mathcal{P}_{m_{n}}^{\mathrm{PH}}=1-\Delta_{\hat{R}}\left(m_{n}\right)$.
Neglecting kinematically suppressed $\mathcal{O}\mbox{\ensuremath{\left(\alpha_{\mathrm{S}}^{2}\right)}}$
terms, as in the case of the MEC, this gives, \begin{equation}
\mathcal{P}_{m_{n}}^{\mathrm{PH}}=\int_{m_{n}}\mathrm{d}\Phi_{R_{1}}\,\frac{\widehat{R}_{1}\left(\Phi_{B},\Phi_{R_{1}}\right)}{B\left(\Phi_{B}\right)}\,.\label{eq:POWHEG_prob}\end{equation}

Comparing $ $Eqs.\,\ref{eq:hard_ME_prob} and \ref{eq:POWHEG_prob},
we see that the emission rates in the high $p_{T}$ region, from \textsf{POWHEG
}and the MEC, are the same up to the scale used in the strong coupling
constant (in the former it is $p_{T}$ while in the latter it is $m_{T}$).
These corresponding emission rates are evident in the $p_{T}$ spectra
in Figs.\,\ref{fig:ggh_pT_spectra} and \ref{fig:ZH_pt_and_jet_rapidity}.
It is also clear that \textsf{MC@NLO} will emit into this region less
often than the other two since, whereas the \textsf{POWHEG} and \textsf{Herwig++}
simulations have an emission probability inversely proportional to
the Born cross section $B\left(\Phi_{B}\right)$, the probability
of emission in \textsf{MC@NLO} is inversely proportional to the, larger,
NLO cross section, $\overline{B}\left(\Phi_{B}\right)$. This argument
is also supported by what we have seen in the $p_{T}$ spectra. Essentially
we have\begin{equation}
\mathcal{P}_{m_{n}}^{\mathrm{HW}}\approx\mathcal{P}_{m_{n}}^{\mathrm{PH}}\approx\mathcal{K}\,\mathcal{P}_{m_{n}}^{\mathrm{NLO}}\label{eq:comparing_leading_probabilities}\end{equation}
 where $\mathcal{K}$ is the relevant NLO K-factor. For processes
with smaller K-factors the differences in the population of the dead
zone in each approach should not differ too much but \textsf{MC@NLO}
will be systematically softer than \textsf{POWHEG} and \HWPP.

Is one of these probabilities more correct than the others? In Ref.\,\cite{Alioli:2008tz}
the authors compared the Higgs boson $p_{T}$ spectrum from their
\textsf{POWHEG} simulation to NNLO predictions \cite{Catani:2007vq,Grazzini:2008tf}
and found that they agreed better than those of \textsf{MC@NLO}. The
amount of radiation produced in the high $p_{T}$ dead zone in the
case of the fixed order, NNLO, Monte Carlo used in Ref.\,\cite{Alioli:2008tz}
should be the relevant fraction of the NNLO cross section\begin{eqnarray}
\mathcal{P}_{m_{n}}^{\mathrm{NNLO}}\left(\Phi_{B}\right) & = & \int_{m_{n}}\mathrm{d}\Phi_{R_{1}}\left[\widehat{R}_{1}\left(\Phi_{B},\Phi_{R_{1}}\right)+R_{1+1}\left(\Phi_{B},\Phi_{R_{1}}\right)+\int\mathrm{d}\Phi_{R_{2}}\, R_{2}\left(\Phi_{B},\Phi_{R_{1}},\Phi_{R_{2}}\right)\right]\nonumber \\
 & \div & \mathrm{d}\sigma_{\mathrm{NNLO}}\left(\Phi_{B}\right)\,,\label{eq:NNLO_prob}\end{eqnarray}
where the sum of the last two terms in the numerator represents the
\emph{finite} combination of the\emph{ }one-loop single emission matrix
element interfering with the tree-level single emission matrix element,
and the squared double emission matrix element respectively, including
PDF and flux factors. Since the numerator of $\mathcal{P}_{m_{n}}^{\mathrm{NNLO}}$
is $\mathcal{O}\left(\alpha_{\mathrm{S}}\right)$, if we omit terms
$\mathcal{O}\left(\alpha_{\mathrm{S}}^{3}\right)$ in $\mathcal{P}_{m_{n}}^{\mathrm{NNLO}}$
we can replace the NNLO differential cross section in the denominator
by the NLO one (Eq.\,\ref{eq:nlo_2_14}), whence it follows that
\begin{equation}
\mathcal{P}_{m_{n}}^{\mathrm{NNLO}}\left(\Phi_{B}\right)=\int_{m_{n}}\mathrm{d}\Phi_{R_{1}}\,\frac{\widehat{R}_{1}\left(\Phi_{B},\Phi_{R_{1}}\right)}{B\left(\Phi_{B}\right)}\left[1-\frac{\overline{B}\left(\Phi_{B}\right)}{B\left(\Phi_{B}\right)}+\frac{\overline{R}_{1}\left(\Phi_{B},\Phi_{R_{1}}\right)}{\widehat{R}_{1}\left(\Phi_{B},\Phi_{R_{1}}\right)}\right]\,,\label{eq:NNLO_prob_O(alphaS^2)}\end{equation}
 where $\overline{B}\left(\Phi_{B}\right)$ was defined in Eq.\,\ref{eq:powheg_2}
and $\overline{R}\left(\Phi_{B},\Phi_{R_{1}}\right)$ is defined analogously
as\begin{eqnarray}
\overline{R}_{1}\left(\Phi_{B},\Phi_{R_{1}}\right) & = & \widehat{R}_{1}\left(\Phi_{B},\Phi_{R_{1}}\right)+R_{1+1}\left(\Phi_{B},\Phi_{R_{1}}\right)+\int\mathrm{d}\Phi_{R_{2}}\, R_{2}\left(\Phi_{B},\Phi_{R_{1}},\Phi_{R_{2}}\right)\,.\label{eq:Rbar_definition}\end{eqnarray}

Whereas the NNLO rate has $\overline{R}_{1}$ in the numerator \emph{all}
of the others just contain the term corresponding to the squared single
emission matrix element $\widehat{R}_{1}$. Replacing $\overline{R}_{1}\rightarrow\widehat{R}_{1}$
in Eq.\,\ref{eq:NNLO_prob_O(alphaS^2)} gives Eq.\,\ref{eq:MC@NLO_prob_alphaS^2}.
We can consider $\overline{B}/B$ and $\overline{R}_{1}/\widehat{R}_{1}$
as K-factors for the processes $gg\rightarrow H$ and $gg\rightarrow H+{\rm jet}$,
differential in $\Phi_{B}$ and $\left\{ \Phi_{B},\Phi_{R_{1}}\right\} $,
respectively. One can see from Eqs.\,\ref{eq:POWHEG_prob} and \ref{eq:NNLO_prob_O(alphaS^2)}
that if these two K-factors coincide so too will the rates for the
\textsf{POWHEG }and NNLO calculations. In fact this seems to be more-or-less
the case for $gg\rightarrow H$ and $gg\rightarrow H+{\rm jet}$ processes,
which have very similar K-factors of around 1.6 / 1.7 \cite{deFlorian:1999zd}.
It then seems quite feasible that the $p_{T}$ spectrum of the Higgs
boson in gluon fusion should be better modelled by the \textsf{POWHEG}
approach, as was found to be the case in Ref.\,\cite{Alioli:2008tz}.

Equation \ref{eq:NNLO_prob_O(alphaS^2)} does not tell us that \textsf{POWHEG}
will \emph{generally} reproduce higher orders better than \textsf{MC@NLO}.
According to Eq.\,\ref{eq:NNLO_prob_O(alphaS^2)} if $\overline{R}_{1}/\widehat{R}_{1}$
is more-or-less one and $\overline{B}/B$ is significantly greater
than one then \textsf{MC@NLO} will give a better estimate of $\mathcal{P}_{m_{n}}^{\mathrm{NNLO}}$
than the \textsf{POWHEG} approach.

\section{Conclusion\label{sec:Conclusion}}

In this work we have fully realized the \textsf{POWHEG} NLO matching
prescription for Higgs boson production via gluon fusion and Higgs-strahlung
processes, within the \HWPP\ Monte Carlo event generator, including
truncated shower effects to correctly include colour coherence.

The cross sections and parton level NLO distributions were found to
be in very good agreement with the MCFM NLO Monte Carlo. The shapes
of the emission spectra from the full simulation, including parton
shower effects, are seen to broadly agree well with the older matrix
element correction method and also \textsf{MC@NLO}.

We observe that the $p_{T}$ spectra from the \textsf{MC@NLO} program
tend to be softer than those of \textsf{POWHEG} and the matrix element
correction method. We ascribe this effect to differences, at the level
of $\mathcal{O}\left(\alpha_{\mathrm{S}}^{2}\right)$ terms, beyond
the stated accuracy of all approaches, in the rate at which they emit
radiation into the so-called \emph{dead zone}, associated to high $p_{T}$
and wide angle emissions. We have been able to estimate the magnitude
of the relative difference with a fair degree of success and, based
on this line of reasoning, we presented an argument for why the \textsf{POWHEG}
method appears to better reproduce the NNLO Higgs boson $p_{T}$ spectrum
in gluon fusion.

Separately, we have shown that a mis-match between terms of this magnitude
manifests itself as marked sensitivity to the unphysical dead zone
partition, for the predictions of the rapidity difference distributions
$\mathrm{y}_{\mathrm{jet}}-\mathrm{y}_{\mathrm{H}}$ and $\mathrm{y}_{\mathrm{jet}}-\mathrm{y}_{\mathrm{VH}}$,
for both the matrix element correction method and \textsf{MC@NLO}.
These distributions acquire irregularities in the central region corresponding
to wide angle emissions. This would seem to rule out the possibility
of these two methods giving a good approximation of unenhanced NNLO
effects in these distributions. Conversely, the \textsf{POWHEG} predictions
for these distributions are smooth and physical in appearance, by
construction, they have no sensitivity to the dead zone partition.

At present there is no known method of consistently including exact NNLO fixed
order calculations within a parton shower simulation. An approximate means of 
doing this, involving the reweighting of \textsf{PYTHIA} \cite{Sjostrand:2006za} 
and \textsf{MC@NLO} events such that they reproduce certain NNLO distributions, 
has been recently carried out in Ref.\,\cite{Davatz:2006ut}. Based on the 
findings discussed above, it is our expectation that applying the same 
reweighting technique to \textsf{POWHEG} events should lead to further 
improvements in those predictions.

Both the gluon fusion and Higgs-strahlung simulations we have presented
here are already available in \HWPP \textsf{2.3}. The algorithm we
have used to implement the \textsf{POWHEG} formalism, specifically
that part concerning the \emph{inverse} \emph{mapping} of the hardest
emission kinematics to a set of shower variables has further applications
in multi-leg matching procedures \emph{e.g.} the CKKW method~\cite{Catani:2001cc,Krauss:2002up}.
Likewise, the general formulae leading to Eqs.\,\ref{eq:nlo_1_14},
\ref{eq:nlo_2_14}, for the phase space and differential cross-section
for $a+b\rightarrow n$, can be readily used to implement other NLO
calculations such as those in Refs.\,\cite{Mele:1990bq,Frixione:1992pj,Frixione:1993yp}.
This work is already at an advanced stage and will appear in the forthcoming
version of \HWPP.

\acknowledgments

We are grateful to all the other members of the \HWPP\ collaboration
for valuable discussions, especially Mike Seymour for carefully
reading the manuscript. Keith\,Hamilton thanks the CERN Theory
group for their kind hospitality in the course of this work, as well
as Fabio Maltoni for sharing his expertise. This work was supported
by the Science and Technology Facilities Council, formerly the Particle
Physics and Astronomy Research Council, the European Union Marie Curie
Research Training Network MCnet under contract MRTN-CT-2006-035606.
Keith\,Hamilton is supported by the Belgian Interuniversity
Attraction Pole, PAI, P6/11.

\appendix
%dummy comment inserted by tex2lyx to ensure that this paragraph is not empty
%dummy comment inserted by tex2lyx to ensure that this paragraph is not empty
%dummy comment inserted by tex2lyx to ensure that this paragraph is not empty
%dummy comment inserted by tex2lyx to ensure that this paragraph is not empty
%dummy comment inserted by tex2lyx to ensure that this paragraph is not empty
%dummy comment inserted by tex2lyx to ensure that this paragraph is not empty

\section{Regularized and unregularized splitting functions\label{sec:Splitting-functions}}

This appendix contains the splitting functions needed in the NLO calculations.
The bare (spin averaged) splitting functions in $n=4-2\epsilon$ dimensions
are of the form \[
\hat{P}_{i,\widetilde{ic}}\left(x;\epsilon\right)=\hat{P}_{i,\widetilde{ic}}\left(x\right)+\epsilon\hat{P}_{i,\widetilde{ic}}^{\epsilon}\left(x\right)\,,\]
 where

\begin{align*}
\hat{P}_{gg}\left(x\right) & =2C_{A}\left[\frac{x}{1-x}+\frac{1-x}{x}+x\left(1-x\right)\right], & \mbox{ }\mbox{ }\mbox{ }\mbox{ }\mbox{ }\mbox{ }\mbox{ }\mbox{ }\mbox{ }\hat{P}_{gg}^{\epsilon}\left(x\right) & =0,\\
\hat{P}_{qq}\left(x\right) & =C_{F}\left[\frac{1+x^{2}}{1-x}\right], & \hat{P}_{qq}^{\epsilon}\left(x\right) & =-C_{F}\left(1-x\right),\\
\hat{P}_{qg}\left(x\right) & =C_{F}\left[\frac{1+\left(1-x\right)^{2}}{x}\right], & \hat{P}_{qg}^{\epsilon}\left(x\right) & =-C_{F}\, x,\\
\hat{P}_{gq}\left(x\right) & =T_{R}\left[1-2x\left(1-x\right)\right], & \hat{P}_{gq}^{\epsilon}\left(x\right) & =-2T_{R}\, x\left(1-x\right)+\mathcal{O}\left(\epsilon\right)\,.\end{align*}
 We write the `customary' regularized Altarelli-Parisi equations kernels
using the $\rho$-distributions as\[
P_{i,\widetilde{ic}}\left(x\right)=P_{i,\widetilde{ic}}^{\rho}\left(x\right)+C_{i,\widetilde{ic}}\left(p_{i,\widetilde{ic}}+4\ln\eta\right)\delta\left(1-x\right)\,,\]
 where\begin{align*}
P_{gg}^{\rho}\left(x\right) & =2C_{A}\left[\frac{x}{\left(1-x\right)_{\rho}}+\frac{1-x}{x}+x\left(1-x\right)\right], & \mbox{ }\mbox{ }\mbox{ }\mbox{ }\mbox{ }\mbox{ }\mbox{ }\mbox{ }\mbox{ }C_{gg} & =C_{A}, & \mbox{ }\mbox{ }\mbox{ }\mbox{ }\mbox{ }\mbox{ }\mbox{ }\mbox{ }\mbox{ }p_{gg} & =\frac{2\pi b_{0}}{C_{A}},\\
P_{qq}^{\rho}\left(x\right) & =C_{F}\left[\frac{1+x^{2}}{\left(1-x\right)_{\rho}}\right], & \mbox{ }\mbox{ }\mbox{ }\mbox{ }\mbox{ }\mbox{ }\mbox{ }\mbox{ }\mbox{ }C_{qq} & =C_{F}, & \mbox{ }\mbox{ }\mbox{ }\mbox{ }\mbox{ }\mbox{ }\mbox{ }\mbox{ }\mbox{ }p_{qq} & =\frac{3}{2},\\
P_{qg}^{\rho}\left(x\right) & =C_{F}\left[\frac{1+\left(1-x\right)^{2}}{x}\right],\\
P_{gq}^{\rho}\left(x\right) & =T_{R}\left[x^{2}+\left(1-x\right)^{2}\right],\end{align*}
 and \[
b_{0}=\frac{1}{4\pi}\left(\frac{11}{3}C_{A}-\frac{4}{3}T_{R}n_{\mathrm{lf}}\right)\,,\]
 with all other $p_{i,\widetilde{ic}}$ and $C_{i,\widetilde{ic}}$
being equal to zero.

\bibliographystyle{JHEP}
\bibliography{Herwig++}

\end{document}